\documentclass[prb,aps,12pt]{revtex4}
\newcommand{\be}{\begin{eqnarray}}
\newcommand{\ee}{\end{eqnarray}}
\usepackage{graphicx}

\begin{document}

\title{Random Diffusion Model}
\author{Gene F. Mazenko }
\address{The James Franck Institute and the Department of Physics\\
The University of Chicago\\
Chicago, Illinois 60637}
\date{\today}
\maketitle
\centerline{  ABSTRACT}
%
We study here the random diffusion model.  This is a continuum model
for a conserved  scalar density field $\phi$ driven by diffusive dynamics. The interesting
feature of the dynamics is that the {\it bare} diffusion coefficient $D$
is density dependent.  In the simplest case $D=\bar{D}+D_{1}\delta \phi $
where $\bar{D}$ is the constant average diffusion constant.  In the case where
the driving effective Hamiltonian is quadratic the model can be treated
using perturbation theory in terms of the single nonlinear coupling $D_{1}$.
We develop perturbation theory to fourth order in $D_{1}$.

The are two ways of analyzing this perturbation theory.  In one approach, developed
by Kawasaki, at one-loop order one finds mode coupling theory with an ergodic-nonergodic
transition.  An alternative more direct interpretation at one-loop order leads
to a slowing down as the nonlinear coupling increases.
Eventually one hits a critical coupling where the time decay becomes algebraic.
Near this critical coupling a weak peak develops at a wavenumber
well above the peak at $q=0$ associated with the conservation law.  
The width of this peak in Fourier space decreases with time
and can be identified with a characteristic kinetic length which grows with a power
law in time. For stronger coupling the system becomes metastable and then unstable.

At two-loop order it is shown that the ergodic-nonergodic transition is not
supported.  It is demonstrated that the {\it critical} properties of the direct
approach survive going to higher order in perturbation theory.


 

\newpage

\section{Introduction}

We study here a dynamical system, the random diffusion model (RDM), under going diffusive
dynamics with a field dependent diffusion coefficient.  This model serves as a very
simple model for the dynamics of the density field in colloidal systems.  The
motivation for studying this model comes from facilitated spin models where the
kinetic coefficient is density dependent and leads to significant slowing down
for dense systems.  Here the expectation is that a properly chosen bare diffusion
coefficient leads to significant slowing down as the density increases.

There has been much speculation but relatively few solid results  
in establishing the existence of a mode coupling theory (MCT)\cite{MCT}  
ergodic-nonergodic (ENE ) 
transition in
field theoretic models of the liquid-glass transition.  The RDM is
a candidate for the simplest such model.
At one-loop order\cite{loop} the self-consistent theory can be organized in two ways.
One approach, involving a rearrangement due to Kawasaki\cite{KKA}, leads to
a conventional ENE transition for a large enough nonlinear coupling.  
A second, more direct approach,
leads, on increasing the same dimensionless coupling, to  
slowing down of the system for wavenumbers well away from zero.
One eventually reaches a coupling where the system produces a peaked
dynamic structure factor at wavenumbers well away from zero.  
We can call this a prepeak since we expect it to show up at wavenumbers 
below those characterizing the first peak in the static structure factor.
The width
of this peak corresponds to a kinetic length which increases 
algebraically with time.

The key point here is that the RDM is simple enough
that the associated perturbation theory, in terms of a single
dimensionless expansion parameter, can be worked out at two-loop
order.  We find that the ENE transition does not survive in the
two-loop theory for reasons that may be generic.  In a direct
self-consistent treatment of the model we find that the new
peaked state is maintained at two-loop
order.

We focus on field theoretical models for the dynamics of 
dense fluids since they
offer the best hope of a self-consistent theory.  This hope  
includes the possibilities
of higher-order computation and the determination of four-point 
correlation functions.  
As mentioned above,
the computation at higher order is necessary to establish the stability
of any ENE transition found at one-loop order.  We also want to compute
multi-point correlation functions since there is speculation\cite{BB} that
they offer information about a growing length as one approaches the
ENE transition.

Despite a number of papers (see below) discussing 
mode coupling theory (MCT) from the point of view
of field theoretical models,
the situation is unclear. We do not really know which models have a transition 
and which do not.
There is work\cite{DMRT} suggesting that nonlinear fluctuating hydrodynamics 
offers viable
kinetic models for studying the dynamics of dense fluids and can lead
to the ENE transition.
Das and Mazenko (DM)\cite{DM}
 introduced a field theoretical model with density and
momentum fields.  They showed from general nonperturbative considerations
and a one loop calculation that the conventional mode coupling transition
is cutoff.
Schmitz et al\cite{Dufty} found a cutoff in a slightly simpler model.
Cates and Ramaswamy\cite{CR}, using heuristic reasoning, argue
that these cutoffs are not effective
in the DM model.

A set of slightly simpler models (involving only the density field) 
were introduced by Dean\cite{Dean} and Kawasaki\cite{kaw}
to describe the  overdamped
diffusive dynamics  in
 colloidal systems.
Miyasaki and  Reichman\cite{RFT}  studied the Dean-Kawasaki (DK)
 model using the MSR method\cite{DDP}.
They found  a nonlinear fluctuation dissipation theorem (FDT)
connecting propagators and correlation functions
which made even the one loop theory difficult to interpret.
Things are complicated by the use of the MSR method which requires
field doubling in carrying out the perturbation theory.
Andreanov, Biroli, and Lefevre (ABL)\cite{ABL}
document that nonlinear terms in the effective hamiltonian 
generate a nonlinear FDT
and make systematic perturbation theory very difficult.  They suggest
introducing auxiliary fields to solve this problem but were unable
to construct a sensible one loop approximation.
Kawasaki and Kim\cite{KK}, taking a similar approach, were able to find
a one-loop approximation which does lead to a ENE transition.

Given the uncertainty in the analysis of the DM and DK models
(Do they have an ENE transition?), we need to analyze a simpler
model which does have an ENE transition at one loop order.
The idea is to check  whether this solution is stable at two-loop
order. 
These technical problems suggest that one needs to study a 
model that has a linear FDT.

The RDM
is related to the DK model. It is the simplest nontrivial
realization of the hindered diffusion model\cite{GFM}
introduced earlier.
The physical motivation for this model is from facilitated 
spin models\cite{FA,chan,WBG1,WBG2,JMS} where the kinetic coefficient 
in a lattice model dynamics depends on the local environment in a 
constraining manner.
In a continuum model, with a conserved density, the analog is a 
density dependent diffusion coefficient.  In both models one can 
have strong kinetic slowing down despite having trivial, 
"noninteracting" static equilibrium behavior.

The random diffusion model model has a single identifiable 
small parameter.  As discussed in reference\onlinecite{GFM}
 the source
 of nonlinearities are in the density dependence of the bare
 diffusion coefficient. In the simplest case the bare diffusion coefficient
 is of the form
 \be
 D(\phi )=D_{0}+D_{1}\phi
 \ee
and the perturbation theory is in powers of $D_{1}$.
In the simplifying case where we assume the static structure, in
our coarse grained system, is a constant up to a cutoff $\Lambda$,
called here the structureless approximation,
we find that the dimensionless coupling constant
is given by
\be
g=\frac{1}{2}(\frac{D_{1}}{D_{0}+\phi_{0}D_{1}})^{2}S
\ee
where $\phi_{0}=\langle \phi \rangle $ is the average density and
\be
S=\langle (\delta \phi )^{2}\rangle
\ee
is the local fluctuation in the density\cite{Scom}.

The RDM 
shows an ENE transition at one loop ($g^{1}$) order but the transition
appears to be inconsistent with the theory at two-loop $g^{2}$ order.
There is an alternate  more direct approach to the perturbative 
treatment of the RDM. 
In this model,
as a function of increasing 
coupling, $g$, one finds a slowing
down.  For coupling  $ g\leq g^{*}$, where $g^{*}$ is the critical coupling, 
there is a cross over from exponential to algebraic time decay
for a band of wave numbers away from zero wavenumber.  Indeed certain 
wavenumber components decay to zero more slowly than others and a small 
amplitude
peak develops in the dynamic structure factor.  
This structural
peak has the form
\be
C_{peak}(Q,t) =Ae^{-B(Q-Q_{0})^{2}}
\nonumber
~~~.
\ee
The width of this
small amplitude peak decreases with time thus giving a length $\sqrt{B}$
which increases algebraically with time.   
The amplitude  $A$
decreases with time and, after a brief initial transient, $Q_{0}$, is 
time-independent.
$A$ shows power-law behavior in time for
$g$ near $g^{*}$.

For $ g \geq g^{*}$ the system is slow but eventually unstable.
The small peak contribution, for long enough time,
begins to grow and the system eventually blows up.
It is not unreasonable to assume that the unstable system
represents the nucleating solid phase.  The model must be extended
with the appropriate static behavior if one is to stabilize the
nucleated solid phase.

It has been traditional to use the MSR method to develop perturbation
theory for dynamical models such as the RDM studied here.  This method has
the distinct advantage that perturbation theory can be developed
in terms of the physical correlation and response functions. In the RDM the
correlation and response functions are linearly related and the calculation
at one-loop order is manageable.  The calculation at two-loop order
is extremely complicated by sums over the labels differentiating fields
from response fields.  The Fokker-Planck description has the advantage that
the bare perturbation theory expansion is formally transparent, the static
behavior is easy to sort out, and one does not have the frequency integrals
found in the MSR method.  The disadvantage is that one has to replace
the bare correlation functions by their full counterparts by hand.  One
is helped by the knowledge from the MSR approach that such a renormalization
(resummation) exists.

\newpage

\section{Random Diffusion Model}

We discuss our model in the context of a Fokker-Planck description.
The equilibrium intermediate dynamic structure factor is
given by
\be
C({\bf q}_{1},{\bf q}_{2};t)=
\int {\cal D}(\phi )W_{\phi}\phi ({\bf q}_{2}) e^{-\tilde{D}_{\phi}t}\phi ({\bf q}_{1})
\nonumber
\ee
\be
=\langle \phi ({\bf q}_{2})e^{-\tilde{D}_{\phi}t}\phi ({\bf q}_{1})\rangle
\ee
where $\phi ({\bf q}_{1})$ is the Fourier transform of the
fundamental field $\delta \phi$ in the theory,
the equilibrium probability distribution is given by
\be
W_{\phi}=\frac{e^{-\beta{\cal H}_{\phi}}}{Z}
\ee
where the effective Hamiltonian ${\cal H}_{\phi}$ can be taken to be quadratic
in $\phi$:
\be
{\cal H}_{\phi}=\int ~ d^{d}x_{1}d^{d}x_{2}\frac{1}{2}
\delta\phi ({\bf x}_{1})\chi^{-1} ({\bf x}_{1}-{\bf x}_{2})\delta\phi ({\bf x}_{2})
\label{eq:48}
\ee
and $\delta\phi ({\bf x}_{1})=\phi ({\bf x}_{1})-\phi_{0}$.
The adjoint Fokker-Planck operator for our model is given by
\be
\tilde{D}_{\phi}=\int d^{d}x \int d^{d}y
\left[\frac{\delta {\cal H}_{\phi}}{\delta \phi ({\bf x})}
-k_{B}T\frac{\delta}{\delta \phi ({\bf x})}\right]
\Gamma_{\phi}({\bf x},{\bf y})
\frac{\delta }{\delta \phi ({\bf y})}
\label{eq:7}
\ee
where \cite{MMN}
\be
\Gamma_{\phi}({\bf x},{\bf y})=
\nabla_{x}\cdot \nabla_{y}
\left(D(\phi ({\bf x}) )
\delta \left({\bf x}-{\bf y}\right)
\right)
\label{eq:30}
\ee
and the bare-diffusion coefficient is taken to be of the
simplest nontrivial form
\be
D(\phi ({\bf x}))=D_{0}+D_{1}\phi ({\bf x})
~~~.
\ee
A more complicated and physical form for $D(\phi )$ was studied in Ref.\onlinecite{GFM}.

Our model can also be written as a field theory of the
MSR\cite{DDP} type.  
The MSR action is given in this case by
\be
A=\int ~ d^{d}x ~dt\left[\beta^{-1} D(\phi )(\nabla \hat{\phi} )^{2}
+i\hat{\phi}\left[\dot{\phi}-\nabla_{i}\left(
D(\phi )\nabla_{i}\frac{\delta{\cal H}_{\phi}}{\delta \phi}\right)\right]\right]
~~~.
\ee
where $\hat{\phi}$ is the MSR auxiliary response field.

\section{Memory Function Formalism}

We use here a memory function formalism in the Fokker-Planck description.
This approach was first fully developed in Ref.\onlinecite{FRKT} for kinetic theory 
and later applied \cite{MRT} to the fluctuating nonlinear 
hydrodynamics of smectic A 
liquid crystals.  A significant advantage of the method is that it allows
one to treat interactions expressed in terms of static averages. Thus, in
the present problem, it is static averages not the bare diffusion coefficient
which appears in the theory.  The structure of this type of theory was
investigated in some detail by Anderson\cite{hca}.

Let us work with the Fourier-Laplace transformed time correlation function
\be
C({\bf q}_{1},{\bf q}_{1};z)=-i\int_{0}^{\infty}dt e^{izt}C({\bf q}_{1},{\bf q}_{1};t)
\nonumber
\ee
\be
=\langle \phi ({\bf q}_{2})R(z)\phi ({\bf q}_{1})\rangle
\label{eq:12}
~~~,
\ee
where the resolvant operator is given by
\be
R(z)=-i\int_{0}^{\infty}dte^{i[z+i\tilde{D}_{\phi}]t}
=[z+i\tilde{D}_{\phi}]^{-1}
~~~.
\label{eq:13}
\ee
Using the identity
\be
zR(z)=1-R(z)i\tilde{D}_{\phi}
\label{eq:42}
\ee
in Eq.(\ref{eq:12}),
leads to the 
kinetic equation
\be
zC({\bf q}_{1},{\bf q}_{2};z)
+\int\frac{d^{d}k_{1}}{(2\pi )^{d}}
K({\bf q}_{1},{\bf k}_{1};z)C({\bf k}_{1},{\bf q}_{2};z)
=\tilde{C}({\bf q}_{1},{\bf q}_{2})
\label{eq:14}
~~~.
\ee
The memory function, $K$, is given by
\be
\Gamma ({\bf q}_{1},{\bf q}_{2};z)=(2\pi )^{d}\delta ({\bf q}_{1}-{\bf q}_{2})
\Gamma ({\bf q}_{1};z)
=\int\frac{d^{d}k_{1}}{(2\pi )^{d}}
K({\bf q}_{1},{\bf k}_{1};z)\tilde{C}({\bf k}_{1},{\bf q}_{2})
\nonumber
\ee
\be
=\Gamma^{(s)} ({\bf q}_{1},{\bf q}_{2})
+\Gamma ^{(d)} ({\bf q}_{1},{\bf q}_{2};z)
~~~.
\ee
The static part of the memory function is
given by
\be
\Gamma^{(s)} ({\bf q}_{1},{\bf q}_{2})
=\langle \phi ({\bf q}_{2})i\tilde{D}_{\phi}\phi ({\bf q}_{1})\rangle
=\langle \phi ({\bf q}_{2})v({\bf q}_{1})\rangle
\ee
where the {\it current} $v$ is defined by
\be
v({\bf q}_{1})=i\tilde{D}_{\phi}\phi ({\bf q}_{1})
~~~,
\ee
and the dynamic part of the memory function is given by
\be
\Gamma^{(d)}({\bf q}_{1},{\bf q}_{2};z)=
\bar{\Gamma}({\bf q}_{1},{\bf q}_{2};z)+
\Gamma_{sub}({\bf q}_{1},{\bf q}_{2};z)
\label{eq:67a}
\ee
where
\be
\bar{\Gamma}({\bf q}_{1},{\bf q}_{2};z)
=-\langle v({\bf q}_{2})R(z)v({\bf q}_{1})\rangle
\label{eq:19}
\ee
and the {\it subtraction} part is given by
\be
\Gamma_{sub}({\bf q}_{1},{\bf q}_{2};z)=
\int\frac{d^{d}k_{2}}{(2\pi )^{d}}\int\frac{d^{d}k_{1}}{(2\pi )^{d}}
W({\bf q}_{2},{\bf k}_{2};z)
C^{-1}({\bf k}_{2},{\bf k}_{1};z)W({\bf q}_{1},{\bf k}_{1})
\label{eq:21}
\ee
where
\be
W({\bf q}_{1},{\bf k}_{1};z)=
\langle \phi ({\bf k}_{1})R(z)v({\bf q}_{1})\rangle
~~~.
\label{eq:21a}
\ee

Using standard arguments we can show that the physical diffusion coefficient is
given by:
\be 
D_{p}=\lim_{z\rightarrow 0}\lim_{q\rightarrow 0}-i\frac{\beta}{q^{2}}
\Gamma (q,z)
~~~.
\ee

\section{Bare Perturbation Theory}

\subsection{Two-time quantities}

In this section we show how to set up perturbation theory
for the dynamic structure factor
\be
C({\bf q}_{1},{\bf q}_{2};z)
=\langle \phi ({\bf q}_{2})R(z)\phi ({\bf q}_{1})\rangle
\ee
where the resolvant operator is defined by Eq.(\ref{eq:13}).
We are then interested in carrying out perturbation theory where the 
FP operator can be written as the sum
\be
\tilde{D}_{\phi} = \tilde{D}_{\phi}^{(0)}+\tilde{D}_{\phi}^{(I)}
~~~,
\ee
where the zeroth-order contribution is given by Eq.(\ref{eq:4})
with $\Gamma_{\phi}$ replaced by
\be
\Gamma_{\phi }^{(0)}(x,y)=\bar{D}
\nabla_{x}\cdot\nabla_{y}\delta ({\bf x}-{\bf y})
\ee
and the {\it interacting} contribution is given by Eq.(\ref{eq:7})
with  $\Gamma_{\phi}$ replaced by $\Delta\Gamma_{\phi}$
given by 
\be
\Delta
\Gamma_{\phi}({\bf x},{\bf y})=
\nabla_{x}\cdot \nabla_{y}
\left(D_{1}\delta \phi ({\bf x}) 
\delta \left({\bf x}-{\bf y}\right)
\right)
~~~.
\label{eq:6a}
\ee
We then use the operator identity
\be
R(z)=[z+i\tilde{D}_{\phi}^{(0)}+i\tilde{D}_{\phi}^{(I)}]^{-1}
\nonumber
\ee
\be
=[z+i\tilde{D}_{\phi}^{(0)}]^{-1}+
[z+i\tilde{D}_{\phi}^{(0)}]^{-1}
(-i\tilde{D}_{\phi}^{(I)})
[z+i\tilde{D}_{\phi}^{(0)}+i\tilde{D}_{\phi}^{(I)}]^{-1}
\nonumber
\ee
\be
=R_{0}(z)+R_{0}(z)(-i\tilde{D}_{\phi}^{(I)})R(z)
\ee
which defines the zeroth-order resovent
\be
R_{0}(z)=[z+i\tilde{D}_{\phi}^{(0)}]^{-1}
~~~.
\ee
Using this result in the correlation function and iterating
gives
\be
R(z)=R_{0}(z)-R_{0}(z)i\tilde{D}_{\phi}^{(I)}R_{0}(z)
+R_{0}(z)i\tilde{D}_{\phi}^{(I)}R_{0}(z)
i\tilde{D}_{\phi}^{(I)}R_{0}(z)+\ldots
\nonumber
\ee
\be
=R_{0}(z)\sum_{n=0}^{\infty}[-i\tilde{D}_{\phi}^{(I)}R_{0}(z)]^{n}
~~~.
\ee
For correlation functions we have the expansion
\be
C({\bf q}_{1},{\bf q}_{2};z)=\langle \phi({\bf q}_{2} )R(z)\phi ({\bf q}_{1})\rangle
\nonumber
\ee
\be
=\sum_{n=0}^{\infty}C^{(n)}({\bf q}_{1},{\bf q}_{2};z)
\ee
where order by order
\be
C^{(0)}({\bf q}_{1},{\bf q}_{2};z)
=\langle \phi ({\bf q}_{2}) )R_{0}(z)\phi ({\bf q}_{1})\rangle
\ee
\be
C^{(1)}({\bf q}_{1},{\bf q}_{2};z)
=\langle \phi({\bf q}_{2} )R_{0}(z)[-i\tilde{D}_{\phi}^{(I)}R_{0}(z)]
\phi ({\bf q}_{1})\rangle
\ee
\be
C^{(2)}({\bf q}_{1},{\bf q}_{2};z)=\langle \phi ({\bf q}_{2})
R_{0}(z)[-i\tilde{D}_{\phi}^{(I)}R_{0}(z)]^{2}
\phi ({\bf q}_{1})\rangle
\ee
\be
C^{(n)}({\bf q}_{1},{\bf q}_{2};z)
=\langle \phi ({\bf q}_{2}) R_{0}(z)
[-i\tilde{D}_{\phi}^{(I)}R_{0}(z)]^{n}
\phi ({\bf q}_{1})\rangle
~~~.
\ee
The first step in the analysis is evaluate $R_{0}(z)\phi ({\bf q}_{1})$.  In
Appendix A we show
\be
R_{0}(z)\phi (q_{1})=T_{0}(q_{1},z)\phi (q_{1})
\ee
where
\be
T_{0}(q_{1},z)=[z+iL_{0}(q_{1})]^{-1}
\ee
and $L_{0}(q)$ is given by Eq.(\ref{eq:84}) below.

On the left in each expression for $C^{(n)}$
we have
\be
\langle \phi (q_{2})R_{0}(z)\ldots \rangle
=\langle [R_{0}(z)\phi (q_{2})]\ldots \rangle
=T_{0}(q_{2},z)\langle \phi (q_{2})\ldots \rangle
~~~.
\ee

At the various orders we have
\be
C^{(0)}(q_{1},q_{2}),z)=T_{0}(q_{1},z)\langle \phi (q_{2})\phi (q_{1})\rangle
\label{eq:38}
\ee
\be
C^{(1)}(q_{1},q_{2}),z)=T_{0}(q_{2},z)\langle \phi(q_{2})
[-i\tilde{D}_{\phi}^{(I)}]\phi (q_{1})\rangle
T_{0}(q_{1},z)
\ee
\be
C^{(n)}(q_{1},q_{2}),z)=T_{0}(q_{2},z)\langle \phi(q_{2})
[-i\tilde{D}_{\phi}^{(I)}R_{0}(z)]^{(n-1)}[-i\tilde{D}_{\phi}^{(I)}]
\phi (q_{1})\rangle T_{0}(q_{1},z)
~~~.
\ee

The zeroth order solution is explicit after identifying
\be
\tilde{C}^{(0)}(q_{1},q_{2}))
=\langle \phi (q_{2})\phi (q_{1})\rangle
=k_{B}T\chi (q_{1})
(2\pi )^{d}\delta ({\bf q}_{1}+{\bf q}_{2})
~~~.
\ee

If we introduce the interaction part of the current,
\be
v^{(I)}(q_{1})=i\tilde{D}_{\phi}^{(I)}\phi (q_{1})
~~~,
\ee
then for the higher-order contributions
\be
C^{(1)}(q_{1},q_{2}),z)=-T_{0}(q_{2},z)\langle
\phi (q_{2})
v^{(I)}(q_{1})\rangle T_{0}(q_{1},z) 
\label{eq:34}
\ee
\be
C^{(2)}(q_{1},q_{2}),z)=T_{0}(q_{2},z)\langle
v^{(I)} (q_{2}) R_{0}(z)
v^{(I)}(q_{1})\rangle T_{0}(q_{1},z)
\label{eq:35}
\ee
and
\be
C^{(n)}(q_{1},q_{2}),z)=T_{0}(q_{2},z)\langle
v^{(I)} (q_{2}) R_{0}(z)
[-i\tilde{D}_{\phi}^{(I)}R_{0}(z)]^{(n-2)}
v^{(I)}(q_{1})\rangle T_{0}(q_{1},z)
~~~.
\ee

Let us look at the nonlinear term $v^{(I)}(q_{1})$. In coordinate space:
\be
v^{(I)}({\bf x}_{1})
=i\int d^{d}x_{2}\Delta\Gamma_{\phi}({\bf x}_{1},{\bf x}_{2})
\frac{\delta}{\delta \phi ({\bf x}_{2})} {\cal H}_{\phi}
\label{eq:70}
\ee
where $\Delta\Gamma_{\phi}$ is given by Eq.(\ref{eq:6a}) and
${\cal H}_{\phi}$ by Eq.(\ref{eq:48}).  Inserting these expressions
into Eq.(\ref{eq:70}) leads to the cubic vertex
\be
v^{(I)}({\bf x}_{1})=-i\sum_{\alpha}\nabla_{x_{1}}^{\alpha}
\left(D_{1}\delta \phi ({\bf x}_{1})
\tilde{\phi}_{\alpha}({\bf x}_{1})\right)
~~~,
\ee
where
\be
\tilde{\phi}_{\alpha}({\bf x}_{1})=\nabla_{x_{1}}^{\alpha}
\int~d^{d}x_{3}\chi^{-1}({\bf x}_{2}-{\bf x}_{3})\delta\phi ({\bf x}_{3})
~~~.
\label{eq:74a}
\ee
Taking the Fourier transform gives the {\it cubic} interaction:
\be 
v^{(I)}(q)=
\int \frac{d^{d}k_{2}}{(2\pi )^{d}}
\int \frac{d^{d}k_{3}}{(2\pi )^{d}}
V({\bf q},{\bf k}_{2},{\bf k}_{3})\phi({\bf k}_{2})\phi({\bf k}_{3})
\label{eq:97}
\ee
where
\be
V(q,k_{2},k_{3})=\frac{i}{2}D_{1}{\bf q}\cdot
\vec{\Lambda}(k_{2},k_{3})
(2\pi )^{d}\delta({\bf q}-{\bf k}_{2}-{\bf k}_{3})
\ee
and
\be
\vec{\Lambda}(k_{2},k_{3})={\bf k}_{2}\chi^{-1}(k_{2})
+{\bf k}_{3}\chi^{-1}(k_{3})
~~~.
\ee
It will also be useful to write
the vertex in the alternative form
\be
V(q,k_{2},k_{3})=V(q,k_{2})
(2\pi )^{d}\delta({\bf q}-{\bf k}_{2}-{\bf k}_{3})
\ee
where
\be
V(q,k_{2})=\frac{i}{2}D_{1}{\bf q}\cdot
\left({\bf k}_{2}\chi^{-1}(k_{2})
+({\bf q}-{\bf k}_{2})\chi^{-1}({\bf q}-{\bf k}_{2})\right)
~~~.
\ee

Since $v^{(I)}(q)$ is even in $\delta\phi$ we have
\be
\langle \phi (q_{2})v^{(I)}(q_{1})\rangle =0
\ee
and from Eq.(\ref{eq:34})
\be
C^{(1)}(q_{1},q_{2},z)=0.
~~~.
\ee

Turning to Eq.(\ref{eq:35}) we have at second order in the coupling:
\be
C^{(2)}(q_{1},q_{2},z)=T_{0}(q_{2},z)
\int \frac{d^{d}k_{1}}{(2\pi )^{d}}
\frac{d^{d}k_{2}}{(2\pi )^{d}}
\frac{d^{d}k_{3}}{(2\pi )^{d}}
\frac{d^{d}k_{4}}{(2\pi )^{d}}
\nonumber
\ee
\be
\times
V(q_{2},k_{3},k_{4})
M^{(2)}(k_{1},k_{2},k_{3},k_{4};z)
V(q_{1},k_{1},k_{2}) T_{0}(q_{1},z)
\label{eq:56}
\ee
and
\be
M^{(2)}(k_{1},k_{2},k_{3},k_{4};z)=
\langle \phi({\bf k}_{3})\phi({\bf k}_{4})R_{0}(z)\phi({\bf k}_{1})\phi({\bf k}_{2})
\rangle
~~~.
\label{eq:50}
\ee
At higher order we have generally
\be
C^{(n)}(q_{1},q_{2},z)=T_{0}(q_{2},z)
\int \frac{d^{d}k_{1}}{(2\pi )^{d}}
\frac{d^{d}k_{2}}{(2\pi )^{d}}
\frac{d^{d}k_{3}}{(2\pi )^{d}}
\frac{d^{d}k_{4}}{(2\pi )^{d}}
\nonumber
\ee
\be
\times
V(q_{2},k_{3},k_{4})
M^{(n)}(k_{1},k_{2},k_{3},k_{4};z)
V(q_{1},k_{1},k_{2}) T_{0}(q_{1},z)
\label{eq:99}
\ee
where
\be
M^{(n)}(k_{1},k_{2},k_{3},k_{4};z)=
\langle \phi({\bf k}_{3})\phi({\bf k}_{4})
R_{0}(z)[-i\tilde{D}_{\phi}^{(I)}R_{0}(z)]^{(n-2)}
\phi({\bf k}_{1})\phi({\bf k}_{2})
\rangle
\label{eq:63}
~~~.
\ee
To go further with $M^{(2)}$ we need the result from Appendix A 
\be
R_{0}(z)\phi ({\bf k}_{1})\phi ({\bf k}_{2})
=T_{0}({\bf k}_{1},{\bf k}_{2};z)[\phi ({\bf k}_{1})\phi ({\bf k}_{2})
-\tilde{C}({\bf k}_{1},{\bf k}_{2})]
+\frac{\tilde{C}({\bf k}_{1},{\bf k}_{2})}{z}
~~~.
\label{eq:53}
\ee
Using this result in $M^{(2)}({\bf k}_{1},{\bf k}_{2},{\bf k}_{3},{\bf k}_{4};z)$,
Eq.(\ref{eq:50}), we can then  
do the static average over gaussian fields to obtain
\be
M^{(2)}({\bf k}_{1},{\bf k}_{2},{\bf k}_{3},{\bf k}_{4};z)=T_{0}({\bf k}_{1},{\bf k}_{2};z)
\nonumber
\ee
\be
\times\Bigg[
[\tilde{C}({\bf k}_{1},{\bf k}_{3})\tilde{C}({\bf k}_{2},{\bf k}_{4})+
\tilde{C}({\bf k}_{1},{\bf k}_{4})\tilde{C}({\bf k}_{2},{\bf k}_{3})]
\nonumber
\ee
\be
+\tilde{C}({\bf k}_{3},{\bf k}_{4})\frac{\tilde{C}({\bf k}_{1},{\bf k}_{2})}{z}
\Bigg]
\label{eq:55}
~~~.
\ee
Putting this result back into Eq.(\ref{eq:56}) and using
the result
\be
V({\bf q},{\bf k}_{1},{\bf k}_{2})\tilde{C}({\bf k}_{1},{\bf k}_{2})
\approx q_{\alpha}\delta ({\bf q}-{\bf k}_{1}-{\bf k}_{2})
\delta ({\bf k}_{1}+{\bf k}_{2})=0
\label{eq:67}
\ee
gives
\be
C^{(2)}({\bf q}_{1},{\bf q}_{2},z)=T_{0}({\bf q}_{2},z)
\int \frac{d^{d}k_{1}}{(2\pi )^{d}}
\frac{d^{d}k_{2}}{(2\pi )^{d}}
\frac{d^{d}k_{3}}{(2\pi )^{d}}
\frac{d^{d}k_{4}}{(2\pi )^{d}}
V({\bf q}_{2},{\bf k}_{3},{\bf k}_{4})
\nonumber
\ee
\be
\times
T_{0}({\bf k}_{1},{\bf k}_{2};z)2
\tilde{C}({\bf k}_{1},{\bf k}_{3})\tilde{C}({\bf k}_{2},{\bf k}_{4})
V(q_{1},k_{1},k_{2}) T_{0}(q_{1},z)
~~~.
\ee
Using the $\delta$-functions in the vertices and static correlation
functions allows one to do three of the ${\bf k}$ integrals and obtain
\be
C^{(2)}(q_{1},q_{2},z)=
2(2\pi )^{d}\delta ({\bf q}_{1}+{\bf q}_{2})
T_{0}(-q_{1},z)\int \frac{d^{d}k_{1}}{(2\pi )^{d}}
V(-{\bf q}_{1},-{\bf k}_{1})
T_{0}({\bf k}_{1},{\bf q}_{1}-{\bf k}_{1};z)
\nonumber
\ee
\be
\times
\tilde{C}({\bf k}_{1})\tilde{C}({\bf q}_{1}-{\bf k}_{1})
V({\bf q}_{1},{\bf k}_{1})T_{0}(q_{1},z)
~~~.
\ee
We return to this expression below.

Going to the third-order contribution we must evaluate
\be
M^{(3)}(k_{1},k_{2},k_{3},k_{4};z)=
\langle \phi({\bf k}_{3})\phi({\bf k}_{4})
R_{0}(z)[-i\tilde{D}_{\phi}^{(I)}R_{0}(z)]
\phi({\bf k}_{1})\phi({\bf k}_{2})
\rangle
~~~.
\ee
since $\tilde{D}_{\phi}^{(I)}$ is odd in $\delta \phi$
it is easy to see that
\be
M^{(3)}(k_{1},k_{2},k_{3},k_{4};z)=0
\ee
and
\be
C^{(3)}(k_{1},k_{2},k_{3},k_{4};z)=0
~~~.
\ee

At fourth order we must evaluate
\be
M^{(4)}(k_{1},k_{2},k_{3},k_{4};z)=
\langle \phi({\bf k}_{3})\phi({\bf k}_{4})
R_{0}(z)[-i\tilde{D}_{\phi}^{(I)}R_{0}(z)]^{2}
\phi({\bf k}_{1})\phi({\bf k}_{2})
\rangle
\nonumber
\ee
\be
=\langle [i\tilde{D}_{\phi}^{(I)}R_{0}(z)\phi({\bf k}_{3})\phi({\bf k}_{4})]
R_{0}(z)
[i\tilde{D}_{\phi}^{(I)}R_{0}(z)\phi({\bf k}_{1})\phi({\bf k}_{2})]
\rangle
~~~.
\label{eq:109}
\ee
We find immediately, using Eq.(\ref{eq:53}), that
\be
i\tilde{D}_{\phi}^{(I)}R_{0}(z)\phi({\bf k}_{1})\phi({\bf k}_{2})
=T_{0}({\bf k}_{1},{\bf k}_{2};z)i\tilde{D}_{\phi}^{(I)}
\phi({\bf k}_{1})\phi({\bf k}_{2})
~~~.
\label{eq:66}
\ee
We have from appendix B
\be
i\tilde{D}_{\phi}^{(I)}\phi({\bf k}_{1})\phi({\bf k}_{2})
=v^{(I)}({\bf k}_{1})\phi({\bf k}_{2})
+v^{(I)}({\bf k}_{2})\phi({\bf k}_{1})+S({\bf k}_{1},{\bf k}_{2})
\label{eq:111}
\ee
where
\be
S({\bf k}_{1},{\bf k}_{2})=2i\beta^{-1}
D_{1}{\bf k}_{1}\cdot{\bf k}_{2}\phi({\bf k}_{1}+{\bf k}_{2})
~~~.
\ee
Using Eq.(\ref{eq:111}) twice in Eq.(\ref{eq:109}) gives
\be
M^{(4)}(k_{1},k_{2},k_{3},k_{4};z)=T_{0}({\bf k}_{1},{\bf k}_{2};z)
T_{0}({\bf k}_{3},{\bf k}_{4};z)N^{(4)}(k_{1},k_{2},k_{3},k_{4};z)
\label{eq:113}
\ee
where
\be
N^{(4)}(k_{1},k_{2},k_{3},k_{4};z)
=\langle [v^{(I)}({\bf k}_{3})\phi({\bf k}_{4})
+v^{(I)}({\bf k}_{4})\phi({\bf k}_{3})+S({\bf k}_{3},{\bf k}_{4})]
\nonumber
\ee
\be
\times R_{0}(z)
[v^{(I)}({\bf k}_{1})\phi({\bf k}_{2})
+v^{(I)}({\bf k}_{2})\phi({\bf k}_{1})+S({\bf k}_{1},{\bf k}_{2})]
\rangle
~~~.
\ee

After a significant amount of algebra
we have the explicit results
for $N^{(4)}$:
\be
N^{(4)}({\bf k}_{1},{\bf k}_{2},{\bf k}_{3},{\bf k}_{4};z)
=N_{sub}(12;34)+N_{R}(12;34)+N_{D}(12;34)
~~~,
\label{eq:74}
\ee
where
we must symmetrize
\be
N_{R,D}(12;34)=\bar{N}_{R,D}(12;34)+\bar{N}_{R,D}(21;34)+
\bar{N}_{R,D}(12;43)+\bar{N}_{R,D}(21;43)
\ee
with
\be
\bar{N}_{R}(12;34)=
\int \frac{d^{d}k_{5}}{(2\pi )^{d}}\frac{d^{d}k_{6}}{(2\pi )^{d}}
\frac{d^{d}k_{7}}{(2\pi )^{d}}\frac{d^{d}k_{8}}{(2\pi )^{d}}
\nonumber
\ee
\be
\times
V({\bf k}_{3},{\bf k}_{7},{\bf k}_{8})
V({\bf k}_{1},{\bf k}_{5},{\bf k}_{6})2
T_{0}({\bf k}_{5},{\bf k}_{6},{\bf k}_{2})
\tilde{C}(24)\tilde{C}(57)\tilde{C}(68)
\label{eq:117}
\ee
where we use the notation $\tilde{C}(68)=\tilde{C}({\bf k}_{6},{\bf k}_{8})$,
\be
\bar{N}_{D}(12;34)=
\int \frac{d^{d}k_{5}}{(2\pi )^{d}}\frac{d^{d}k_{6}}{(2\pi )^{d}}
\frac{d^{d}k_{7}}{(2\pi )^{d}}\frac{d^{d}k_{8}}{(2\pi )^{d}}
\nonumber
\ee
\be
\times
V({\bf k}_{3},{\bf k}_{7},{\bf k}_{8})
V({\bf k}_{1},{\bf k}_{5},{\bf k}_{6})2
T_{0}({\bf k}_{5},{\bf k}_{6},{\bf k}_{2})
2\tilde{C}(27)\tilde{C}(46)\tilde{C}(58)
\label{eq:118}
\ee
and
\be
N_{sub}(12;34)
=\int \frac{d^{d}k_{5}}{(2\pi )^{d}}(-4)T_{0}({\bf k}_{5})
V({\bf k}_{5},{\bf k}_{1},{\bf k}_{2})
V(-{\bf k}_{5},{\bf k}_{3},{\bf k}_{4})
\nonumber
\ee
\be
\times
\tilde{C}^{-1}({\bf k}_{5})
\tilde{C} ({\bf k}_{1})\tilde{C} ({\bf k}_{2})
\tilde{C}({\bf k}_{3})\tilde{C} ({\bf k}_{4})
~~~.
\label{eq:78}
\ee
Put Eqs.(\ref{eq:117}), (\ref{eq:118}) and (\ref{eq:78}) into 
(\ref{eq:74}); in turn put Eq.(\ref{eq:74}) into Eq.(\ref{eq:113})
and Eq.(\ref{eq:113}) back into Eq.(\ref{eq:99}) with $n=4$ to obtain
an explicit expression for $C^{(4)}({\bf q}_{1},{\bf q}_{2};z)$. 

Thus we have explicit expressions for 
$C^{(n)}({\bf q}_{1},{\bf q}_{2},z)$ for $n < 5$.
We use these results below.

\newpage

\section{Evaluation of Memory Function in Perturbation Theory}

\subsection{Static part of Memory Function}

We want to determine the memory function $K$ in a perturbation theory in powers of
$D_{1}$.  
We find that the static part of the memory function is
of zeroth order in $D_{1}$, while the dynamic part of the memory function
begins at second order in $D_{1}$.

The static part of the memory function is determined by the equilibrium
average
\be
\int d^{d}w~ K^{(s)}({\bf x},{\bf w})\tilde{C}({\bf w},{\bf y})=
\Gamma^{(s)}({\bf x},{\bf y})=\langle \delta\phi ({\bf y})i\tilde{D}_{\phi}
\delta\phi ({\bf x})
\rangle
.
\ee
In evaluating this static average it is very useful to use the identity:
\be
\langle B\tilde{D}_{\phi}A\rangle
=\beta^{-1}\int d^{d}x_{1}d^{d}x_{2}
\langle\frac{\delta B}{\delta \phi ({\bf x}_{1})}
\Gamma_{\phi}({\bf x}_{1},{\bf x}_{2})
\frac{\delta A}{\delta \phi ({\bf x}_{2})}\rangle
\ee
and we obtain 
\be
\Gamma^{(s)}({\bf x},{\bf y})=i\beta^{-1}
\langle\Gamma_{\phi}({\bf x},{\bf y})\rangle
~~~.
\ee
It is easy to show, using Eq.(\ref{eq:30}) for $\Gamma_{\phi}$, that
\be
\Gamma^{(s)}({\bf x},{\bf y})
=i\beta^{-1}\nabla_{x}\cdot\nabla_{y}
\left[\bar{D}\delta({\bf x}-{\bf y})\right]
\label{eq:59}
\ee
where the average diffusion coefficient is given by
\be
\bar{D}=\langle D(\phi )\rangle = D_{0}+D_{1}\phi_{0}
\ee
where $\phi_{0}=\langle \phi\rangle$.

Taking the Fourier transform of Eq.(\ref{eq:59}) and multiplying by
$\tilde{C}^{-1}(k)$ gives the static part of the memory function:
\be
K^{(s)}(k)=i\beta^{-1}k^{2}\bar{D}\tilde{C}^{-1}(k)
=ik^{2}\bar{D}\chi^{-1} (k)\equiv i L_{0}(k)
~~~.
\label{eq:84}
\ee
Putting this result back into Eq.(\ref{eq:14}), dropping the dynamic part of
the memory function and inverting the Laplace
transform, gives the zeroth order approximation for the
density-density time correlation function
\be
C_{0}(k,t)=e^{-k^{2}\bar{D}\chi^{-1} (k)t}\tilde{C}(k)=e^{-L_{0}(k)t}\tilde{C}(k)
\label{eq:66}
\ee
which agrees with the lowest-order result found previously.

\subsection{Dynamic Part of Memory Function}

The dynamic part of the memory function for the dynamic
structure factor is the sum of
two pieces:
\be
\Gamma^{(d)}({\bf q}_{1},{\bf q}_{2};z)=
\bar{\Gamma}({\bf q}_{1},{\bf q}_{2};z)+
\Gamma_{sub}({\bf q}_{1},{\bf q}_{2};z)
\label{eq:67a}
\ee
where the direct contribution is given by\cite{1PI}
\be
\bar{\Gamma}({\bf q}_{1},{\bf q}_{2};z)
=-\langle v^{I}({\bf q}_{2})R(z)v^{I}({\bf q}_{1})\rangle
\ee
and the {\it subtraction} part is given by Eq.(\ref{eq:21}) with
\be
W({\bf q}_{1},{\bf k}_{1};z)=
\langle \phi ({\bf k}_{1})R(z)v^{I}({\bf q}_{1})\rangle
\ee
and
\be
v^{I}({\bf q}_{1})=i\tilde{D}^{I}_{\phi}\phi ({\bf k}_{1})
~~~.
\ee
We see that the dynamic part of the memory function vanishes
at zeroth and first order in $D_{1}$.

We show here how to evaluate $\Gamma^{(d)}$ in
perturbation theory up to fourth order.

\subsection{Direct contribution}

Focussing first on the direct contribution to the memory
function we have
\be
\bar{\Gamma}({\bf q}_{1},{\bf q}_{2};z)
=-\int \frac{d^{d}k_{1}}{(2\pi )^{d}}
\frac{d^{d}k_{2}}{(2\pi )^{d}}
\frac{d^{d}k_{3}}{(2\pi )^{d}}
\frac{d^{d}k_{4}}{(2\pi )^{d}}
\nonumber
\ee
\be
\times
V({\bf q}_{2},{\bf k}_{3},{\bf k}_{4})
V({\bf q}_{1},{\bf k}_{1},{\bf k}_{2})
M({\bf k}_{1},{\bf k}_{2},{\bf k}_{3},{\bf k}_{4};z)
\ee
where $M$ is defined by Eq.(\ref{eq:63}).  In perturbation theory
\be
\bar{\Gamma}^{(n)}({\bf q}_{1},{\bf q}_{2};z)
=-\int \frac{d^{d}k_{1}}{(2\pi )^{d}}
\frac{d^{d}k_{2}}{(2\pi )^{d}}
\frac{d^{d}k_{3}}{(2\pi )^{d}}
\frac{d^{d}k_{4}}{(2\pi )^{d}}
\nonumber
\ee
\be
\times
V({\bf q}_{2},{\bf k}_{3},{\bf k}_{4})
V({\bf q}_{1},{\bf k}_{1},{\bf k}_{2})
M^{(n)}({\bf k}_{1},{\bf k}_{2},{\bf k}_{3},{\bf k}_{4};z)
~~~.
\ee
At second order $M^{(2)}$ is given by Eq.(\ref{eq:55}).
Using Eq.(\ref{eq:21}), this reduces to
\be
\bar{\Gamma}^{(2)}({\bf q}_{1},{\bf q}_{2};z)
=-\int \frac{d^{d}k_{1}}{(2\pi )^{d}}
\frac{d^{d}k_{2}}{(2\pi )^{d}}
\frac{d^{d}k_{3}}{(2\pi )^{d}}
\frac{d^{d}k_{4}}{(2\pi )^{d}}
V({\bf q}_{2},{\bf k}_{3},{\bf k}_{4})
V({\bf q}_{1},{\bf k}_{1},{\bf k}_{2})
\nonumber
\ee
\be
\times
T_{0}({\bf k}_{1},{\bf k}_{2};z)
2
\tilde{C}({\bf k}_{1},{\bf k}_{3})\tilde{C}({\bf k}_{2},{\bf k}_{4})
\label{eq:139}
~~~.
\ee
The  third-order contribution vanishes, while the
fourth-order contribution is given by
\be
\bar{\Gamma}^{(4)}({\bf q}_{1},{\bf q}_{2};z)
=-\int \frac{d^{d}k_{1}}{(2\pi )^{d}}
\frac{d^{d}k_{2}}{(2\pi )^{d}}
\frac{d^{d}k_{3}}{(2\pi )^{d}}
\frac{d^{d}k_{4}}{(2\pi )^{d}}
\nonumber
\ee
\be
\times
V({\bf q}_{2},{\bf k}_{3},{\bf k}_{4})T_{0}({\bf k}_{3},{\bf k}_{4};z)
V({\bf q}_{1},{\bf k}_{1},{\bf k}_{2})T_{0}({\bf k}_{1},{\bf k}_{2};z)
\nonumber
\ee
\be
\times
N^{(4)}({\bf k}_{1},{\bf k}_{2},{\bf k}_{3},{\bf k}_{4};z)
~~~.
\ee
$N^{(4)}$ is given by Eq.(\ref{eq:74}).
This contribution  divides naturally into three pieces.  One piece, $\bar{\Gamma}_{sub}^{(4)}$,
when added to $\Gamma_{sub}^{(4)}$, vanishes.  We have then that the
fourth-order contribution is given by
\be
\Gamma^{(4)}({\bf q}_{1},{\bf q}_{2};z)
=
\bar{\Gamma}^{(4)}_{R}({\bf q}_{1},{\bf q}_{2};z)
+
\bar{\Gamma}^{(4)}_{D}({\bf q}_{1},{\bf q}_{2};z)
\ee
where
\be
\bar{\Gamma}^{(4)}_{R}({\bf q}_{1},{\bf q}_{2};z)
=-\int \frac{d^{d}k_{1}}{(2\pi )^{d}}
\frac{d^{d}k_{2}}{(2\pi )^{d}}
\frac{d^{d}k_{3}}{(2\pi )^{d}}
\frac{d^{d}k_{4}}{(2\pi )^{d}}
\frac{d^{d}k_{5}}{(2\pi )^{d}}
\frac{d^{d}k_{6}}{(2\pi )^{d}}
\frac{d^{d}k_{7}}{(2\pi )^{d}}
\frac{d^{d}k_{8}}{(2\pi )^{d}}
\nonumber
\ee
\be
\times
V({\bf q}_{2},{\bf k}_{3},{\bf k}_{4})T_{0}({\bf k}_{3},{\bf k}_{4};z)
V({\bf q}_{1},{\bf k}_{1},{\bf k}_{2})T_{0}({\bf k}_{1},{\bf k}_{2};z)
\nonumber
\ee
\be
\times
4V({\bf k}_{3},{\bf k}_{7},{\bf k}_{8})
V({\bf k}_{1},{\bf k}_{5},{\bf k}_{6})2
T_{0}({\bf k}_{5},{\bf k}_{6},{\bf k}_{2})
\tilde{C}(24)\tilde{C}(57)\tilde{C}(68)
\label{eq:95}
\ee
and
\be
\bar{\Gamma}^{(4)}_{D}({\bf q}_{1},{\bf q}_{2};z)
=-\int \frac{d^{d}k_{1}}{(2\pi )^{d}}
\frac{d^{d}k_{2}}{(2\pi )^{d}}
\frac{d^{d}k_{3}}{(2\pi )^{d}}
\frac{d^{d}k_{4}}{(2\pi )^{d}}
\frac{d^{d}k_{5}}{(2\pi )^{d}}
\frac{d^{d}k_{6}}{(2\pi )^{d}}
\frac{d^{d}k_{7}}{(2\pi )^{d}}
\frac{d^{d}k_{8}}{(2\pi )^{d}}
\nonumber
\ee
\be
\times
V({\bf q}_{2},{\bf k}_{3},{\bf k}_{4})T_{0}({\bf k}_{3},{\bf k}_{4};z)
V({\bf q}_{1},{\bf k}_{1},{\bf k}_{2})T_{0}({\bf k}_{1},{\bf k}_{2};z)
\nonumber
\ee
\be
\times
4V({\bf k}_{1},{\bf k}_{5},{\bf k}_{6})
V({\bf k}_{3},{\bf k}_{7},{\bf k}_{8})2
T_{0}({\bf k}_{5},{\bf k}_{6},{\bf k}_{2})
2\tilde{C}(27)\tilde{C}(46)\tilde{C}(58)
~~~.
\label{eq:96}
\ee

\section{Bare Perturbation Theory at Second order}

\subsection{General Form}

Now that we have the perturbation theory results
we need to see the physical consequences.  We
begin with bare perturbation theory at second order.
For a general static structure factor,
$\Gamma^{(2)}(q,z)$ is given by Eq.(\ref{eq:139}).
After integrating over the
$\delta$-functions we have
\be
\Gamma^{(2)}(q,z)
=-2\int \frac{d^{d}k}{(2\pi )^{d}}
\left[V({\bf q},{\bf k})\right]^{2}
\frac{\tilde{C}({\bf k})\tilde{C}({\bf q}-{\bf k})}
{z+iL_{0}({\bf k})+iL_{0}({\bf q}-{\bf k})}
\label{eq:97}
\ee
where
\be
V({\bf q},{\bf k})=\frac{i}{2}D_{1}
{\bf q}\cdot\left[{\bf k}\chi^{-1}({\bf k})
+({\bf q}-{\bf k})\chi^{-1}({\bf q}-{\bf k})\right]
~~~.
\ee

\subsection{Structureless approximation}

In the structureless approximation we assume that the
static susceptibility is independent of wavenumber
\be
\chi^{-1}({\bf k})=r
\nonumber
\ee
and
introduce a large wavenumber cutoff $\Lambda$.
This approximation (model) is appealing for two reasons.
First, in this case,
the vertex simplifies to the form
\be
V({\bf q},{\bf k})=\frac{i}{2}D_{1}rq^{2}
~~~.
\ee
Second this model corresponds to a coarse-grained
system where one has integrated out short-distance
degrees of freedom including the first peak in the static structure factor.
Eq.(\ref{eq:97}) then becomes
\be
\Gamma^{(2)}(q,z)=\frac{1}{2}D_{1}^{2}q^{4}\beta^{-2}
\int^{\Lambda} \frac{d^{d}k}{(2\pi )^{d}}
\frac{1}{z+i\bar{D}r[k^{2}+({\bf q}-{\bf k})^{2}]}
~~~.
\ee
Letting ${\bf k}=\frac{{\bf q}}{2}+{\bf p}$ in the integral
gives
\be
\Gamma^{(2)}(q,z)=\frac{1}{2}D_{1}^{2}q^{4}\beta^{-2}
\int^{\Lambda} \frac{d^{d}p}{(2\pi )^{d}}
\frac{1}{z+i\bar{D}r[2p^{2}+q^{2}/2]}
\nonumber
~~~.
\ee
Doing the angular integral
\be
\Gamma^{(2)}(q,z)=
\frac{1}{2}D_{1}^{2}q^{4}\beta^{-2}K_{d}
\int_{0}^{\Lambda}p^{d-1}dp
\frac{1}{z+i\bar{D}r[2p^{2}+q^{2}/2]}
\ee
where
\be
K_{d}=\int \frac{d^{d}k}{(2\pi )^{d}} \delta (k-1)
~~~.
\ee
At this point we move to dimensionless variables.  If
we set $p=\Lambda x$ in the integral then
\be
\Gamma^{(2)}(q,z)=
\frac{1}{2}D_{1}^{2}q^{4}\beta^{-2}K_{d}\Lambda^{d}
\int_{0}^{1}x^{d-1}dx
\frac{1}{z+i\bar{D}r\Lambda^{2}[2x^{2}+Q^{2}/2]}
\ee
where $Q=q/\Lambda$ and we introduce the time 
$\tau =\frac{1}{\bar{D}r\Lambda^{2}}$. Then
we have
\be
\Gamma^{(2)}(q,z)=-i\frac{g}{\tau}Q^{4}N_{0}(\Omega )
\label{eq:14a}
\ee
where
\be
N_{0}(\Omega )=K_{d}I_{d}(\Omega )
\ee
and
\be
I_{d}(\Omega )
=\int_{0}^{1}
\frac{x^{d-1}dx}{\Omega +2x^{2}}
~~~,
\label{eq:105}
\ee
with
\be
\Omega =-iz\tau +Q^{2}/2
~~~.
\ee
The dimensionless coupling $g$ is given by
\be
g=\frac{1}{2}\left(\frac{D_{1}}{\bar{D}}\right)^{2}\tilde{C}\Lambda^{d}
\nonumber
\ee
\be
=\frac{1}{2}\left(\frac{D_{1}}{\bar{D}}\right)^{2}S
\ee
where 
\be
S=\tilde{C}\Lambda^{d}=\langle (\delta \rho )^{2}\rangle
~~~.
\ee
To see that  $g$ is dimensionless, note that
$D_{1}$ has dimensions  of
$\bar{D}/\phi_{0} $ where $\phi_{0}$ is the 
equilibrium particle density which
has dimensions of  $\Lambda^{d}$.  Finally the fourier
transform of the static structure factor has dimensions
$\tilde{C}\approx L^{d}\rho^{2}_{0} \approx \Lambda^{d}$.
Because of the $Q^{4}$ factor in Eq.(\ref{eq:14a}) we see that there is no second-order
contribution to the diffusion coefficient.

To go further we must evaluate the dimensionless integrals in
Eq.(\ref{eq:105}).
In two dimensions  we find
\be
I_{2}(\Omega )
=\frac{1}{4} ln\left(\frac{\Omega +2}{\Omega }\right)
~~~.
\ee
In three dimensions we have the explicit result
\be
I_{3}(\Omega )
=\frac{1}{2}\left[1-
\sqrt{\frac{\Omega}{2}}tan^{-1}\sqrt{\frac{2}{\Omega}}\right]
~~~.
\label{eq:111b}
\ee

In the small $q$ and $z$ limit 
we have for general $d$:
\be
I_{d}(0)=\frac{1}{2}\int_{0}^{1}x^{d-3}dx
=\frac{1}{2(d-2)}
\nonumber
\ee
which is well defined for $d > 2$ and
\be
\Gamma^{(2)}(q,0)=-i\frac{g}{\tau}Q^{4}\tilde{C}\frac{K_{d}}{2(d-2)}
~~~.
\ee
The kinetic equation in bare second-order perturbation 
theory is given by
\be
\left[z+iL_{0}(q)+K^{(d)}(q,z)\right]C(q,z)=\frac{k_{B}T}{r}
~~~.
\ee
This
can be written in dimensionless form
\be
[\nu +iQ^{2}D(Q,\nu )]C(Q,\nu)
=\frac{k_{B}T\tau}{r}
\ee
where $\nu =z\tau$ and the damping is given by the real part of
\be
D(Q,\nu )=1-gQ^{2}K_{d}I_{d}(\Omega )
~~~.
\ee
The dynamic  structure factor is given by
\be
S(Q,\nu)=-2\pi Im \left[\frac{\beta^{-1}\tau}{r}
\frac{1}{\nu' +i\nu''}\right]
\nonumber
\ee
\be
=\frac{2\pi\beta^{-1}\tau}{r}\frac{\nu ''}{(\nu')^{2}+(\nu'')^{2}}
\ee
where
\be
\nu''=Q^{2}(1-gQ^{2}K_{d}I')
\ee
\be
\nu'=\nu (1+Q^{4}gK_{d}J)
~~~.
\ee
For $d=3$ we have the simple integrals
\be
I'=\int_{0}^{1}x^{2}dx\frac{x_{1}}{\nu^{2}+x_{1}^{2}}
\ee
\be
J=\int_{0}^{1}x^{2}dx\frac{1}{\nu^{2}+x_{1}^{2}}
~~~,
\ee
where $x_{1}=\frac{Q^{2}}{2}+2x^{2}$.

\begin{figure}
\begin{center}\includegraphics[width=6in,scale=1.0]{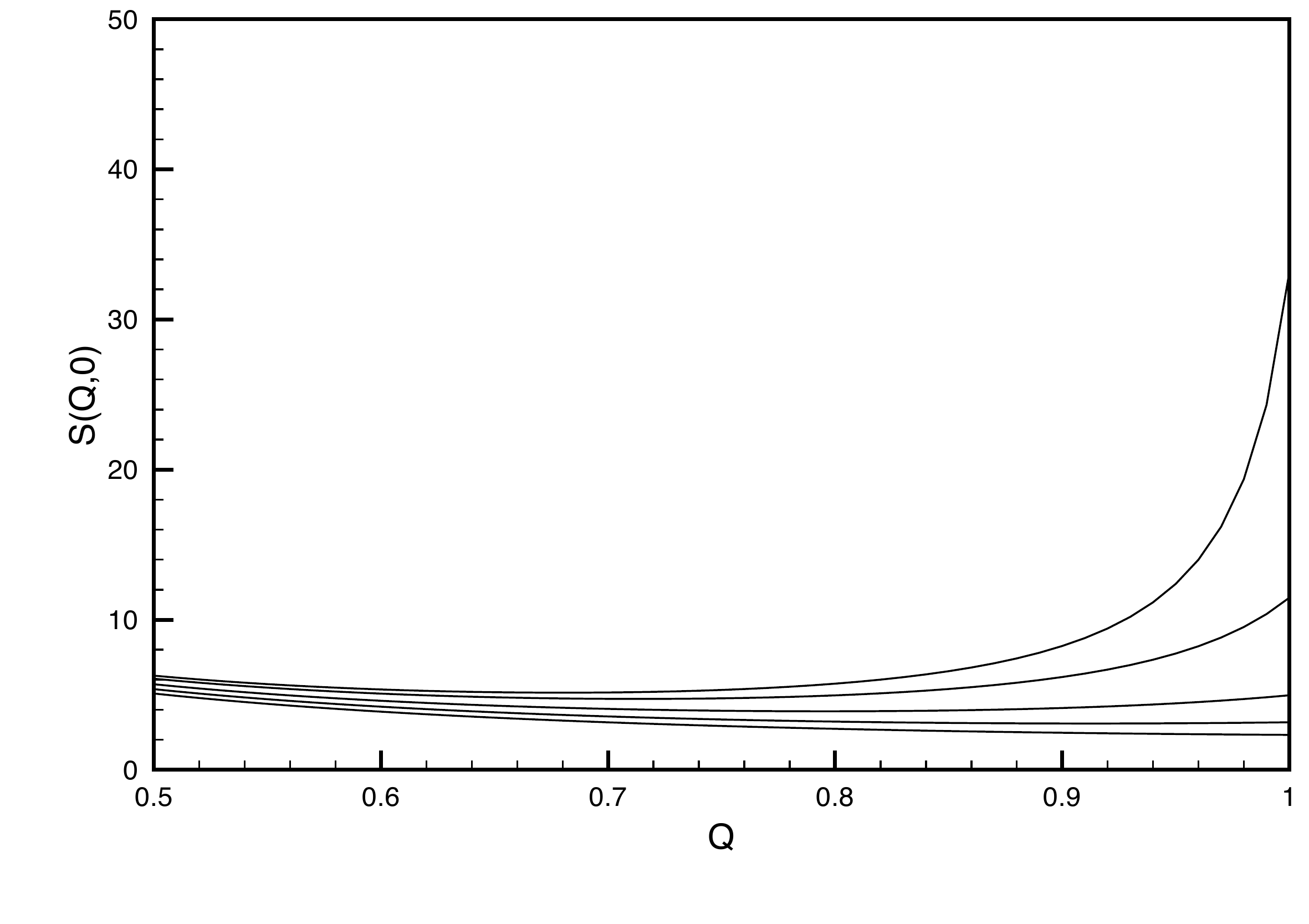}\end{center}
\caption{\small  Plot of the dynamic  structure factor versus wavenumber for zero
frequency for different values of the coupling: 
g=50, 60, 70, 80 and 85 from bottom 
to top.}
\label{fig:1}
\end{figure}

We 
plot the dynamic structure factor in Fig.\ref{fig:1} for $d=3$.  
The conservation law dominates
the structure for small wavenumbers.  However for large wavenumbers 
one sees the development
of an instability.   In this approximation the instability comes 
from short distances, $Q=1$, and low frequencies as seen in Fig.\ref{fig:1}.

\begin{figure}
\begin{center}\includegraphics[width=6in,scale=1.0]{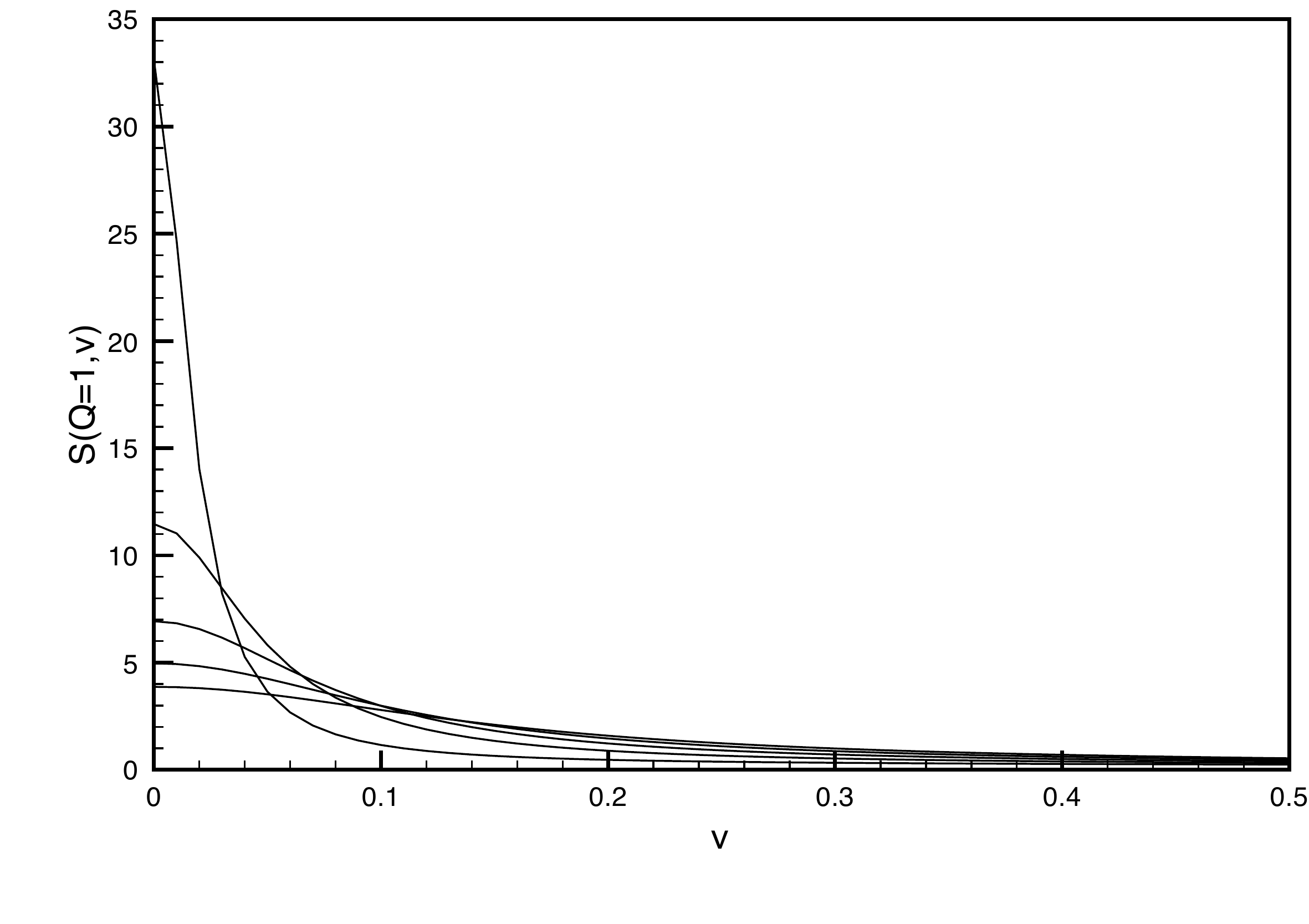}\end{center}
\caption{\small  Plot of dynamic structure factor versus frequency for $Q=1$ 
for different values of the coupling: g=65, 70, 75, 80, and 85 from bottom 
to top at $\nu =0$.}
\label{fig:2}
\end{figure}

\noindent In Fig.\ref{fig:2} we plot the dynamic structure factor at 
$Q=1$ versus frequency.  The instability is
manifest at low frequencies.

\noindent The instability in the problem sets in for the coupling
$g$ where $Re D$ first becomes negative.  In three dimensions the 
critical $g$ is given 
by
\be
1-g^{*}K_{3}I'(Q=1,\nu =0)=0 
~~~.
\ee 
This has a solution
\be
g^{*}=\frac{2\pi^{2}}{I'(Q=1,\nu =0)}
~~~.
\ee
We have from Eq.(\ref{eq:111})
\be
I'(Q=1,\nu =0)=\frac{1}{2}[1-\frac{1}{2}tan^{-1}(2)]
~~~.
\ee
Given that $tan^{-1}(2)=1.107\ldots$ we find 
$g^{*}=88.42\ldots$.

In two dimensions 
\be
I'(Q,0)=\int_{0}^{1}\frac{xdx}{Q^{2}/2 +2x^{2}}
=\frac{1}{4}ln\left(\frac{2+Q^{2}}{Q^{2}}\right)
\ee
and the damping is given by
\be
D(Q^{2},0)=1-gQ^{2}\frac{1}{8\pi}ln\left(\frac{2+Q^{2}}{Q^{2}}\right)
~~~.
\ee
which leads to the instability coupling
\be
g^{*}_{min}=\frac{8\pi}{ln 3}
~~~.
\ee

\subsection{Gaussian Structure in Hydrodynamical limit}

How strongly does our result depend on the cut off $\Lambda$.
To see this consider the case where $\chi (q)$ falls
smoothly to zero for large $q$.  It is simplest to look at the problem 
in the small $q$ and $z$ regime.  In this case, to leading order in $q$,
the vertex takes the form
\be
V({\bf q},{\bf k})=\frac{i}{2}D_{1}\chi^{-1}({\bf k})
q_{i}q_{j}\left[\delta_{ij}-k_{i}k_{j}\sigma (k)\right]
\ee
where sums over i and j are implied,
\be
\sigma (k)=\frac{\chi '(k)}{k\chi (k)}
~~~,
\ee
and from Eq.(\ref{eq:97}) for small $q$:
\be
\Gamma^{(2)}(q,z)
=\frac{D_{1}^{2}}{2}q_{i}q_{j}q_{k}q_{m}
\int \frac{d^{d}k}{(2\pi )^{d}}
\left[\delta_{ij}-k_{i}k_{j}\sigma (k)\right]
\left[\delta_{km}-k_{k}k_{m}\sigma (k)\right]
\frac{\beta^{-2}}{z+2iL_{0}(k)}
~~~.
\label{eq:129}
\ee
A practical choice for the static susceptibility is given by
\be
\chi (k)=\chi_{0}e^{-\frac{1}{2}(k\ell )^{2}}
\ee
where $\ell$ is the characteristic length and
\be
\sigma =\frac{\chi '}{k\chi}=-\ell^{2}
~~~.
\ee
In the $z=0$ limit, after considerable algebra, Eq.(\ref{eq:129}) reduces to
\be
\Gamma^{(2)}(q,0)=-i\tilde{g}Q^{4}\tilde{\tau}^{-1} \tilde{C}(0)
\gamma_{d}
\ee
where
\be
\gamma_{d}=\frac{K_{d}}{2}2^{(d-4)/2}\Gamma\left(\frac{d}{2}\right)
\frac{(d^{3}+12d^{2}-20)}{(d-2)d(d+2)}
~~~,
\ee
where
\be
\frac{1}{\tilde{\tau}}=\frac{\bar{D}\chi_{0}^{-1}}{\ell^{2}}
~~~,
\ee
\be
\tilde{g}=\frac{1}{2}
\left(\frac{D_{1}}{\bar{D}}\right)^{2}\frac{\tilde{C}(0)}{\ell^{d}}
~~~,
\ee
and $Q=q\ell$.
We see that the results for $\Gamma^{(2)}(q,0)$ are very similar
for the two different choices for static  structure factor if we make the correspondence
$\Lambda\rightarrow 1/\ell$ and $r\rightarrow \chi_{0}^{-1}$.

\section{Bare Perturbation Theory at Fourth Order}

Here we look at the reduction of the two-loop contributions 
to the memory-function in more detail.  There are two contributions.

\subsection{General Reduction of $\bar{\Gamma}^{(4)}_{D}$}

We have from Eq.(\ref{eq:96}),
\be
\bar{\Gamma}^{(4)}_{D}({\bf q}_{1},{\bf q}_{2};z)
=-\int\frac{d^{d}k_{1}}{(2\pi )^{d}}
\frac{d^{d}k_{2}}{(2\pi )^{d}}
\frac{d^{d}k_{3}}{(2\pi )^{d}}
\frac{d^{d}k_{4}}{(2\pi )^{d}}
\frac{d^{d}k_{5}}{(2\pi )^{d}}
\frac{d^{d}k_{6}}{(2\pi )^{d}}
\frac{d^{d}k_{7}}{(2\pi )^{d}}
\frac{d^{d}k_{8}}{(2\pi )^{d}}
\nonumber
\ee
\be
\times
V({\bf q}_{2},{\bf k}_{3},{\bf k}_{4})T_{0}({\bf k}_{3},{\bf k}_{4};z)
V({\bf q}_{1},{\bf k}_{1},{\bf k}_{2})T_{0}({\bf k}_{1},{\bf k}_{2};z)
\nonumber
\ee
\be
\times
4V({\bf k}_{1},{\bf k}_{5},{\bf k}_{6})
V({\bf k}_{3},{\bf k}_{7},{\bf k}_{8})2
T_{0}({\bf k}_{5},{\bf k}_{6},{\bf k}_{2})
2\tilde{C}(27)\tilde{C}(46)\tilde{C}(58)
~~~.
\ee
First do the integrations over the $\delta$-functions
associated with the $\tilde{C}$, then over those 
associated with the cubic vertices.  This leads to the result
\be
\bar{\Gamma}^{(4)}_{D}({\bf q}_{1},{\bf q}_{2};z)
=D_{1}^{4}(2\pi )^{d}\delta\left({\bf q}_{1}+{\bf q}_{2}\right)
\int \frac{d^{d}k_{1}}{(2\pi )^{d}}
\int \frac{d^{d}k_{3}}{(2\pi )^{d}}
 (-1){\bf q}_{1}\cdot \vec{\Lambda}({\bf k}_{3},-{\bf q}_{1}-{\bf k}_{3})
\nonumber
\ee
\be
\times
T_{0}({\bf k}_{3},-{\bf q}_{1}-{\bf k}_{3};z)
{\bf q}_{1}\cdot \vec{\Lambda}({\bf k}_{1},{\bf q}_{1}-{\bf k}_{1})
T_{0}({\bf k}_{1},{\bf q}_{1}-{\bf k}_{1};z)
\nonumber
\ee
\be
\times
{\bf k}_{3}\cdot \vec{\Lambda}({\bf q}_{1}-{\bf k}_{1},{\bf k}_{1}-{\bf q}_{1}-{\bf k}_{3})
{\bf k}_{1}\cdot \vec{\Lambda}({\bf k}_{1}-{\bf q}_{1}-{\bf k}_{3},{\bf k}_{3}+{\bf q}_{1})
\nonumber
\ee
\be
\times
T_{0}({\bf k}_{1}-{\bf q}_{1}-{\bf k}_{3},{\bf k}_{3}+{\bf q}_{1},{\bf q}_{1}-{\bf k}_{1})
\tilde{C}({\bf q}_{1}-{\bf k}_{1})\tilde{C}(-{\bf q}_{1}-{\bf k}_{3})
\tilde{C}({\bf k}_{1}-{\bf q}_{1}-{\bf k}_{3})
~~~.
\label{eq:137}
\ee

\subsection{General Reduction of $
\bar{\Gamma}^{(4)}_{R}$}

We have from Eq.(\ref{eq:95}))
\be
\bar{\Gamma}^{(4)}_{R}({\bf q}_{1},{\bf q}_{2};z)=
-\int \frac{d^{d}k_{1}}{(2\pi )^{d}}
\int \frac{d^{d}k_{2}}{(2\pi )^{d}}
\int \frac{d^{d}k_{3}}{(2\pi )^{d}}
\int \frac{d^{d}k_{4}}{(2\pi )^{d}}
\int \frac{d^{d}k_{5}}{(2\pi )^{d}}
\int \frac{d^{d}k_{6}}{(2\pi )^{d}}
\int \frac{d^{d}k_{7}}{(2\pi )^{d}}
\int \frac{d^{d}k_{8}}{(2\pi )^{d}}
\nonumber
\ee
\be
\times V({\bf q}_{2},{\bf k}_{3},{\bf k}_{4})T_{0}({\bf k}_{3},{\bf k}_{4};z)
V({\bf q}_{1},{\bf k}_{1},{\bf k}_{2})T_{0}({\bf k}_{1},{\bf k}_{2};z)
\nonumber
\ee
\be
\times
4V({\bf k}_{3},{\bf k}_{7},{\bf k}_{8})
V({\bf k}_{1},{\bf k}_{5},{\bf k}_{6})2
T_{0}({\bf k}_{5},{\bf k}_{6},{\bf k}_{2})
\tilde{C}(24)\tilde{C}(57)\tilde{C}(68)
~~~.
\ee
Doing the integrations over the internal $\delta$-functions leads to
the result
\be
\bar{\Gamma}^{(4)}_{R}({\bf q}_{1},{\bf q}_{2};z)
=-\frac{1}{2}D_{1}^{4}(2\pi )^{d}\delta ({\bf q}_{1}+{\bf q}_{2})
\int \frac{d^{d}k_{1}}{(2\pi )^{d}}
\left[{\bf q}_{1}\cdot \vec{\Lambda}({\bf k}_{1},{\bf q}_{1}-{\bf k}_{1})
\right]^{2}
\nonumber
\ee
\be
\times
T_{0}^{2}({\bf k}_{1},{\bf q}_{1}-{\bf k}_{1};z)
\tilde{C}({\bf q}_{1}-{\bf k}_{1})
\gamma ({\bf k}_{1},z+iL_{0}({\bf q}_{1}-{\bf k}_{1}))
\label{eq:138}
\ee
where the insertion $\gamma$ is defined
\be
\gamma ({\bf k}_{1},z)
=\int \frac{d^{d}k_{5}}{(2\pi )^{d}}
\left[{\bf k}_{1}\cdot \vec{\Lambda}({\bf k}_{5},{\bf k}_{1}-{\bf k}_{5})
\right]^{2}
T_{0}({\bf k}_{5},{\bf k}_{1}-{\bf k}_{5},{\bf q}_{1}-{\bf k}_{1})
\tilde{C}({\bf k}_{5})\tilde{C}({\bf k}_{1}-{\bf k}_{5})
~~~.
\ee

\subsection{$
\bar{\Gamma}^{(4)}_{D}$ In the structureless approximation}

In the structureless approximation the interaction vertices simplify
significantly and Eq.(\ref{eq:137}) becomes
\be
\bar{\Gamma}^{(4)}_{D}({\bf q}_{1},{\bf q}_{2};z)
=(2\pi)^{d}\delta\left({\bf q}_{1}+{\bf q}_{2}\right)
\bar{\Gamma}^{(4)}_{D}({\bf q}_{1},z)
\ee
where
\be
\bar{\Gamma}^{(4)}_{D}({\bf q}_{1},z)=
-D_{1}^{4}\beta^{-3}rq_{1}^{4}
\int \frac{d^{d}k_{1}}{(2\pi )^{d}}
\int \frac{d^{d}k_{3}}{(2\pi )^{d}} k_{1}^{2}k_{3}^{2}
T_{0}({\bf k}_{3},-{\bf q}_{1}-{\bf k}_{3};z)
\nonumber
\ee
\be
\times
T_{0}({\bf k}_{1},{\bf q}_{1}-{\bf k}_{1};z)
T_{0}({\bf k}_{1}-{\bf q}_{1}-{\bf k}_{3},{\bf k}_{3}+{\bf q}_{1},{\bf q}_{1}-{\bf k}_{1})
\nonumber
~~~.
\ee

As a first check on this result
let us look at the small $q_{1}$ and $z$ limit where we find
\be
\bar{\Gamma}^{(4)}_{D}({\bf q}_{1},0)=
-ig^{2}\frac{\tilde{C}}{\tau}Q^{4}K_{d}^{2}\tilde{J}_{d}
\ee
where
\be
\tilde{J}_{d}=\frac{1}{4(d-1)}\int_{0}^{1}y^{d/2-1}dy
\left[\frac{1}{2y+\frac{3}{2}}
+\frac{1}{2+\frac{3}{2}y}\right]
~~~.
\label{eq:199}
\ee

\subsection{$
\bar{\Gamma}^{(4)}_{R}$ In the structureless approximation}

Similarly
we evaluate $\bar{\Gamma}^{(4)}_{R}$, given by Eq.(\ref{eq:138})
in the structureless approximation
and obtain in the long time and distance regime
\be
\bar{\Gamma}^{(4)}_{R}({\bf q}_{1},{\bf q}_{2};z)
=(2\pi)^{d}\delta ({\bf q}_{1}+{\bf q}_{2})
\bar{\Gamma}^{(4)}_{R}({\bf q}_{1};z)
\ee
and
\be
\bar{\Gamma}^{(4)}_{R}({\bf q}_{1};0)
=-ig^{2}Q^{4}\frac{\tilde{C}}{\tau}\frac{K_{d}^{2}\tilde{J}_{d}}{2}
\ee
where $\tilde{J}_{d}$ is given by Eq.(\ref{eq:199}).

\subsection{Summary of Bare Perturbation Theory Results}

Combining the small $q$ and $z$ limits for terms up to fourth
order we have
\be 
\Gamma^{(d)}(q,0)=
-igQ^{4}\frac{\tilde{C}}{\tau}K_{d}\gamma
\label{eq:267}
\ee
where
\be
\gamma=
\frac{1}{2(d-2)}
+\frac{3}{2}g\tilde{J}_{d}K_{d}
\nonumber
\ee
One can interpret this in terms of an effective coupling
\be
g_{eff}=g(1+3(d-2)g\tilde{J}_{d}K_{d})
~~~.
\ee

For perturbation theory to make sense we require that the coefficient
\be
{\cal C}_{d}=3[d-2]\tilde{J}_{d}K_{d}
\ee
be small.  In three dimensions ${\cal C}_{3}
=\frac{3\tilde{J}_{3}}{2\pi^{2}}$ where
\be
\tilde{J}_{3}=\frac{1}{4}
\int_{0}^{1}dx x^{2}\left[\frac{2}{4x^{2}+3}+\frac{2}{4+3x^{2}}\right]
=0.0622\ldots
\ee
and ${\cal C}_{3}=0.00946\ldots$.

\section{Self-Consistent Perturbation Theory}

A key ingredient of MCT is that it is a self-consistent theory
where the memory function is a function of the full correlation
function.  Here we show how this is arranged through two-loop
order in our development here.
Elsewhere we discuss how this is naturally
carried out in the MSR formulation.

\subsection{Second Order Theory}

One wants to replace bare correlation functions by renormalized
correlation functions.  We begin with the bare second-order
memory function given by Eq.(\ref{eq:139})
\be
\Gamma^{(2)}({\bf q}_{1},{\bf q}_{2};z)
=-\int\frac{d^{d}k_{1}}{(2\pi)^{d}}
\frac{d^{d}k_{2}}{(2\pi)^{d}}
\frac{d^{d}k_{3}}{(2\pi)^{d}}
\frac{d^{d}k_{4}}{(2\pi)^{d}}
\nonumber
\ee
\be
\times
V({\bf q}_{1},{\bf k}_{1},{\bf k}_{2}) 
 V({\bf q}_{2},{\bf k}_{3},{\bf k}_{4})T_{0}({\bf k}_{1},{\bf k}_{2};z)
 \nonumber
 \ee
 \be
 \times
 \left[\tilde{C}({\bf k}_{1},{\bf k}_{3})\tilde{C}({\bf k}_{2},{\bf k}_{4})
 +\tilde{C}({\bf k}_{1},{\bf k}_{4})\tilde{C}({\bf k}_{2},{\bf k}_{3})\right]
~~~.
 \ee
 We first write this in terms of the
  bare two-point correlation functions.  We have
 \be
 T_{0}({\bf k}_{1},{\bf k}_{2};z)
 \left[\tilde{C}({\bf k}_{1},{\bf k}_{3})\tilde{C}({\bf k}_{2},{\bf k}_{4})
 +\tilde{C}({\bf k}_{1},{\bf k}_{4})\tilde{C}({\bf k}_{2},{\bf k}_{3})\right]
 \nonumber
 \ee
 \be
 =-i\int_{0}^{\infty} dt e^{izt} e^{-L_{0}({\bf k}_{1})t}e^{-L_{0}({\bf k}_{2})t}
 \left[\tilde{C}({\bf k}_{1},{\bf k}_{3})\tilde{C}({\bf k}_{2},{\bf k}_{4})
 +\tilde{C}({\bf k}_{1},{\bf k}_{4})\tilde{C}({\bf k}_{2},{\bf k}_{3})\right]
 \nonumber
 \ee
 \be
 =-i\int_{0}^{\infty} dt e^{izt}
 \left[C_{0}({\bf k}_{1},{\bf k}_{3};t)C_{0}({\bf k}_{2},{\bf k}_{4};t)
 +C_{0}({\bf k}_{1},{\bf k}_{4};t)C_{0}({\bf k}_{2},{\bf k}_{3};t)\right]
 \label{eq:136a}
 ~~~.
 \ee
At this order we can replace $C_{0}\rightarrow C$, and the last equation is replaced by
 \be
 =-i\int_{0}^{\infty} dt e^{izt}
 \left[C({\bf k}_{1},{\bf k}_{3};t)C({\bf k}_{2},{\bf k}_{4};t)
 +C({\bf k}_{1},{\bf k}_{4};t)C({\bf k}_{2},{\bf k}_{3};t)\right]
\label{eq:136b}
 \ee
 and we have for the memory function at second order
 \be
\Gamma^{(2)}_{R}({\bf q}_{1},{\bf q}_{2};z)
=-\int\frac{d^{d}k_{1}}{(2\pi)^{d}}
\frac{d^{d}k_{2}}{(2\pi)^{d}}
\frac{d^{d}k_{3}}{(2\pi)^{d}}
\frac{d^{d}k_{4}}{(2\pi)^{d}}
V({\bf q}_{1},{\bf k}_{1},{\bf k}_{2}) 
 V({\bf q}_{2},{\bf k}_{3},{\bf k}_{4})
\nonumber
\ee
\be
\times\left[
 -i\int_{0}^{\infty} dt e^{izt}
 \left[C({\bf k}_{1},{\bf k}_{3};t)C({\bf k}_{2},{\bf k}_{4};t)
 +C({\bf k}_{1},{\bf k}_{4};t)C({\bf k}_{2},{\bf k}_{3};t)\right]
\right]
\label{eq:142a}
~~~.
 \ee
 This form will generate contributions at 4th order in perturbation
 theory. Generating the $4th$-order contribution from 
this result,  requires
generating the second-order contribution to the correlation function. 
Iterating Eq.(\ref{eq:14})
 \be
 C({\bf q}_{1},{\bf q}_{2};z)=T_{0}({\bf q}_{1};z)
\tilde{C}({\bf q}_{1},{\bf q}_{2})
 -T_{0}({\bf q}_{1};z)\int\frac{d^{d}k_{1}}{(2\pi)^{d}}K^{(2)}({\bf q}_{1},{\bf k}_{1})
 \tilde{C}({\bf k}_{1},{\bf q}_{2})T_{0}({\bf q}_{2};z)
 \nonumber
 \ee
 \be
 =T_{0}({\bf q}_{1};z)\tilde{C}({\bf q}_{1},{\bf q}_{2}
 -T_{0}({\bf q}_{1};z)\Gamma^{(2)}({\bf q}_{1},{\bf q}_{2})
 T_{0}({\bf q}_{2};z)
~~~.
 \nonumber
 \ee
 Taking the inverse Laplace transform to go to the time domain gives
 \be
 C({\bf q}_{1},{\bf q}_{2};t)=e^{-L_{0}({\bf q})t}
\tilde{C}({\bf q}_{1},{\bf q}_{2})
 \nonumber
 \ee
 \be
 +\int_{0}^{t}ds~ e^{-L_{0}({\bf q}_{1})(t-s)}\int_{0}^{s}d\tau
 \Gamma^{(2)}({\bf q}_{1},{\bf q}_{2},s-\tau)e^{-L_{0}({\bf q}_{2})\tau }
 \nonumber
 \ee
 We then substitute this result into Eq.(\ref{eq:136b}) 
and keep terms of $4th$
 order. We find
\be
\Delta\Gamma^{(4)}_{R}({\bf q}_{1},{\bf q}_{2};t)
=\int\frac{d^{d}k_{1}}{(2\pi)^{d}}
\frac{d^{d}k_{2}}{(2\pi)^{d}}
\frac{d^{d}k_{3}}{(2\pi)^{d}}
\frac{d^{d}k_{4}}{(2\pi)^{d}}
V({\bf q}_{1},{\bf k}_{1},{\bf k}_{2})
 V({\bf q}_{2},{\bf k}_{3},{\bf k}_{4})
\nonumber
\ee
\be
\times
(2i)\int_{0}^{\infty} dt e^{izt}
[2C_{0}({\bf k}_{2},{\bf k}_{4};t)
\nonumber
\ee
\be
\times \int_{0}^{t}ds e^{-L_{0}({\bf k}_{1})(t-s)}\int_{0}^{s}d\tau
 \Gamma^{(2)}({\bf k}_{1},{\bf k}_{3},s-\tau)e^{-L_{0}({\bf k}_{3})\tau }
 \nonumber
 \ee
\be
=\int\frac{d^{d}k_{1}}{(2\pi)^{d}}
\frac{d^{d}k_{2}}{(2\pi)^{d}}
\frac{d^{d}k_{3}}{(2\pi)^{d}}
\frac{d^{d}k_{4}}{(2\pi)^{d}}
V({\bf q}_{1},{\bf k}_{1},{\bf k}_{2})
 V({\bf q}_{2},{\bf k}_{3},{\bf k}_{4})
 \nonumber
 \ee
 \be
 \times
(2i)\int_{0}^{\infty} dt e^{izt}e^{-L_{0}({\bf k}_{3})t}
2\tilde{C}_{0}({\bf k}_{2},{\bf k}_{4})
\nonumber
\ee
\be
\times
\int_{0}^{t}ds e^{-L_{0}({\bf k}_{1})(t-s)}\int_{0}^{s}d\tau
V({\bf k}_{1},{\bf k}_{5}{\bf k}_{6})
V({\bf k}_{3},{\bf k}_{7}{\bf k}_{8})
\nonumber
\ee
\be
\times
e^{-(L_{0}({\bf k}_{5})+L_{0}({\bf k}_{6}))(s-\tau )}
2\tilde{C}_{0}({\bf k}_{5},{\bf k}_{7})
\tilde{C}_{0}({\bf k}_{6},{\bf k}_{8})
e^{-L_{0}({\bf k}_{3})\tau)}
\nonumber
\ee
\be
=\int\frac{d^{d}k_{1}}{(2\pi)^{d}}
\frac{d^{d}k_{2}}{(2\pi)^{d}}
\frac{d^{d}k_{3}}{(2\pi)^{d}}
\frac{d^{d}k_{4}}{(2\pi)^{d}}
V({\bf q}_{1},{\bf k}_{1},{\bf k}_{2})
 V({\bf q}_{2},{\bf k}_{3},{\bf k}_{4})
2\tilde{C}_{0}({\bf k}_{2},{\bf k}_{4})
\nonumber
\ee
\be
\times
(2i)\int_{0}^{\infty} dt e^{izt}
\int_{0}^{t}ds~e^{-(L_{0}({\bf k}_{1})+L_{0}({\bf k}_{2}))(t-s)}
V({\bf k}_{1},{\bf k}_{5},{\bf k}_{6})
V({\bf k}_{3},{\bf k}_{7},{\bf k}_{8})
\nonumber
\ee
\be
\times
\int_{0}^{s}d\tau 
e^{-(L_{0}({\bf k}_{5})+L_{0}({\bf k}_{6})+L_{0}({\bf k}_{2}))(s-\tau )}
2\tilde{C}_{0}({\bf k}_{5},{\bf k}_{7})
\tilde{C}_{0}({\bf k}_{6},{\bf k}_{8})
\nonumber
\ee
\be
\times
e^{-(L_{0}({\bf k}_{3})+L_{0}({\bf k}_{4}))\tau)}
\nonumber
\ee
\be
=(-4)\int\frac{d^{d}k_{1}}{(2\pi)^{d}}
\frac{d^{d}k_{2}}{(2\pi)^{d}}
\frac{d^{d}k_{3}}{(2\pi)^{d}}
\frac{d^{d}k_{4}}{(2\pi)^{d}}
V({\bf q}_{1},{\bf k}_{1},{\bf k}_{2})
 V({\bf q}_{2},{\bf k}_{3},{\bf k}_{4})
\tilde{C}_{0}({\bf k}_{2},{\bf k}_{4})T_{0}({\bf k}_{1},{\bf k}_{2};z)
\nonumber
\ee
\be
\times
V({\bf k}_{1},{\bf k}_{5}{\bf k}_{6})
V({\bf k}_{3},{\bf k}_{7}{\bf k}_{8})
T_{0}({\bf k}_{5},{\bf k}_{6},{\bf k}_{2};z)
2\tilde{C}_{0}({\bf k}_{5},{\bf k}_{7})
\tilde{C}_{0}({\bf k}_{6},{\bf k}_{8})
\nonumber
\ee
which agrees with $\bar{\Gamma}_{R}^{(4)}({\bf q}_{1},{\bf q}_{2},z)$
given by Eq.(\ref{eq:95}).
So $\Gamma_{R}^{(4)}$ is generated by expanding $\Gamma^{(2)}_{R}$.

\subsection{Two Loop self-consistent theory}

We want to replace the $4th$ order bare contribution with a self-
consistent form which depends on the full correlation functions.
We begin with the bare contribution
\be
\bar{\Gamma}^{(4)}_{D}({\bf q}_{1},{\bf q}_{2};z)
=-16\int \frac{d^{d}k_{1}}{(2\pi )^{d}}
\frac{d^{d}k_{2}}{(2\pi )^{d}}
\frac{d^{d}k_{3}}{(2\pi )^{d}}
\frac{d^{d}k_{4}}{(2\pi )^{d}}
\frac{d^{d}k_{5}}{(2\pi )^{d}}
\frac{d^{d}k_{6}}{(2\pi )^{d}}
\frac{d^{d}k_{7}}{(2\pi )^{d}}
\frac{d^{d}k_{8}}{(2\pi )^{d}}
\nonumber
\ee
\be
\times
V({\bf q}_{1},{\bf k}_{1},{\bf k}_{2})T_{0}({\bf k}_{1},{\bf k}_{2};z)
V({\bf q}_{2},{\bf k}_{3},{\bf k}_{4})T_{0}({\bf k}_{3},{\bf k}_{4};z)
\nonumber
\ee
\be
\times
V({\bf k}_{1},{\bf k}_{5},{\bf k}_{6})
V({\bf k}_{3},{\bf k}_{7},{\bf k}_{8})
T_{0}({\bf k}_{5},{\bf k}_{6},{\bf k}_{2})
\tilde{C}(27)\tilde{C}(46)\tilde{C}(58)
\nonumber
\ee
\be
=-16\int \frac{d^{d}k_{1}}{(2\pi )^{d}}
\frac{d^{d}k_{2}}{(2\pi )^{d}}
\frac{d^{d}k_{3}}{(2\pi )^{d}}
\frac{d^{d}k_{4}}{(2\pi )^{d}}
\frac{d^{d}k_{5}}{(2\pi )^{d}}
\frac{d^{d}k_{6}}{(2\pi )^{d}}
\frac{d^{d}k_{7}}{(2\pi )^{d}}
\frac{d^{d}k_{8}}{(2\pi )^{d}}
\nonumber
\ee
\be
\times
V({\bf q}_{1},{\bf k}_{1},{\bf k}_{2})
V({\bf q}_{2},{\bf k}_{3},{\bf k}_{4})
V({\bf k}_{1},{\bf k}_{5},{\bf k}_{6})
V({\bf k}_{3},{\bf k}_{7},{\bf k}_{8})
\nonumber
\ee
\be
\times
\tilde{C}(27)\tilde{C}(46)\tilde{C}(58)
\nonumber
\ee
\be
\times
\frac{1}{[z+iL_{0}(1)+iL_{0}(2)]}
\frac{1}{[z+iL_{0}(3)+iL_{0}(4)]}
\frac{1}{[z+iL_{0}(2)+iL_{0}(4)+iL_{0}(5)]}
\nonumber
~~~.
\ee

We can then use the following result based on the pole structure
of the zeroth order correlation function:
\be
\int\frac{d\omega_{1}}{2\pi}\frac{C_{0}(k_{1},\omega_{1})}
{z-\omega_{1}+iL_{0}(2)}
=\frac{\tilde{C}(k_{1})}{z+iL_{0}(1)+iL_{0}(2)}
\nonumber
~~~.
\ee
Using essentially this result five times we find
\be
\bar{\Gamma}^{(4)}_{D}({\bf q}_{1},{\bf q}_{2};z)
=-16\int \frac{d^{d}k_{1}}{(2\pi )^{d}}
\frac{d^{d}k_{2}}{(2\pi )^{d}}
\frac{d^{d}k_{3}}{(2\pi )^{d}}
\frac{d^{d}k_{4}}{(2\pi )^{d}}
\frac{d^{d}k_{5}}{(2\pi )^{d}}
\nonumber
\ee
\be
\times
V({\bf q}_{2},{\bf k}_{3},{\bf k}_{4})
V({\bf q}_{1},{\bf k}_{1},{\bf k}_{2})
V({\bf k}_{1},{\bf k}_{5},-{\bf k}_{4})
V({\bf k}_{3},-{\bf k}_{2},-{\bf k}_{5})
\nonumber
\ee
\be
\times
\int\frac{d\omega_{1}}{2\pi}C_{0}(k_{1},\omega_{1})
\int\frac{d\omega_{2}}{2\pi}C_{0}(k_{2},\omega_{2})
\int\frac{d\omega_{3}}{2\pi}C_{0}(k_{3},\omega_{3})
\int\frac{d\omega_{4}}{2\pi}C_{0}(k_{4},\omega_{4})
\int\frac{d\omega_{5}}{2\pi}C_{0}(k_{5},\omega_{5})
\nonumber
\ee
\be
\times
\tilde{C}^{-1}(k_{1})
\tilde{C}^{-1}(k_{3})
\frac{1}{[z-\omega_{1}-\omega_{2}]}
\frac{1}{[z-\omega_{3}-\omega_{4}]}
\frac{1}{[z-\omega_{2}-\omega_{4}-\omega_{5}]}
\nonumber
\ee

To obtain the self-consistent generalization to this order
we replace
\be
C_{0}\rightarrow C
~~~.
\ee
We then have
\be
\Gamma^{(4)}_{D}({\bf q}_{1},{\bf q}_{2};z)
=-16\int \frac{d^{d}k_{1}}{(2\pi )^{d}}
\frac{d^{d}k_{2}}{(2\pi )^{d}}
\frac{d^{d}k_{3}}{(2\pi )^{d}}
\frac{d^{d}k_{4}}{(2\pi )^{d}}
\frac{d^{d}k_{5}}{(2\pi )^{d}}
\nonumber
\ee
\be
\times
V({\bf q}_{2},{\bf k}_{3},{\bf k}_{4})
V({\bf q}_{1},{\bf k}_{1},{\bf k}_{2})
\tilde{C}^{-1}(k_{1})
V({\bf k}_{1},{\bf k}_{5},-{\bf k}_{4})
\tilde{C}^{-1}(k_{3})
V({\bf k}_{3},-{\bf k}_{2},-{\bf k}_{5})
\nonumber
\ee
\be
\times
\int\frac{d\omega_{2}}{2\pi}C_{R}(k_{1},z-\omega_{2})
C(k_{2},\omega_{2})
\int\frac{d\omega_{4}}{2\pi}C_{R}(k_{3},z-\omega_{4})
C(k_{4},\omega_{4})
C_{R}(k_{5},z-\omega_{2}-\omega_{4})
\nonumber
\ee
where the retarded correlation functions are defined by
\be
C_{R}(k_{1},z)=
\int\frac{d\omega_{1}}{2\pi}\frac{C(k_{1},\omega_{1})}
{z-\omega_{1}}
\ee

If we define a vertex
\be
\bar{V}({\bf q}_{1},{\bf k}_{1},{\bf k}_{2})=\tilde{C}^{-1}(q_{1})
V({\bf q}_{1},{\bf k}_{1},{\bf k}_{2})
\nonumber
\ee
then
\be
\Gamma^{(4)}_{D}({\bf q}_{1},{\bf q}_{2};z)
=-16\tilde{C}({\bf q}_{1})
\tilde{C}({\bf q}_{2})
\int \frac{d^{d}k_{1}}{(2\pi )^{d}}
\frac{d^{d}k_{2}}{(2\pi )^{d}}
\frac{d^{d}k_{3}}{(2\pi )^{d}}
\frac{d^{d}k_{4}}{(2\pi )^{d}}
\frac{d^{d}k_{5}}{(2\pi )^{d}}
\nonumber
\ee
\be
\times
\bar{V}({\bf q}_{2},{\bf k}_{3},{\bf k}_{4})
\bar{V}({\bf q}_{1},{\bf k}_{1},{\bf k}_{2})
\bar{V}({\bf k}_{1},{\bf k}_{5},-{\bf k}_{4})
\bar{V}({\bf k}_{3},-{\bf k}_{2},-{\bf k}_{5})
\nonumber
\ee
\be
\times
\int\frac{d\omega_{2}}{2\pi}C_{R}(k_{1},z-\omega_{2})
C(k_{2},\omega_{2})
\int\frac{d\omega_{4}}{2\pi}C_{R}(k_{3},z-\omega_{4})
C(k_{4},\omega_{4})
C_{R}(k_{5},z-\omega_{2}-\omega_{4})
\nonumber
~~~.
\ee

The most useful form for our purposes is
\be
\Gamma^{(4)}_{D}({\bf q}_{1},{\bf q}_{2};z)
=-16\tilde{C}({\bf q}_{1})
\tilde{C}({\bf q}_{2})
\int \frac{d^{d}k_{1}}{(2\pi )^{d}}
\frac{d^{d}k_{2}}{(2\pi )^{d}}
\frac{d^{d}k_{3}}{(2\pi )^{d}}
\frac{d^{d}k_{4}}{(2\pi )^{d}}
\frac{d^{d}k_{5}}{(2\pi )^{d}}
\nonumber
\ee
\be
\times
\bar{V}({\bf q}_{2},{\bf k}_{3},{\bf k}_{4})
\bar{V}({\bf q}_{1},{\bf k}_{1},{\bf k}_{2})
\bar{V}({\bf k}_{1},{\bf k}_{5},-{\bf k}_{4})
\bar{V}({\bf k}_{3},-{\bf k}_{2},-{\bf k}_{5})
\nonumber
\ee
\be
\times
\int\frac{d\omega_{1}}{2\pi}C(k_{1},\omega_{1})
\int\frac{d\omega_{2}}{2\pi}C(k_{2},\omega_{2})
\int\frac{d\omega_{3}}{2\pi}C(k_{3},\omega_{3})
\int\frac{d\omega_{4}}{2\pi}C(k_{4},\omega_{4})
\int\frac{d\omega_{5}}{2\pi}C(k_{5},\omega_{5})
\nonumber
\ee
\be
\times
\frac{1}{[z-\omega_{1}-\omega_{2}]}
\frac{1}{[z-\omega_{3}-\omega_{4}]}
\frac{1}{[z-\omega_{2}-\omega_{4}-\omega_{5}]}
\nonumber
\ee

We need to invert the Laplace transform and obtain this contribution in the time regime.
The key result we need is
\be
\int\frac{dz}{2\pi i}e^{-izt}
\frac{1}{[z-\omega_{1}-\omega_{2}]}
\frac{1}{[z-\omega_{3}-\omega_{4}]}
\frac{1}{[z-\omega_{2}-\omega_{4}-\omega_{5}]}
\nonumber
\ee
\be
=-\int_{0}^{t}dt_{1}e^{-i(\omega_{1}+\omega_{2})(t-t_{1})}
\int_{0}^{t_{1}}dt_{2}e^{-i(\omega_{2}+\omega_{4}+\omega_{5})(t_{1}-t_{2})}
e^{-i(\omega_{3}+\omega_{4})t_{2}}
\nonumber
\ee 
which is a product of convolutions.  One can the do the frequency integrals easily to take
one fully to the time regime:
\be
\Gamma^{(4)}_{D}({\bf q}_{1},{\bf q}_{2};t)
=16\tilde{C}({\bf q}_{1})
\tilde{C}({\bf q}_{2})
\int \frac{d^{d}k_{1}}{(2\pi )^{d}}
\frac{d^{d}k_{2}}{(2\pi )^{d}}
\frac{d^{d}k_{3}}{(2\pi )^{d}}
\frac{d^{d}k_{4}}{(2\pi )^{d}}
\frac{d^{d}k_{5}}{(2\pi )^{d}}
\nonumber
\ee
\be
\times
\bar{V}({\bf q}_{2},{\bf k}_{3},{\bf k}_{4})
\bar{V}({\bf q}_{1},{\bf k}_{1},{\bf k}_{2})
\bar{V}({\bf k}_{1},{\bf k}_{5},-{\bf k}_{4})
\bar{V}({\bf k}_{3},-{\bf k}_{2},-{\bf k}_{5})
\nonumber
\ee
\be
\times \int_{0}^{t}dt_{1}
\int_{0}^{t_{1}}dt_{2}
C(k_{1},t-t_{1})
C(k_{2},t-t_{2})
C(k_{3},t_{2})
C(k_{4},t_{1})
C(k_{5},t_{1}-t_{2 })
\nonumber
~~~.
\ee

In the structureless approximation this reduces to
\be
\Gamma^{(4)}_{D}({\bf q}_{1},{\bf q}_{2};t)=(2\pi )^{d}
\delta (({\bf q}_{1}+{\bf q}_{2})\Gamma^{(4)}_{D}({\bf q}_{1};t)
\ee
and, in terms of dimensionless variables,
\be
\Gamma^{(4)}_{D}({\bf q}_{1};t)=4\frac{g^{2}}{\tau^{2}}\tilde{C}Q^{4}
\int \frac{d^{d}K_{1}}{(2\pi )^{d}}
\frac{d^{d}K_{3}}{(2\pi )^{d}}K_{1}^{2}K_{3}^{2} \int_{0}^{t}dt_{1}
\int_{0}^{t_{1}}dt_{2}f(K_{1},T-T_{1})
f(Q-K_{1},T-T_{2})
\nonumber
\ee
\be
\times
f(K_{3},T_{2})
f(Q+K_{3},T_{1})
f(-Q+K_{1}-K_{3},T_{1}-T_{2 })
~~~.
\label{eq:205}
\ee
We show elsewhere that this same structure is found in the time regime
at two-loop order using the MSR formulation.

\newpage	

\subsection{Self-Consistent Kinetic Equation at Second Order}

The second  order memory function in terms of full correlation functions
is given by
\be
\Gamma^{(2)}_{R}({\bf q}_{1},{\bf q}_{2};z)
=-\int\frac{d^{d}k_{1}}{(2\pi)^{d}}
\frac{d^{d}k_{2}}{(2\pi)^{d}}
\frac{d^{d}k_{3}}{(2\pi)^{d}}
\frac{d^{d}k_{4}}{(2\pi)^{d}}
V({\bf q}_{1},{\bf k}_{1},{\bf k}_{2}) 
 V({\bf q}_{2},{\bf k}_{3},{\bf k}_{4})
\nonumber
\ee
\be
\times\left[
 -i\int_{0}^{\infty} dt e^{izt}
 \left[C({\bf k}_{1},{\bf k}_{3};t)C({\bf k}_{2},{\bf k}_{4};t)
 +C({\bf k}_{1},{\bf k}_{4};t)C({\bf k}_{2},{\bf k}_{3};t)\right]
\right]
\nonumber
~~~.
 \ee
In the structureless approximation this reduces to
 \be
 \Gamma_{R}^{(2)}(q,t)=
K^{(d,2)}(Q,t)
 \tilde{C}
 \ee 
and
 \be
 K^{(d,2)}(Q,t)
=Q^{4}\frac{1}{\tau^{2}}N_{0}(K,t)
\ee
where
\be
N_{0}(Q,T)=g\int\frac{d^{d}K}{(2\pi )^{d}}f(K,T)f({\bf Q}-{\bf K},T)
\nonumber
\ee
where 
\be
f(Q,T)=C(Q,T)/\tilde{C}(Q)
\ee
and  we have introduced the same dimensionless variables as in the case
of bare perturbation theory.
The kinetic equation reduces to
\be
\frac{\partial f(Q,T)}{\partial T}=-Q^{2}f (Q,T)
+Q^{4}\int_{0}^{T}dS N_{0}(Q,T-S)f(Q,S)
~~~.
\label{eq:159}
\ee 

\begin{figure}
\begin{center}\includegraphics[width=6in,scale=1.0]{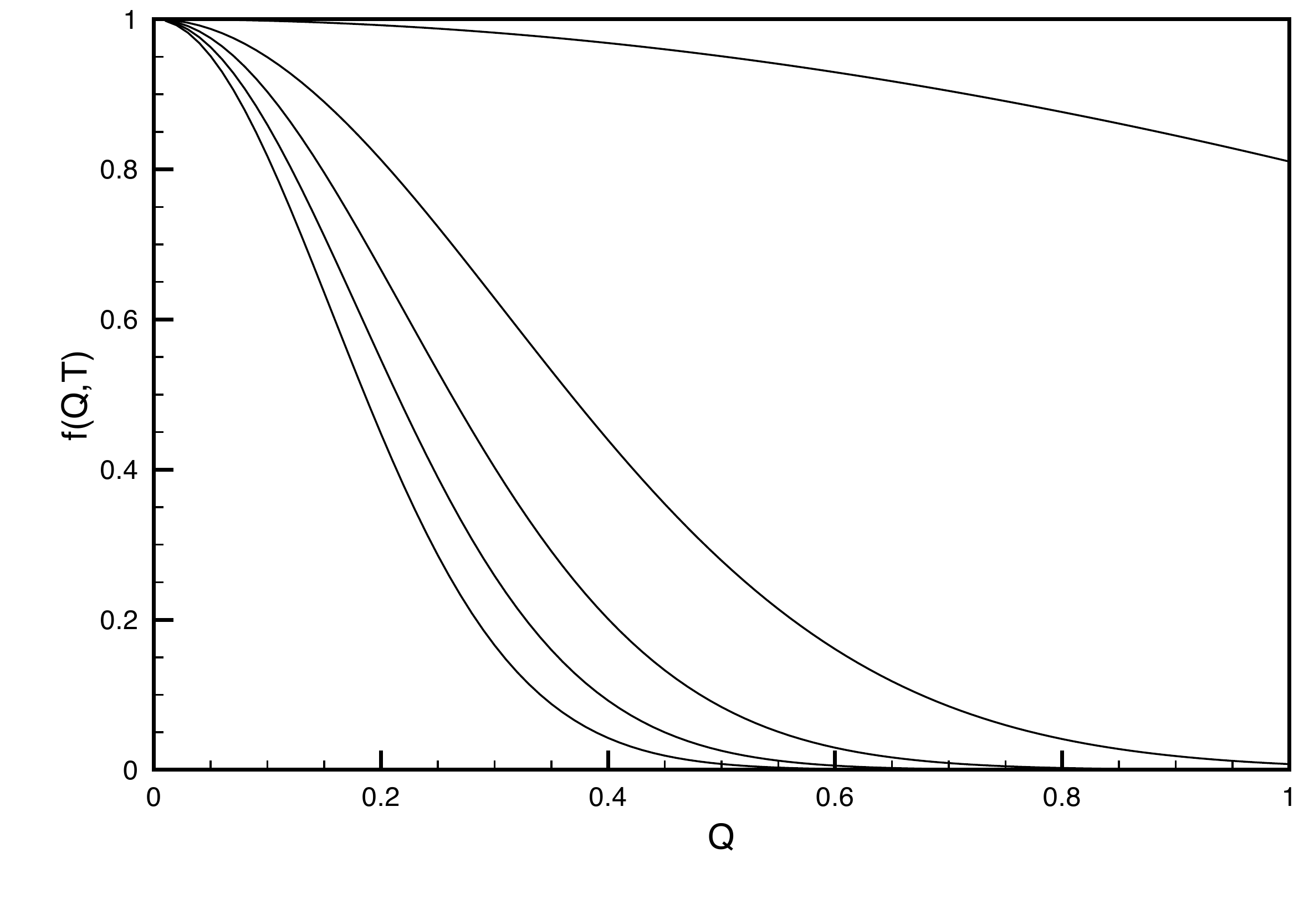}\end{center}
\caption{\small  Plot of normalized intermediate structure factor 
versus wavenumber 
for different times $T= 0.1, 5.0, 10.0, 15.0$, and $20.0$ 
from  top to bottom.  The coupling $g=10.0$ is relatively weak.}     
\label{fig:3}
\end{figure}

\begin{figure}
\begin{center}\includegraphics[width=6in,scale=1.0]{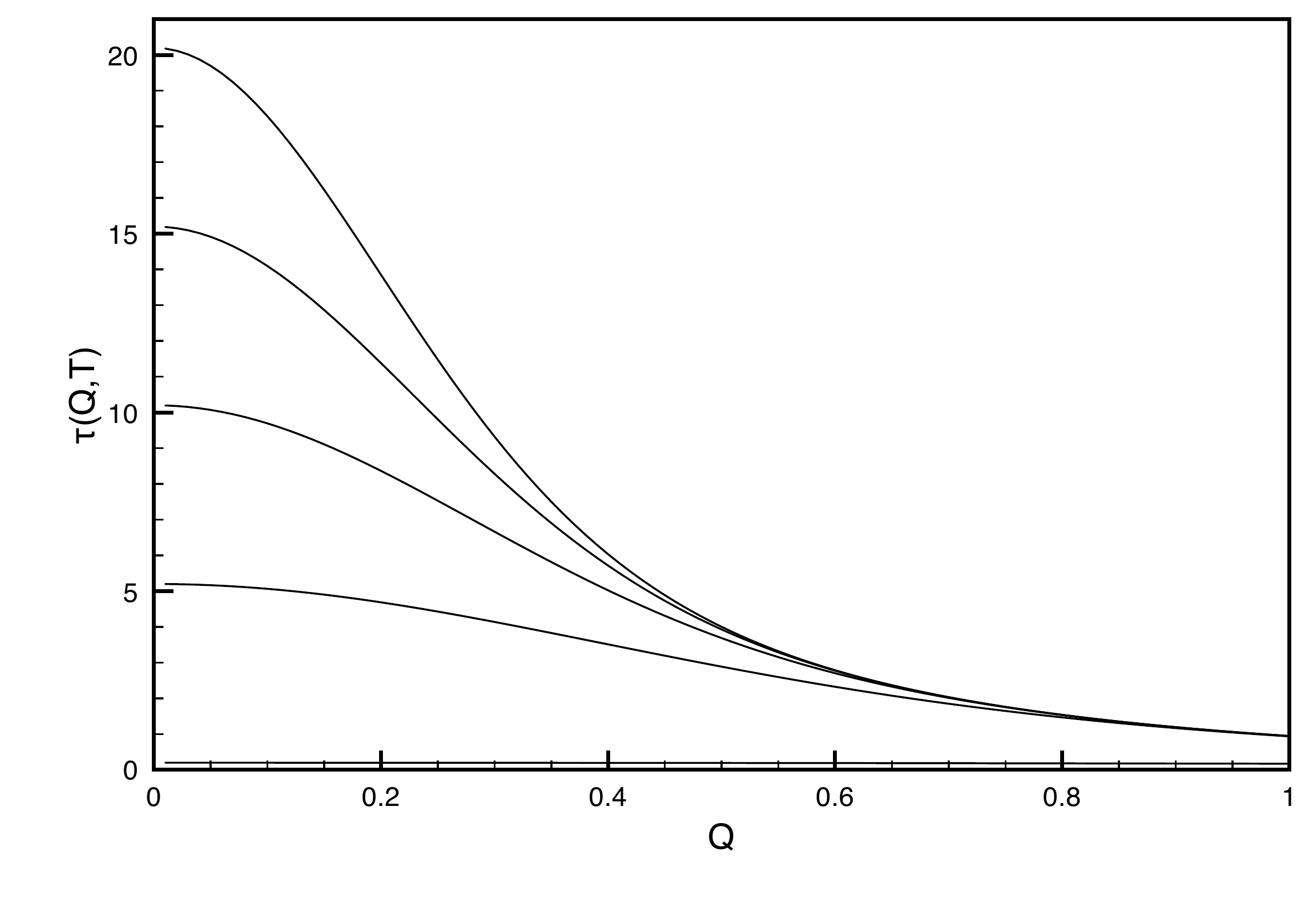}\end{center}
\caption{\small  Plot of $\tau (Q,T)$ versus wavenumber for different times T= 0.1, 5.0, 10.0, 15.0, and 20.0 
from  bottom to top.  The coupling $g=10.0$.}     
\label{fig:4}
\end{figure}

\begin{figure}
\begin{center}\includegraphics[width=6in,scale=1.0]{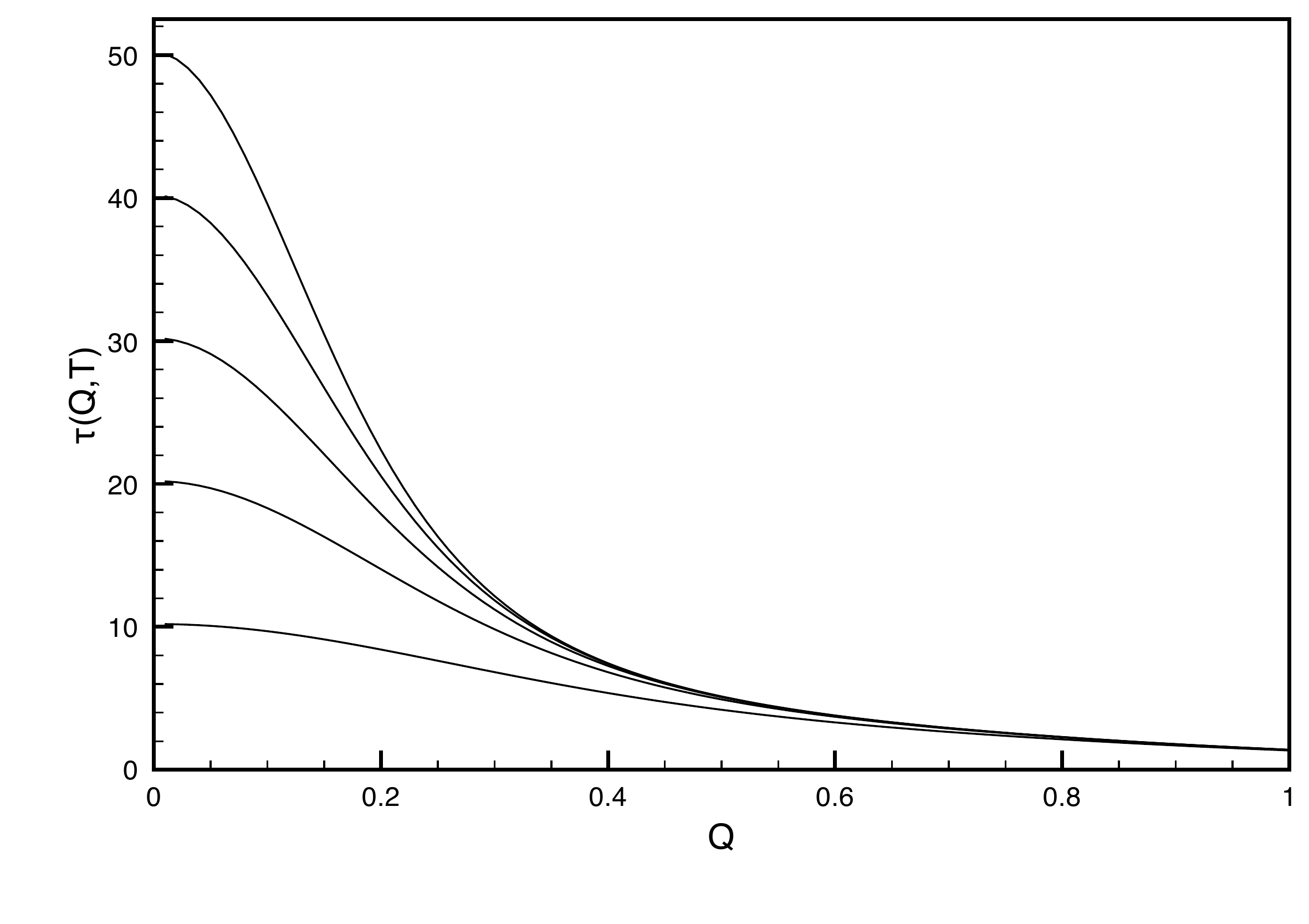}\end{center}
\caption{\small  Plot of $\tau (Q,T)$ versus wavenumber for different times T= 10.0, 20.0, 30.0, 40.0 and 50.0 
from   bottom to top.  The coupling $g=60.0$.}     
\label{fig:5}
\end{figure}

\begin{figure}
\begin{center}\includegraphics[width=6in,scale=1.0]{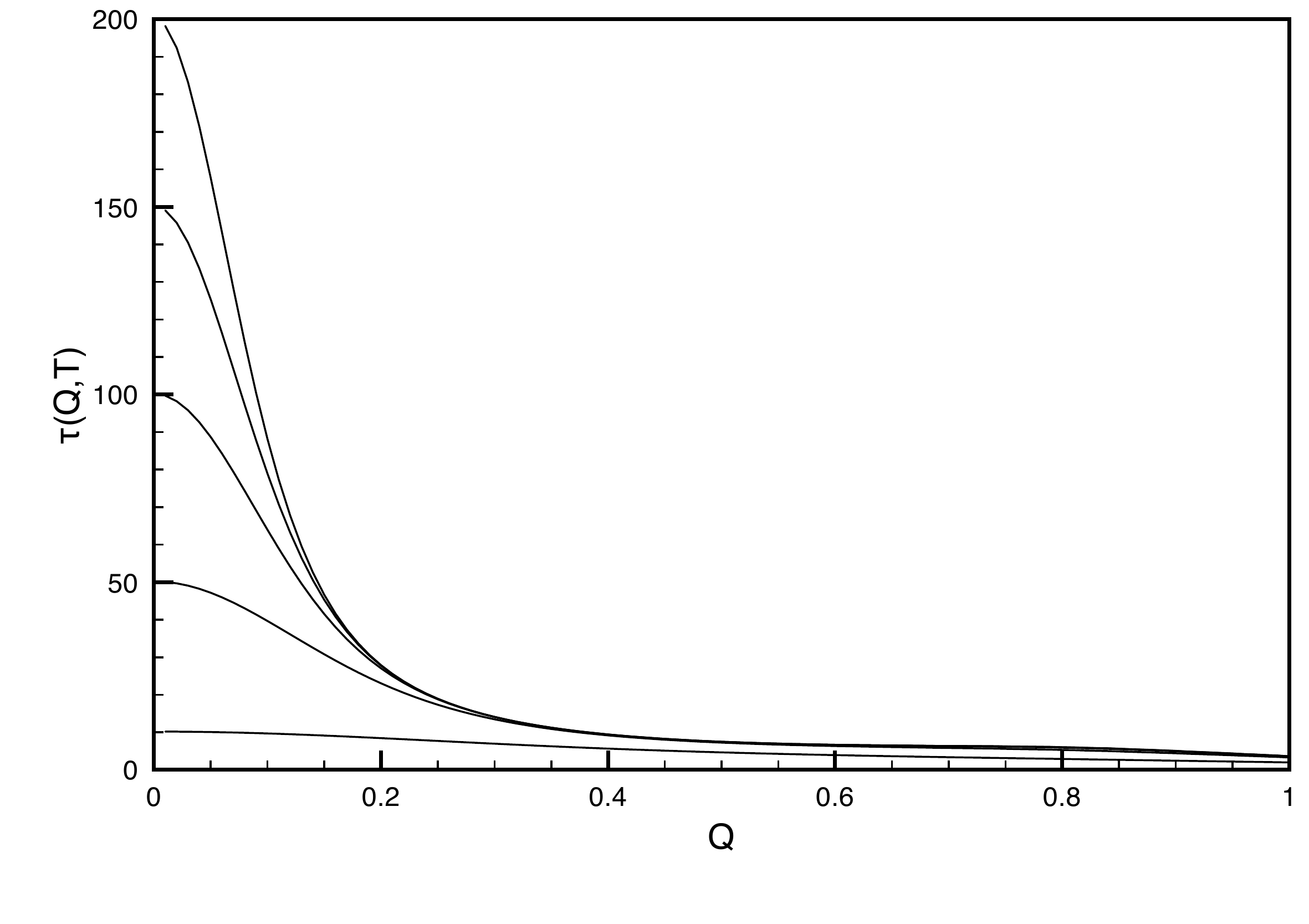}\end{center}
\caption{\small  Plot of $\tau (Q,T)$ versus wavenumber for different times T= 10.0, 50.0, 100.0, 150.0 and 200.0 
from  bottom to top.  The coupling $g=90.0$.}     
\label{fig:6}
\end{figure}

\begin{figure}
\begin{center}\includegraphics[width=6in,scale=1.0]{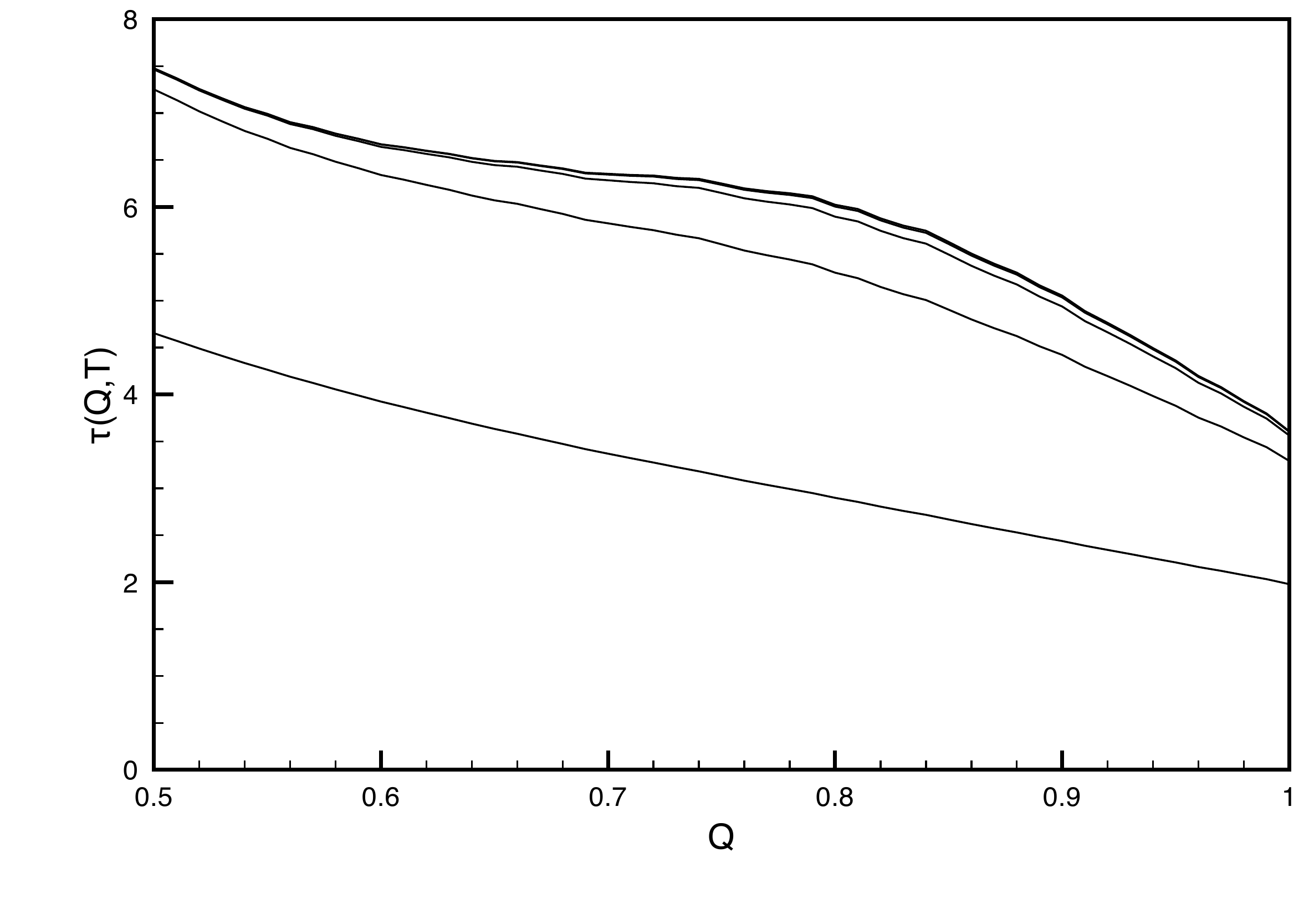}\end{center}
\caption{\small  Plot of $\tau (Q,T)$ versus wavenumber for different times 
$T= 10.0, 50.0, 100.0, 150.0$ and $200.0$ 
from  bottom to top.  The coupling $g=90.0$.}     
\label{fig:7}
\end{figure}

\begin{figure}
\begin{center}\includegraphics[width=6in,scale=1.0]{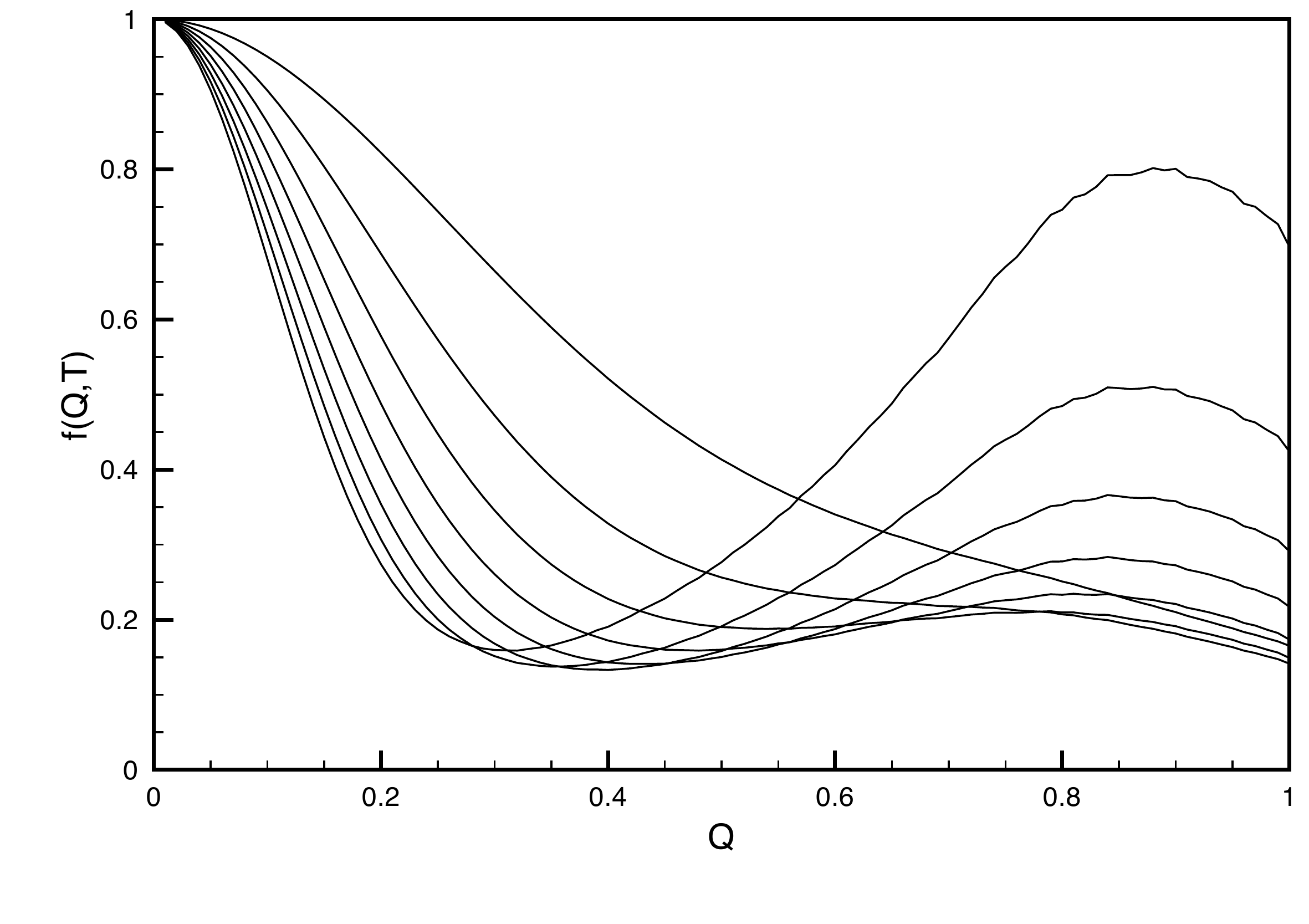}\end{center}
\caption{\small  Plot of normalized intermediate structure factor 
versus high wavenumber for different times $T= 5.0, 10.0, 15.0, 20.0, 25.0,
30.0, 35.0$ and $40.0$ 
from  bottom to top  at $Q=1$. The coupling $g=100.0$ is above the critical value.}     
\label{fig:8}
\end{figure}

\begin{figure}
\begin{center}\includegraphics[width=6in,scale=1.0]{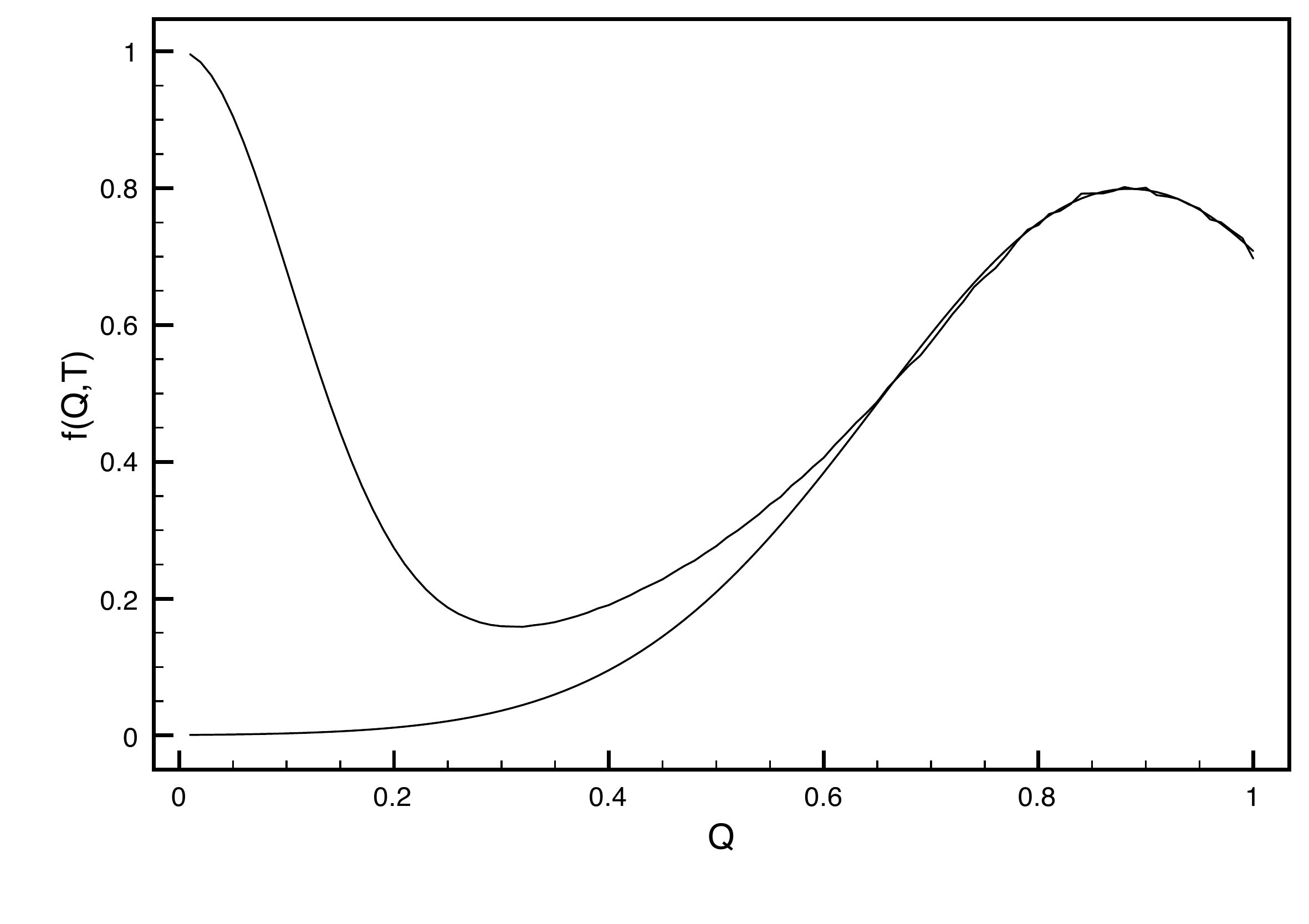}\end{center}
\caption{\small  Plot of the normalized intermediate structure factor
versus wavenumber for  time $T= 40.0$.
The coupling $g=100.0$ is above the critical value. 
Shown also is a fit to a gaussian of the form given by Eq.(\ref{eq:210}). }
\label{fig:9}
\end{figure}

\begin{figure}
\begin{center}\includegraphics[width=6in,scale=1.0]{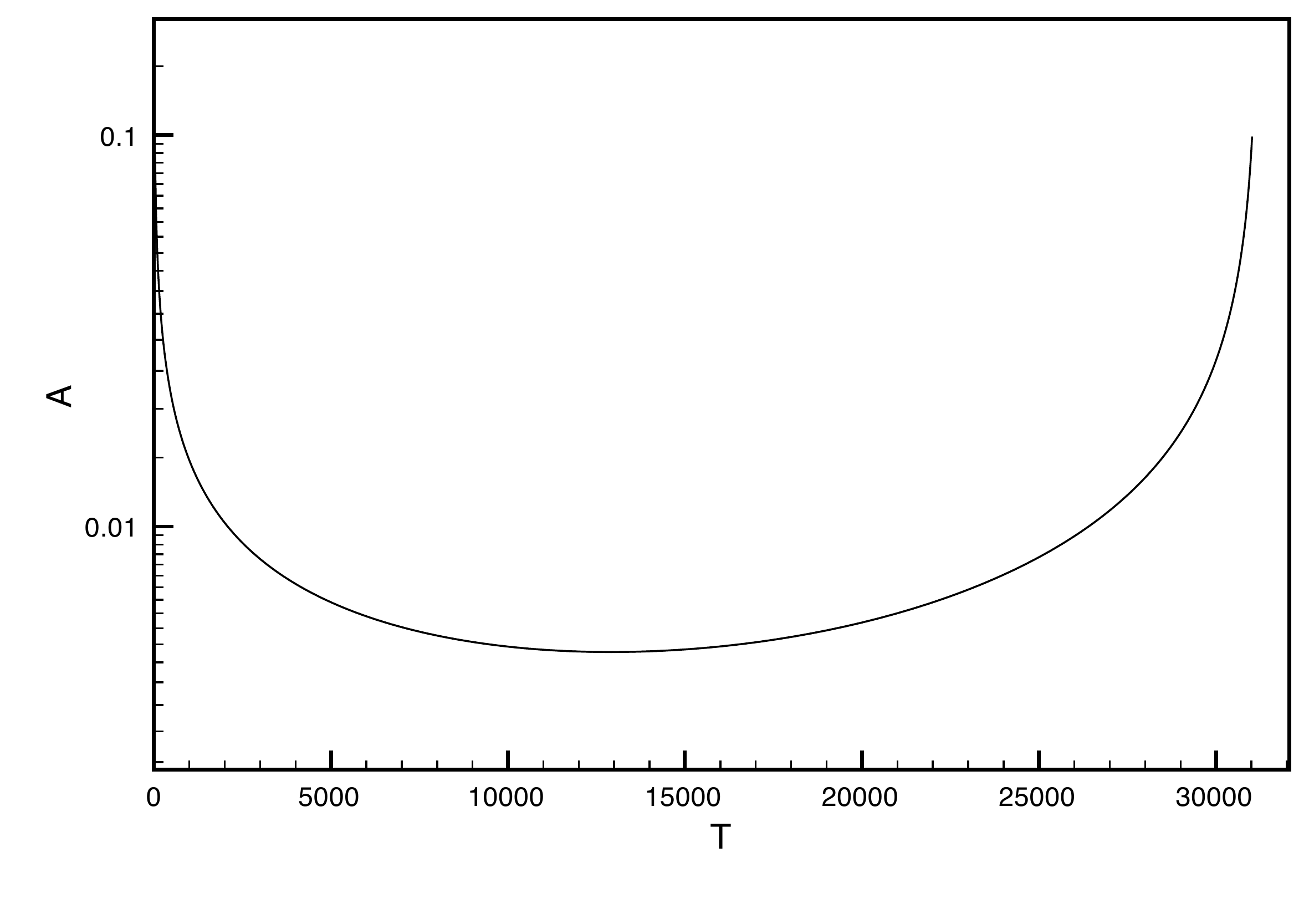}\end{center}
\caption{\small  Typical behavior of peak amplitude  $A$
versus time for
coupling $g > g^{*}$. This example is for $g=93.005$.}
\label{fig:13}
\end{figure}

\begin{figure}
\begin{center}\includegraphics[width=6in,scale=1.0]{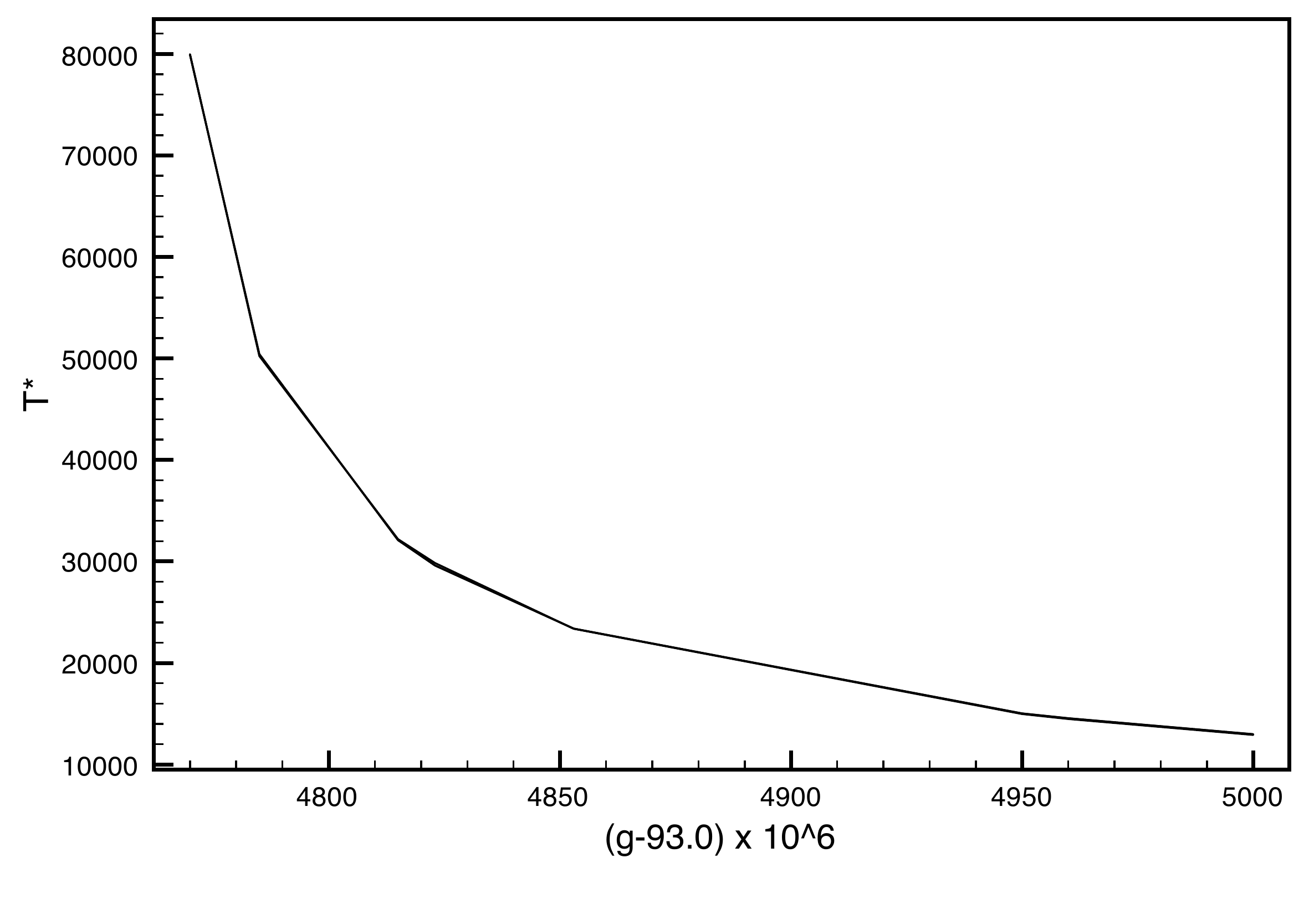}\end{center}
\caption{\small  Plot of the time $T^{*}$ where the peak amplitude $A$ hits a minimum
before going unstable versus $g$.  There is an excellent fit 
to $T^{*}=435445/(g-g^{*})^{0.6389}$
with $g^{*}=93.0047558$.}
\label{fig:14}
\end{figure}

\begin{figure}
\begin{center}\includegraphics[width=6in,scale=1.0]{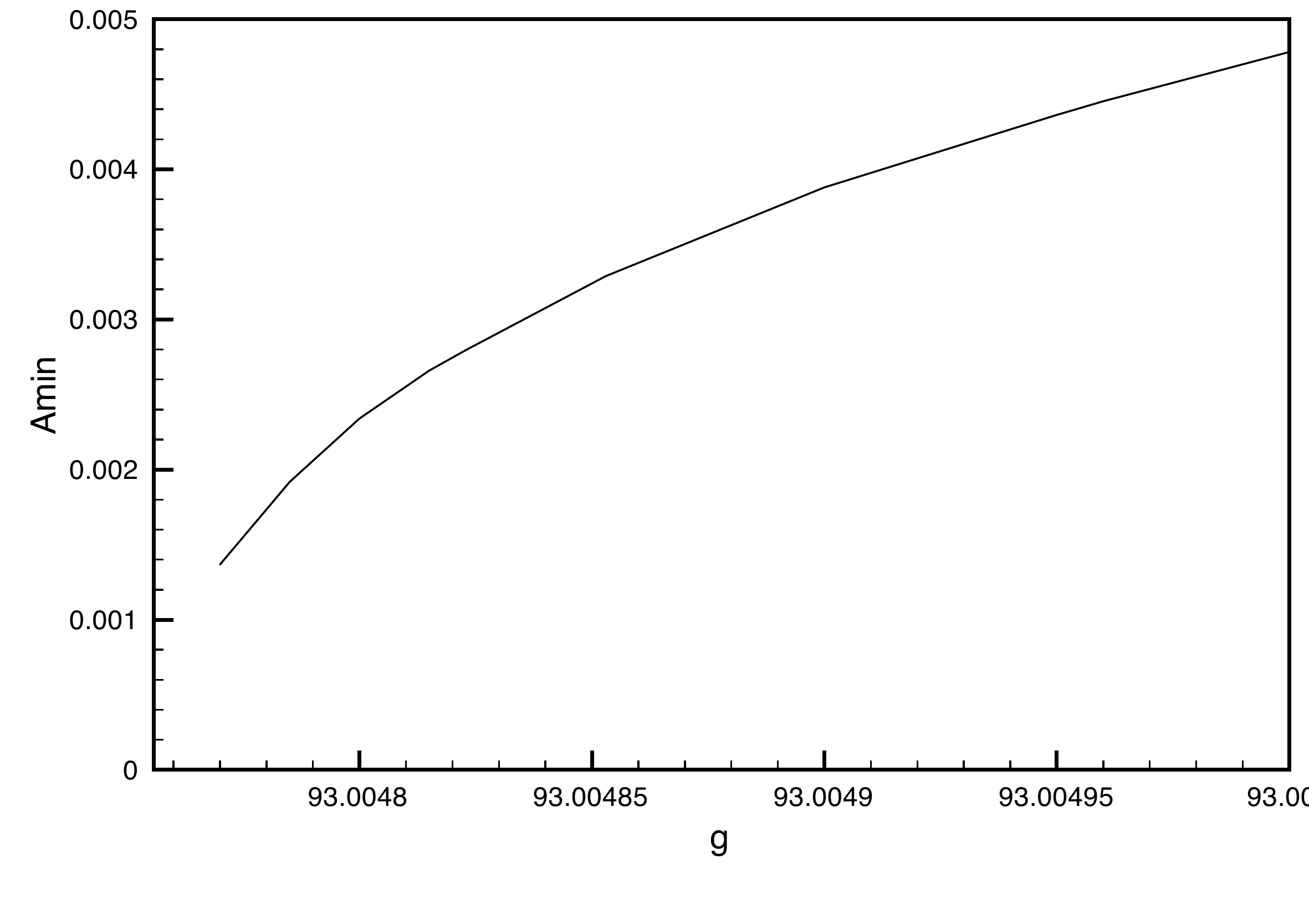}\end{center}
\caption{\small  Plot  of $A_{min}$, the peak amplitude  minimum
before going unstable, as a function of $g$.}  
\label{fig:15}
\end{figure}

\begin{figure}
\begin{center}\includegraphics[width=6in,scale=1.0]{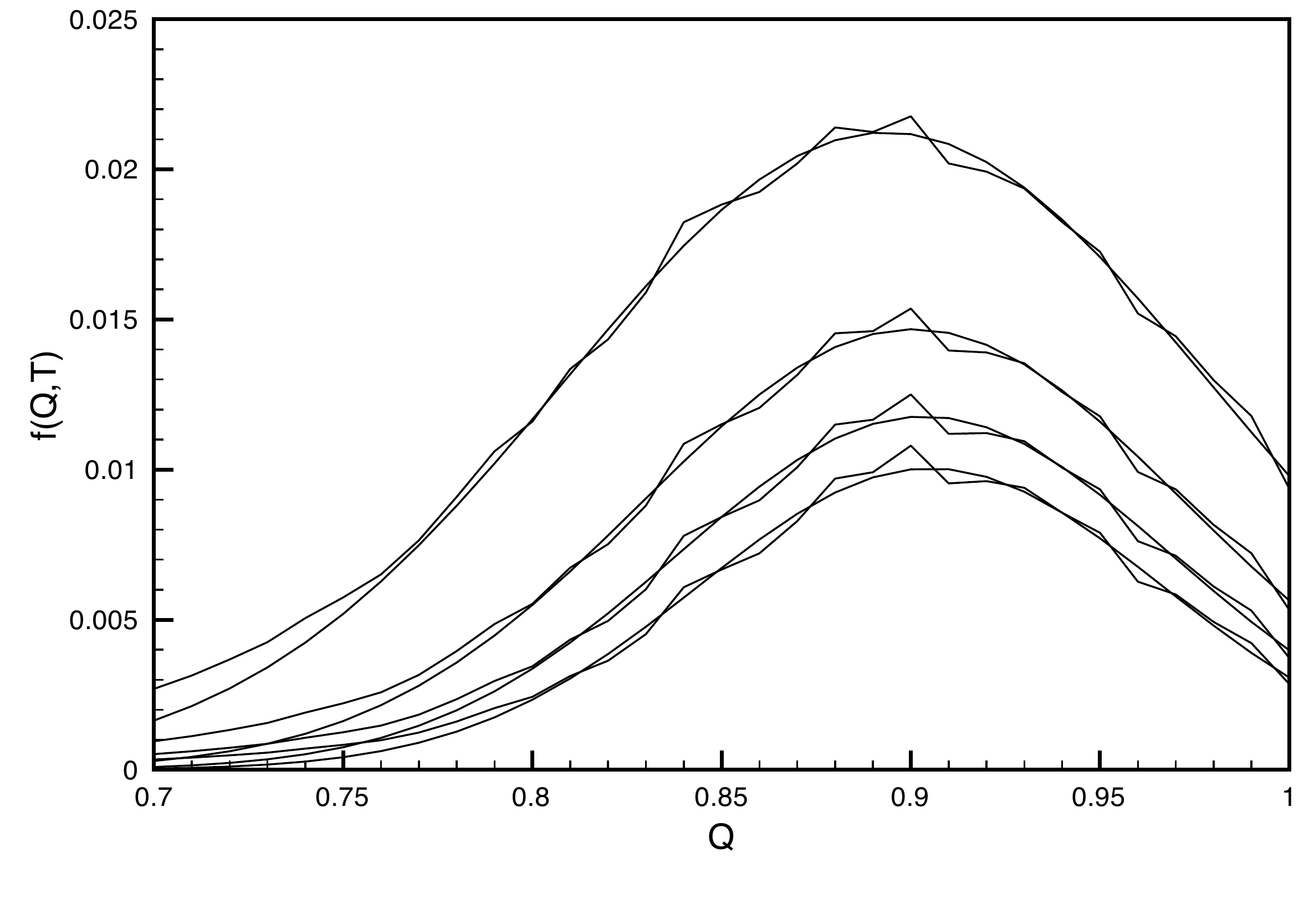}\end{center}
\caption{\small  Plot  of the normalized intermediate
structure factor for high wavenumbers at the critical
coupling $g=g^{*}$ for T=500, 1000, 1500, and 2000. 
Shown also is the gaussian fit to the
structural peak.}
\label{fig:16}
\end{figure}

\begin{figure}
\begin{center}\includegraphics[width=6in,scale=1.0]{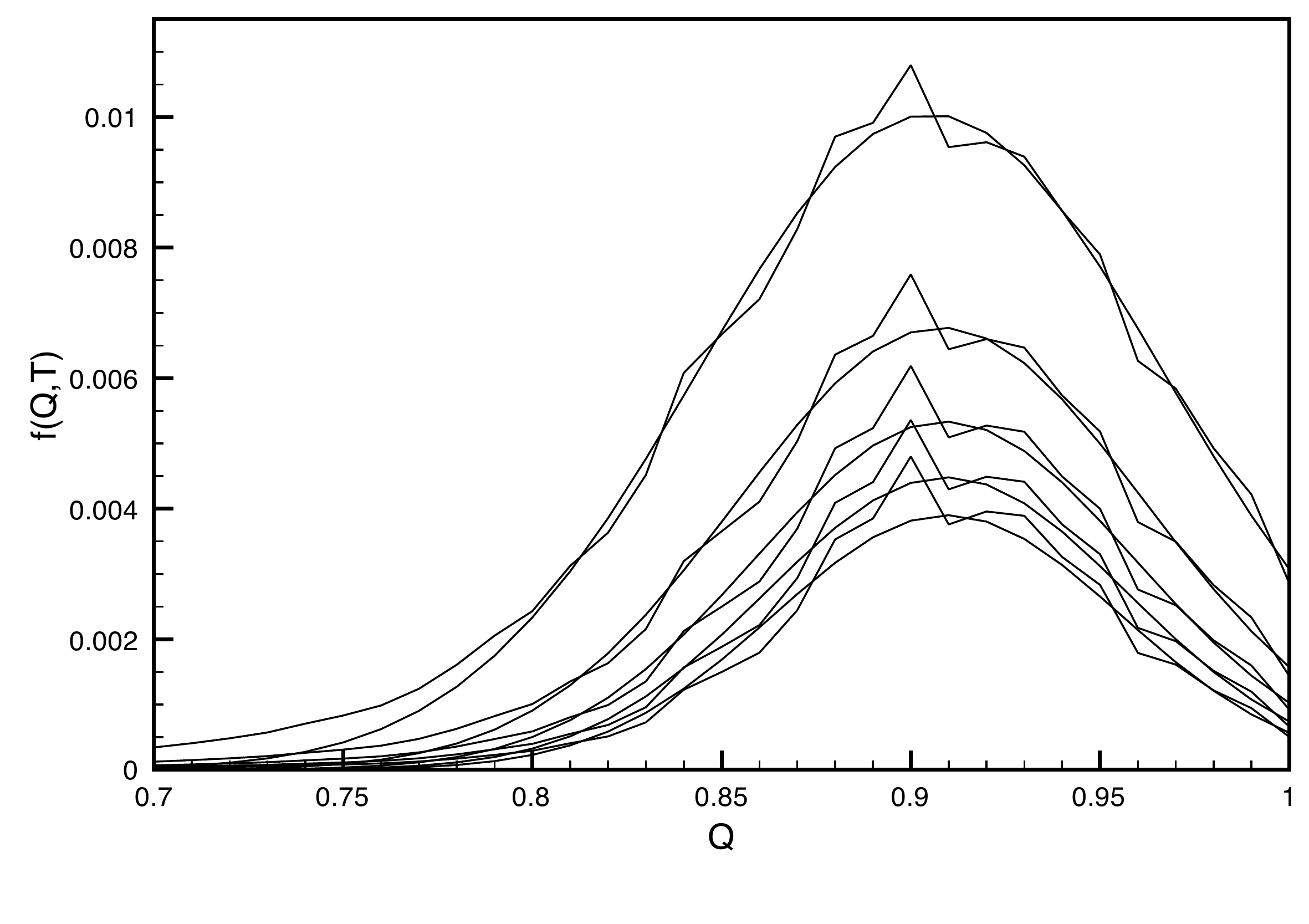}\end{center}
\caption{\small  Plot  of the normalized intermediate 
structure factor for high wavenumbers at the critical
coupling $g=g^{*}$ for $T=2000, 4000, 6000, 8000$,  and $10000$. 
Shown also is the gaussian fit to the
structural peak.}
\label{fig:17}
\end{figure}

\begin{figure}
\begin{center}\includegraphics[width=6in,scale=1.0]{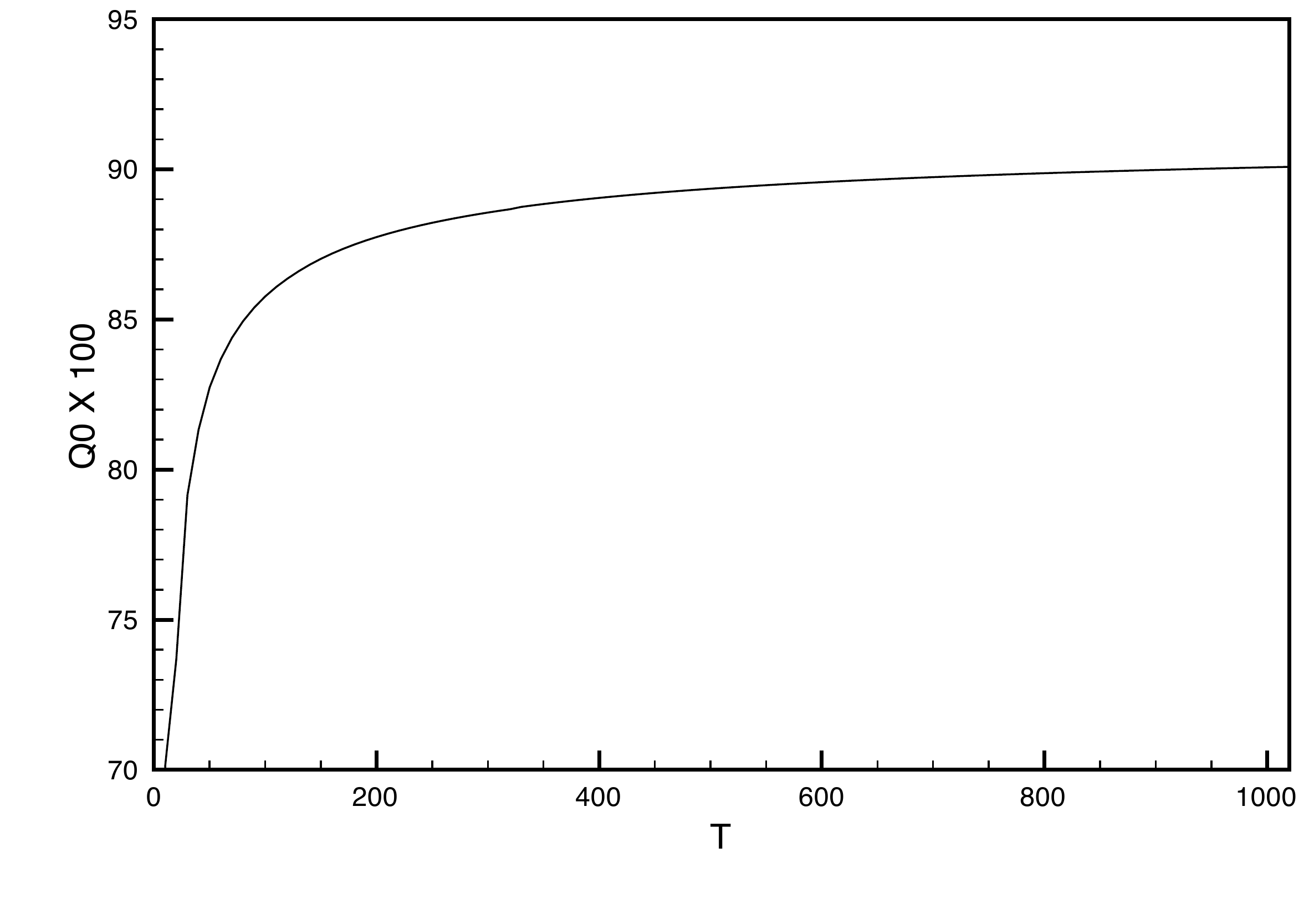}\end{center}
\caption{\small  Plot  of the ordering wavenumber (X 100) versus time  at the critical
coupling $g=g^{*}$.}
\label{fig:18}
\end{figure}

\begin{figure}
\begin{center}\includegraphics[width=6in,scale=1.0]{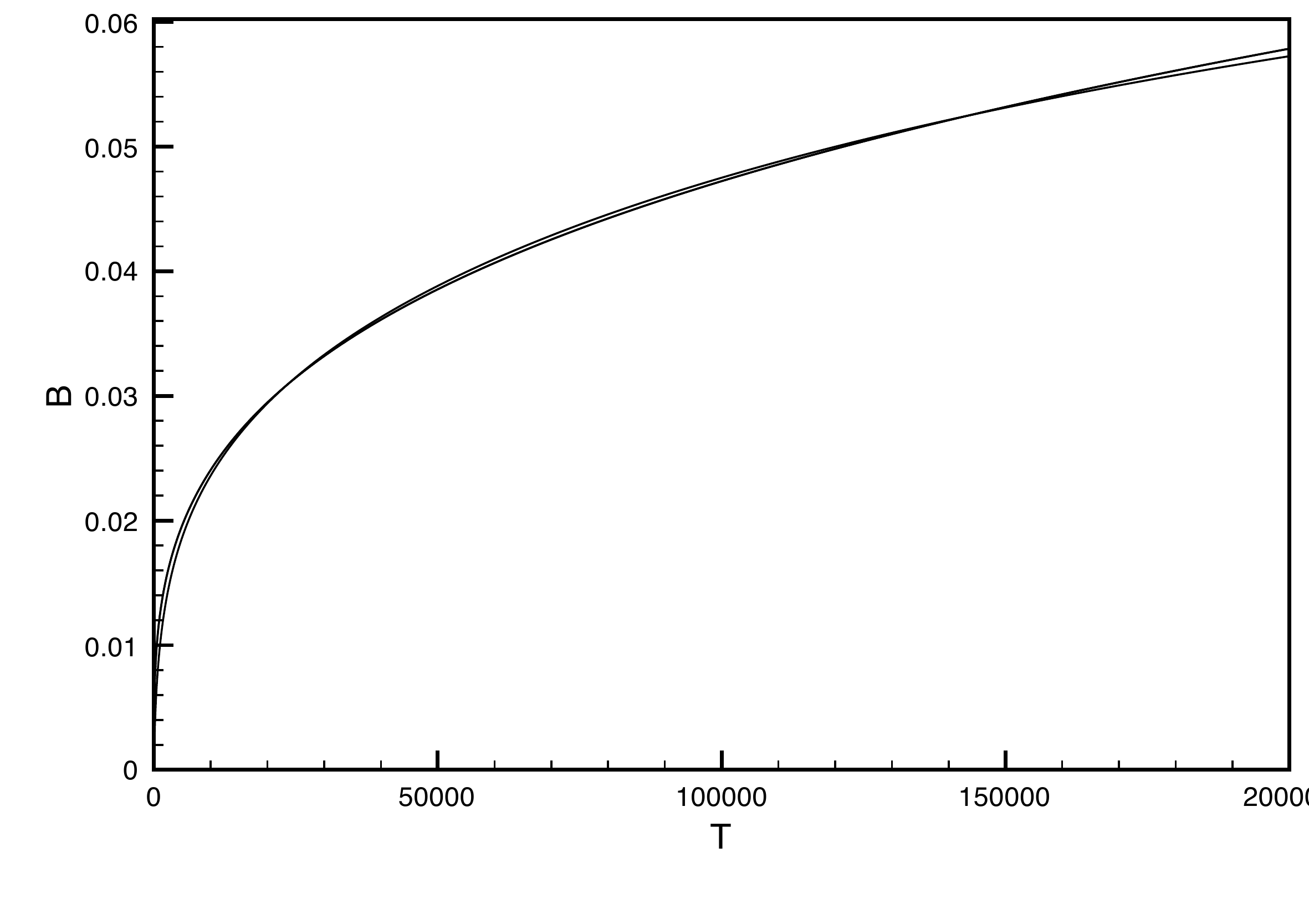}\end{center}
\caption{\small  Plot  of the peak width versus time  at the critical
coupling $g=g^{*}$. Fit is shown to a power law form $B=16.21 (T+121.2)^{0.293}$.}
\label{fig:19}
\end{figure}

\begin{figure}
\begin{center}\includegraphics[width=6in,scale=1.0]{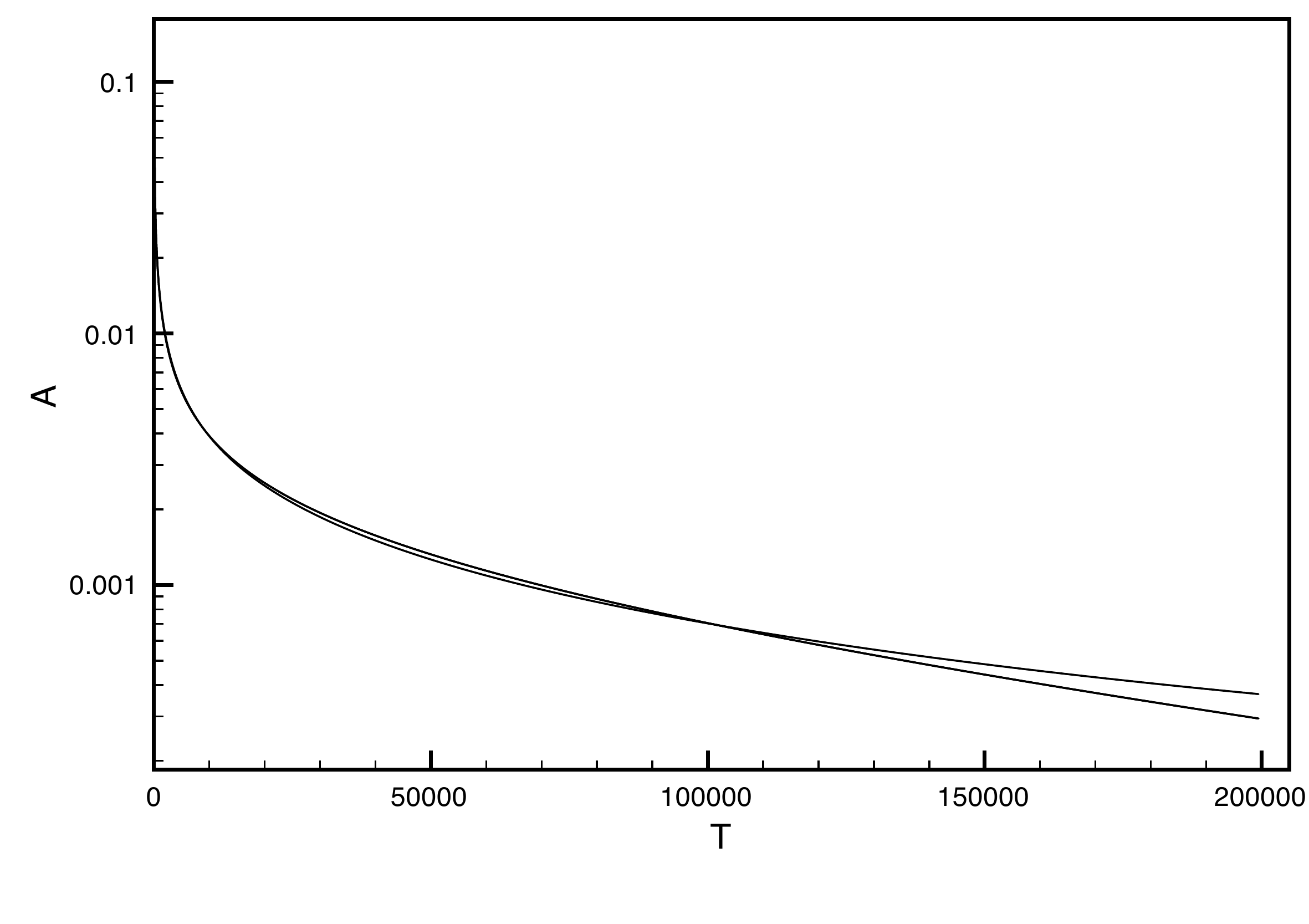}\end{center}
\caption{\small  Plot  of the peak amplitude  versus time  at the critical
coupling $g=g^{*}$. Fit shown is to the form given be Eq.(\ref{eq:288}) with
$A_{0}=0.672$, $E=4.70\times 10^{-6}$, $t_{0}=9.00$ and $\alpha =0.55$.}
\label{fig:20}
\end{figure}

\noindent
We first look at the solution for $f(Q,T)$ numerically.  We begin, in 
Fig.\ref{fig:3},
 with small $g$ and
find, as in bare perturbation theory, near exponential decay with time for fixed wavenumber.
Another way of characterizing the data is in terms of a running relaxation time
\be
\tau (Q,T)=\int_{0}^{T}dt f(Q,t)
~~~.
\label{eq:286}
\ee
This quantity is shown in Fig.\ref{fig:4} for a series of short times and 
coupling
$g=10$.  Notice that $\tau (Q,T)$ has approached $\tau (Q)=\tau (Q,\infty )$ for the largest times shown for $Q > 0.75$ but not for smaller $Q$.  If we increase $g$ to $60.0$ we
see substantial slowing down as shown in Fig.\ref{fig:5}.  These trends are continued as
we move to $g=90.0$ as shown in Fig.\ref{fig:6} with the new feature,
shown more clearly in Fig.\ref{fig:7}, that we see the development
of a weak peak near $Q\approx 0.80$ which saturates at $T=200.0$.  Finally, with
$g=100.0$ the system, shown in Fig.\ref{fig:8}, is rendered unstable and a peak at $Q=0.88$
grows rapidly with time in the structure factor.
We show in Fig.\ref{fig:9} that this large wavenumber peak can be fit to a gaussian form
\be
f_{p}(Q,T)=Ae^{-B(Q-Q_{0})^{2}}
\label{eq:210}
\ee
where $A$ is the peak amplitude,
$B$ is a new growing length squared in the problem and $Q_{0}$ is an ordering wavenumber.

The central $Q=0$ peak
can also be fit to a gaussian with $Q_{0}=0$ and $A=1$ and 
another growing length squared $B_{0}$.
We find $B_{0}\gg B$ and both grow as a power law with time.

We see that as $g$ goes from $90$ to $100$ that the system goes from stable to unstable.
How do we find the transition value $g^{*}$?
A good way of determining $g^{*}$ is to work in the unstable phase where
the structure peak  amplitude $A$ has a minimum at time $T^{*}$,
as a function of $T$ as shown, 
for example,
in Fig.\ref{fig:13}. 
The closer $g$ is to $g^{*}$
the longer $T^{*}$.   If one plots, as in Fig.\ref{fig:14}, $T^{*}$ versus $g$ and fits $T^{*}$
to a power-law diverging at $g=g^{*}$ one obtains an accurate estimate of $g^{*}$.
A good fit to the data shown in Fig.\ref{fig:14} gives an estimate $g^{*}=93.0047558$.
If we plot the value of the amplitude minimum, $A_{min}$, for $g > g^{*}$, we see 
in Fig.\ref{fig:15}, as $g\rightarrow g^{*}$ from
above, $A_{min}$ appears to go to zero.

\begin{figure}
\begin{center}\includegraphics[width=6in,scale=1.0]{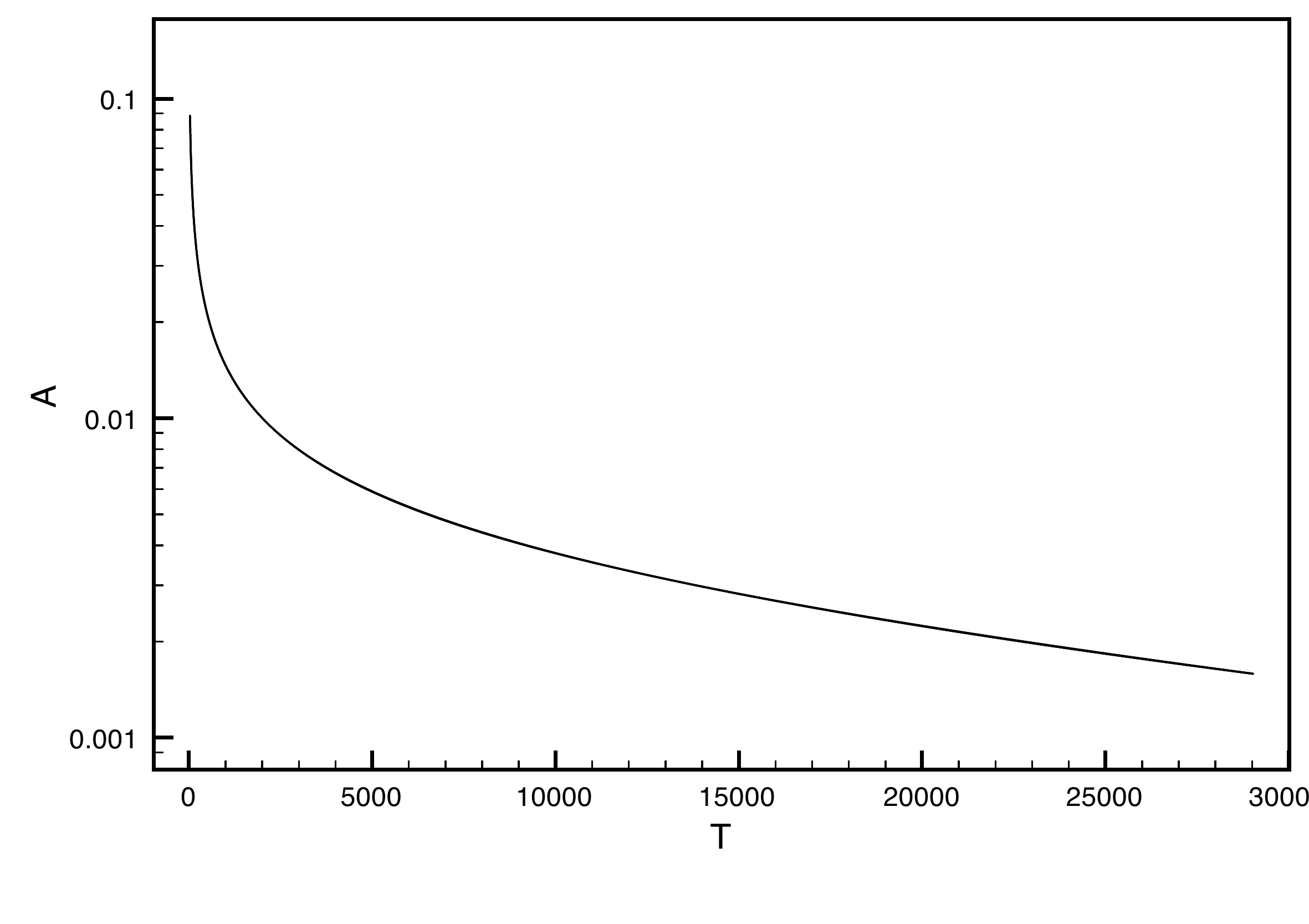}\end{center}
\caption{\small  Plot  of the peak amplitude  versus times  at coupling
$g=93.004720$. Fit is shown to the form given by Eq.(\ref{eq:288}).  The
fit is indistinguishable from the numerical data.}
\label{fig:21}
\end{figure}

\begin{figure}
\begin{center}\includegraphics[width=6in,scale=1.0]{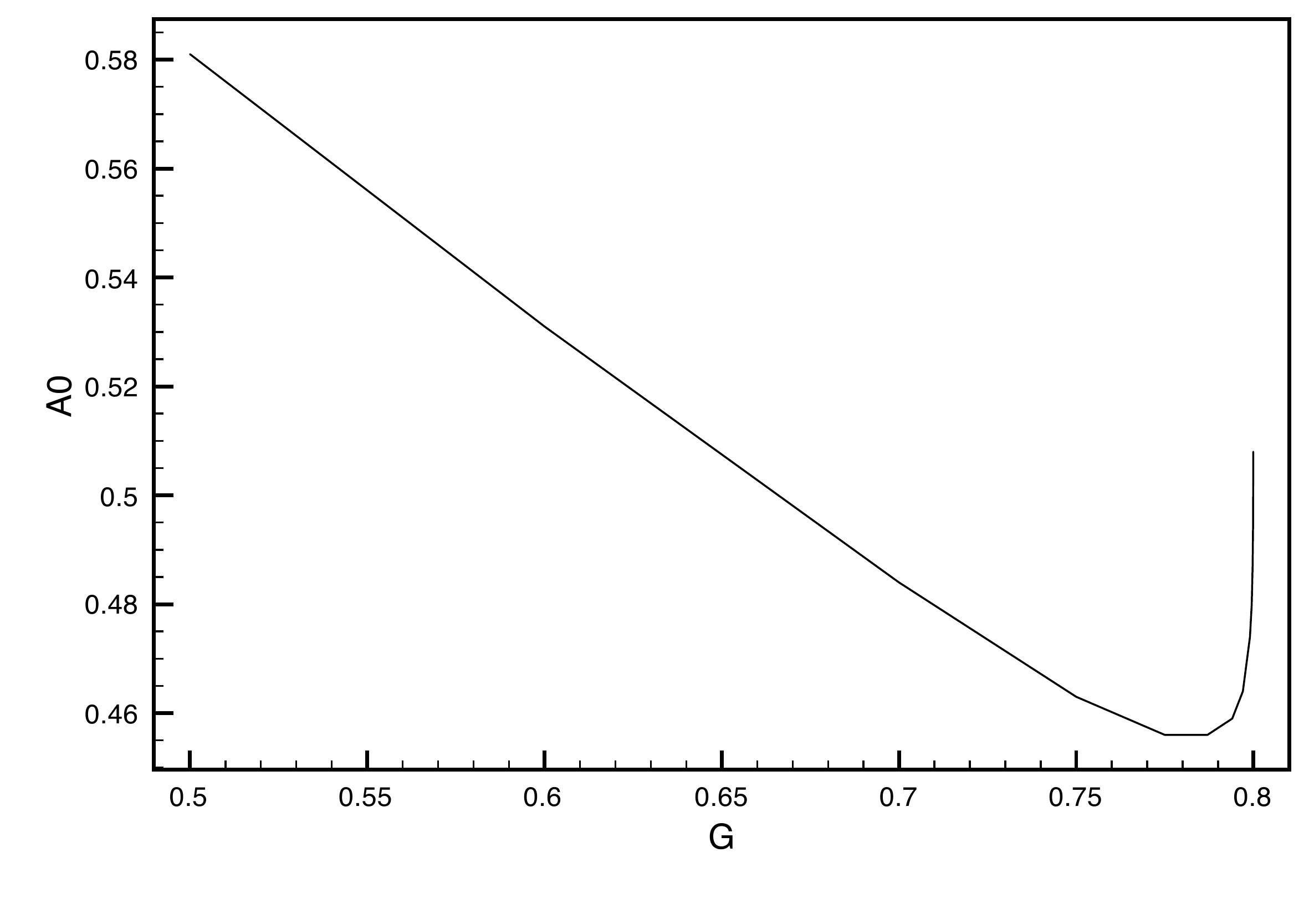}\end{center}
\caption{\small  Plot  of the parameter $A_{0}$ as a function of $g$.
Fit is given by $0.194/(93.0048 -g)^{0.104}$.}
\label{fig:22}
\end{figure}

\begin{figure}
\begin{center}\includegraphics[width=6in,scale=1.0]{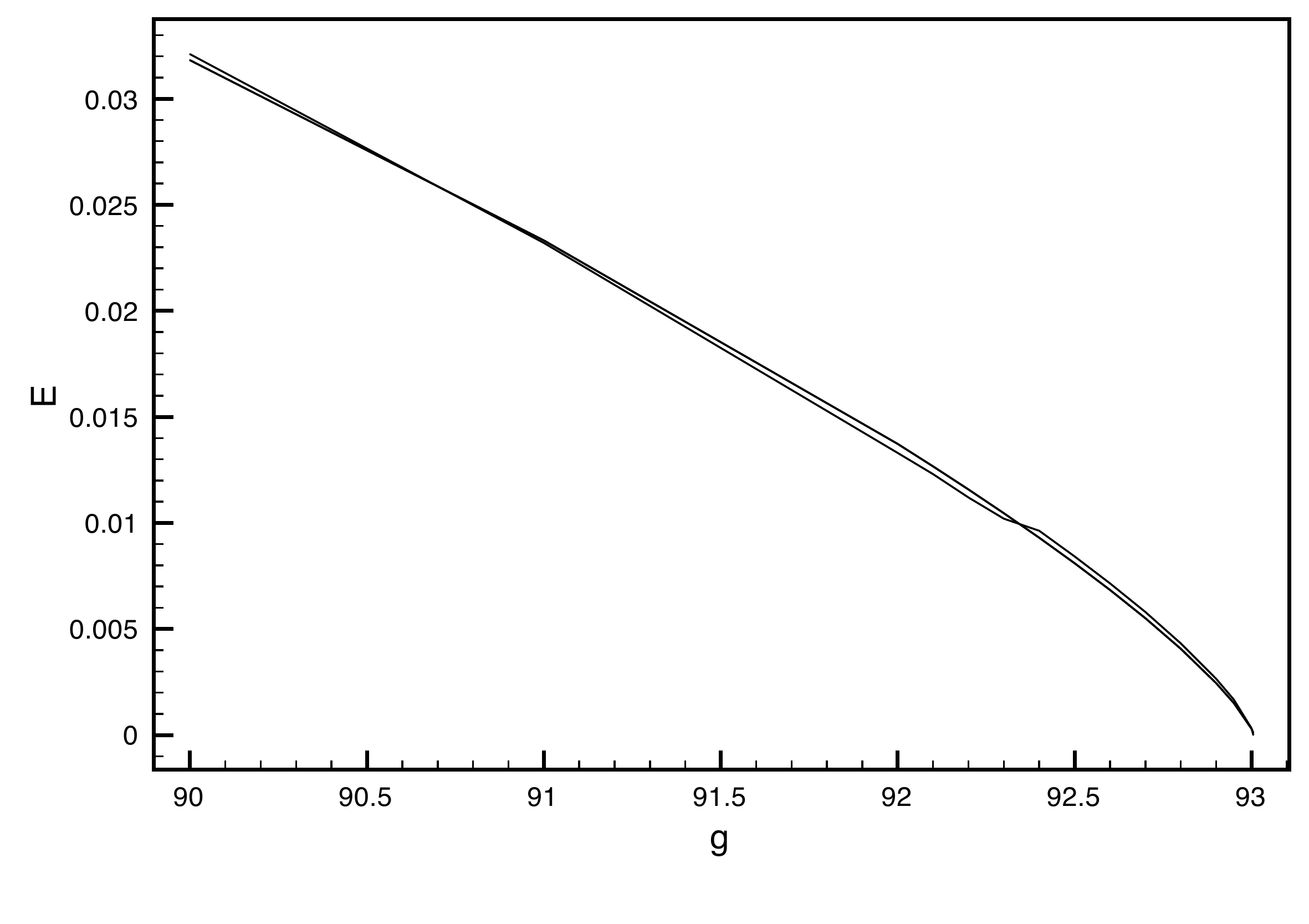}\end{center}
\caption{\small  Plot  of the parameter $E$  as a function of $g$.
The fit is to the form given by Eq.(\ref{eq:289}) with $A_{E}=0.0137$, $x_{B}=0.768$, $g^{*}=93.0067$.}
\label{fig:24}
\end{figure}

\begin{figure}
\begin{center}\includegraphics[width=6in,scale=1.0]{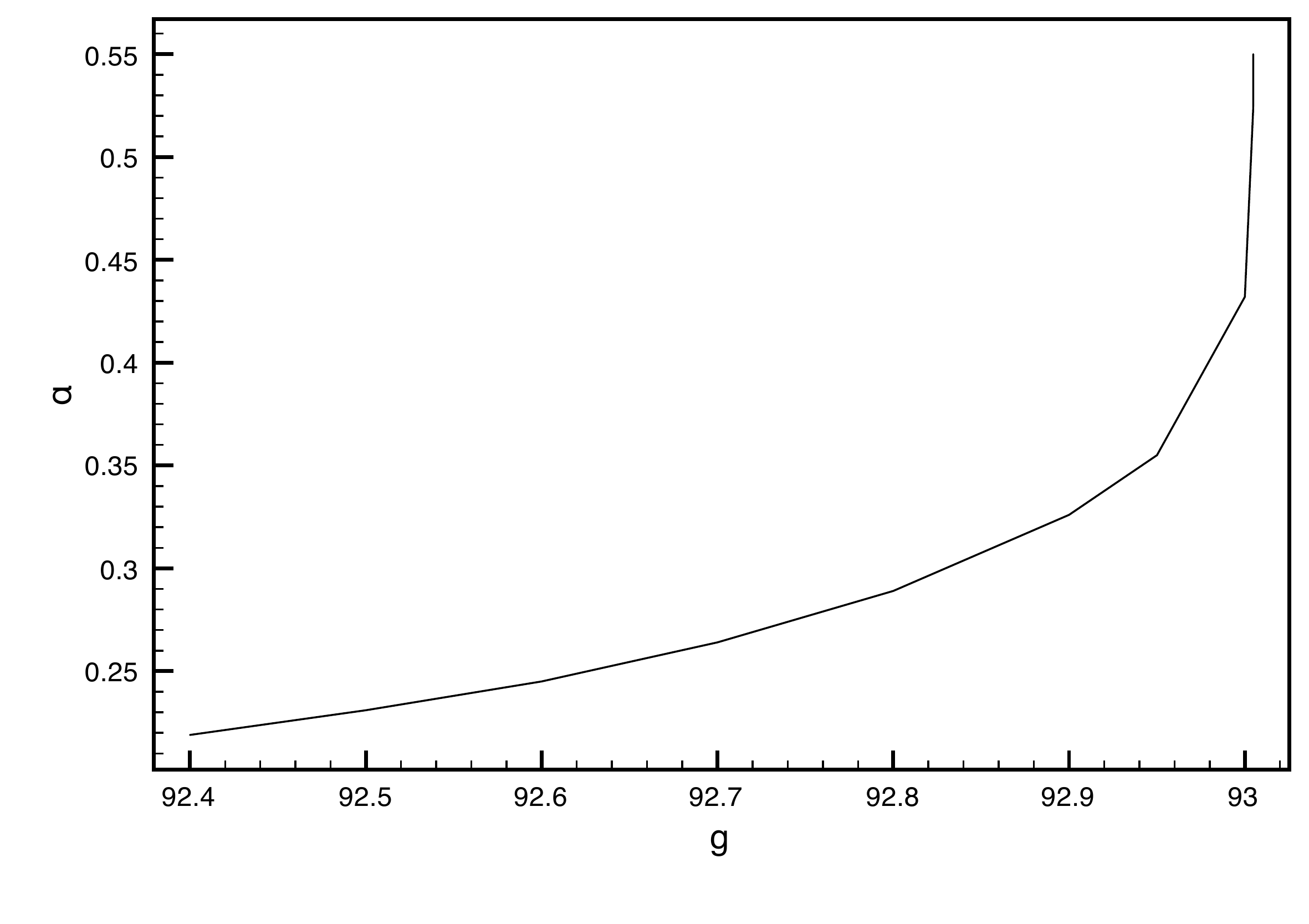}\end{center}
\caption{\small  Plot  of the parameter $\alpha$ in Eq.(\ref{eq:288}),  as a function of $g$.} 
\label{fig:25}
\end{figure}

Next we look at the behavior of the intermediate structure factor peak for
$g = g^{*}$.  In Figs.\ref{fig:16} and \ref{fig:17} we plot the structural peak for
intermediate times and show the gaussian fits. The fit parameters,
$Q_{0}$, $B$ , and $A$, are shown as functions of time in 
Figs.\ref{fig:18}, \ref{fig:19} and \ref{fig:20}.  
$Q_{0}$ orders
rapidly, while $B$ can be fit to a simple power-law form.  
For $g < g^{*}$,  $A$ can reasonably be fit using
\be
A=A_{0}\frac{e^{-E T}}{(T+t_{0})^{\alpha}}
~~~.
\label{eq:288}
\ee
Such fits are shown in Figs.\ref{fig:20} and \ref{fig:21}.  
The fit in Fig.\ref{fig:20} is over a very long time scale and is breaking down for
the longest time.
Carrying out fits using Eq.(\ref{eq:288}) for a range of
values of the coupling constant, we find  
the parameter $A_{0}$
shown as a function of $g$ in 
Fig.\ref{fig:22}.
The parameters $E$ and $\alpha$
are more interesting.  $E$, as a function of $g$, shown in Fig.\ref{fig:24},
 can be fit to the form
\be
E=A_{E}(g^{*}-g)^{x_{E}}
\label{eq:289}
\ee
with fitted parameters $A_{E}=0.0137$,  $x_{E}=0.768$ and $g^{*}=93.0067$.
It is clear physically that the critical point corresponds to $E\rightarrow 0$.
The exponent $\alpha$ is plotted in Fig.\ref{fig:25} versus $g$.

\begin{figure}
\begin{center}\includegraphics[width=6in,scale=1.0]{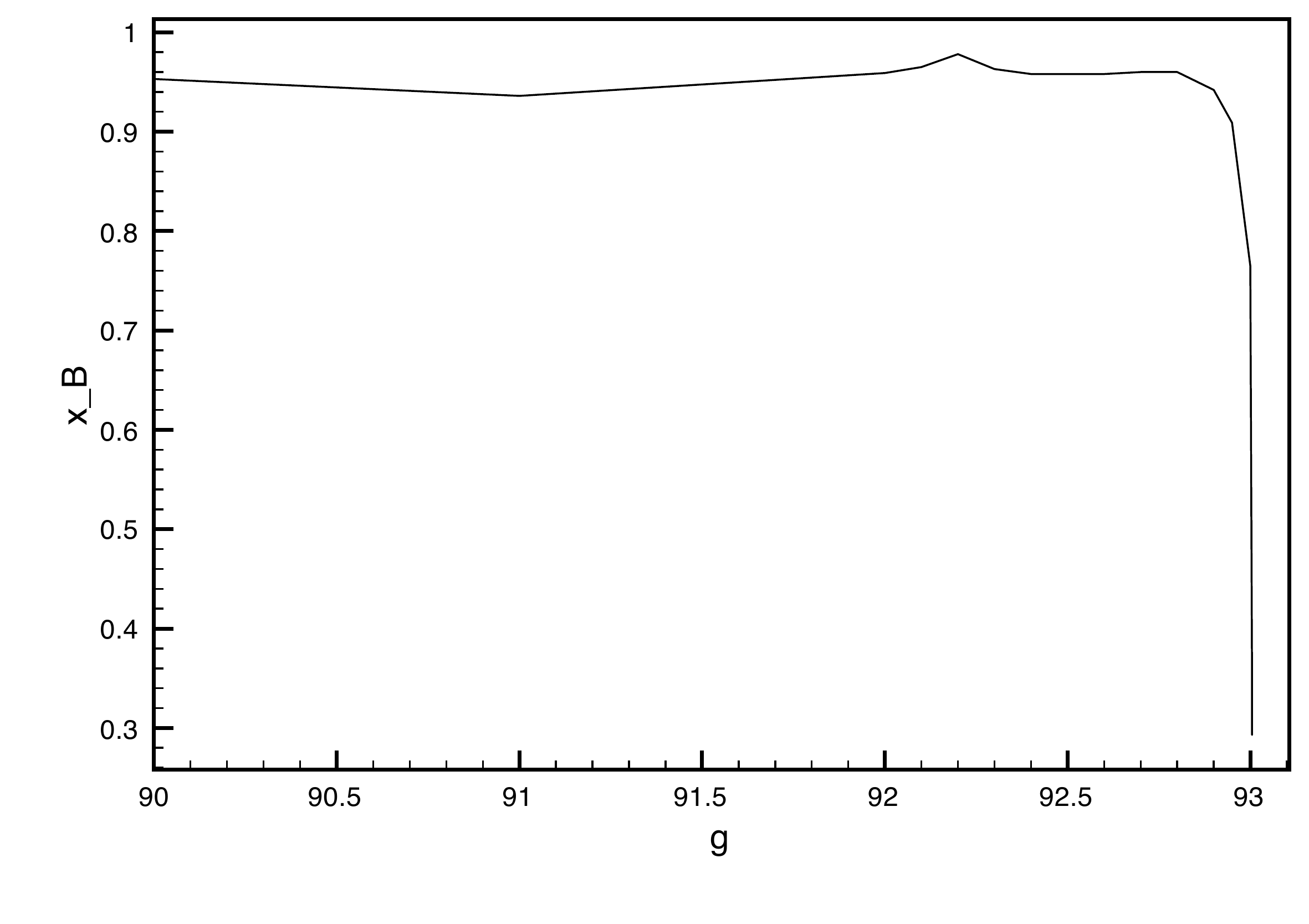}\end{center}
\caption{\small  Plot  of the parameter $x_{B}$ in Eq.(\ref{eq:289}),  as a function of $g$.}
\label{fig:26}
\end{figure}

The new kinetic length in the problem is $\sqrt{B}$.  As shown in 
Fig.\ref{fig:19},
for the critical coupling,
$B$ can be fit to the form
\be
B=B_{0}T^{x_{B}}
\ee
The exponent $x_{B}$ is shown as a function of $g$ in 
Fig.\ref{fig:26}.  Away from the critical point
$B$ grows nearly linearly with time but it crosses over to $1/3$ as
$g\rightarrow g^{*}$.

\newpage

\newpage

\pagebreak

\begin{figure}
\begin{center}\includegraphics[width=6in,scale=1.0]{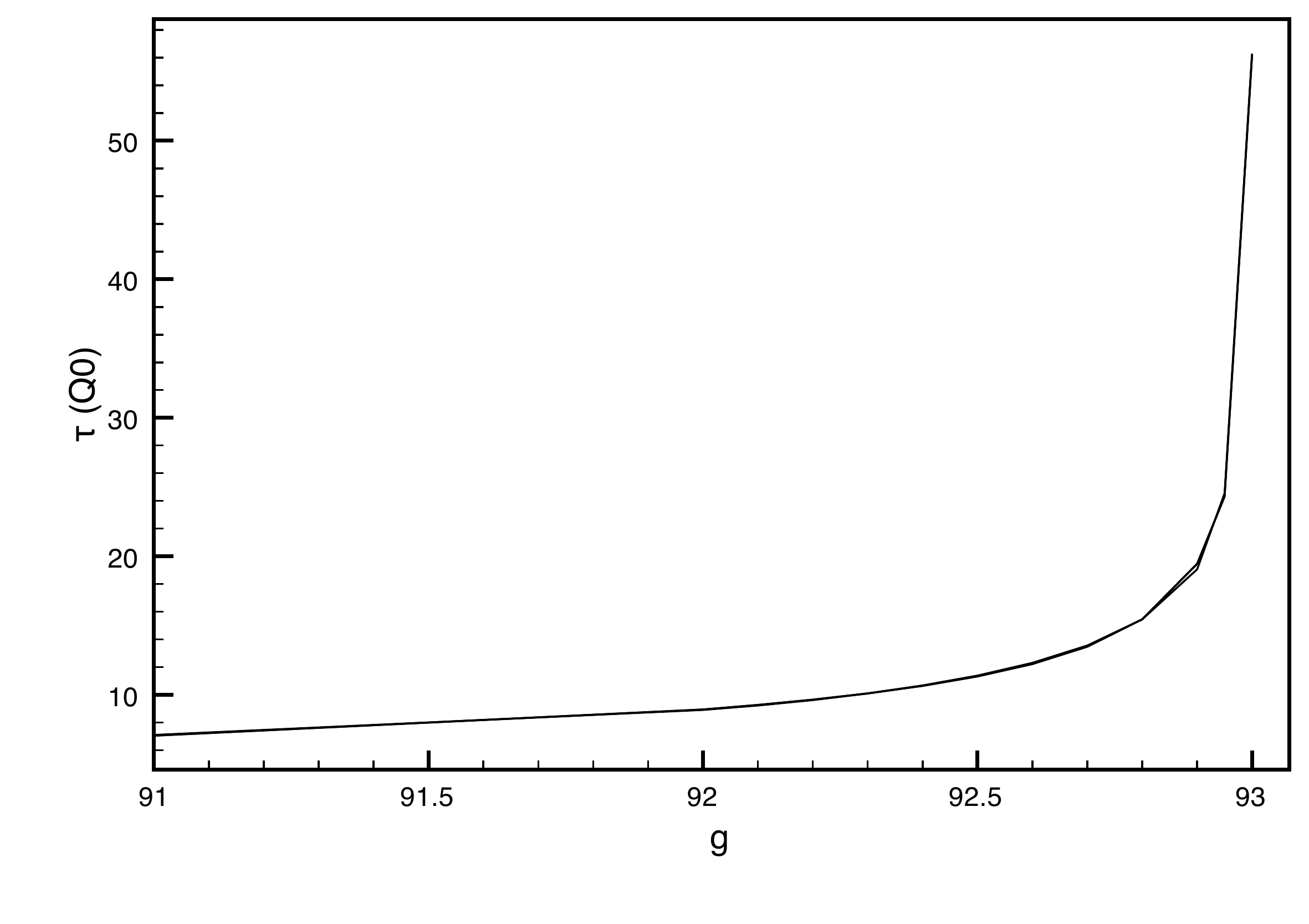}\end{center}
\caption{\small  Plot  of $\tau (Q_{0})$   as a function of $g$.
Fit is to form $\tau_{0}/(g^{*}-g)^{x_{\tau}}$ with values
$\tau_{0}=8.96$, $g^{*}=93.0048$ and $x_{\tau}=0.344$.}
\label{fig:28}
\end{figure}

One can return to the relaxation time defined by Eq.(\ref{eq:286})
with $\tau (Q)=\tau (Q,\infty )$.
In Fig.\ref{fig:28} we plot the the maximum ($Q>0.5$) of $\tau (Q)$ versus coupling
$g$.  We see a very strong dependence on $g$ as it approaches $g^{*}$.

\pagebreak

One can estimate $\tau (Q_{0})$ using Eq.(\ref{eq:288}) for the peak
contribution to the structure factor,
\be
\tau (Q_{0})=\int_{0}^{\infty}dt~A_{0}\frac{e^{-Et}}{(t+t_{0})^{\alpha}}
\nonumber
\ee
\be
=A_{0}E^{\alpha -1}\int_{0}^{\infty}dy\frac{e^{-y}}{(y+t_{0}E)^{\alpha}}
~~~.
\ee
Assuming as $E\rightarrow 0$, that $t_{0}E$ goes to zero, we have
\be
\tau (Q_{0})=A_{0}E^{\alpha -1}\Gamma (1-\alpha )
~~~.
\ee
The results  $E\approx (g^{*}-g)^{3/4}$ and $\tau (Q_{0}) \approx (g^{*}-g)^{-1/3}$
implies that $\alpha =5/9$ which is compatible with the value for $\alpha$
given by the fit in Fig.\ref{fig:20}.

We conclude that in this model there is the development of unanticipated
structure in the intermediate structure factor for $g$ near $g^{*}$.
In the stable regime the associated peak narrows thus giving a growing
length$\sqrt{B}$ in the problem.  In the next section we show how some
approximate analytical progress can be made looking at this peak formation.

\newpage

\newpage

\section{Peak Amplitude Equation}

\subsection{Mapping onto Amplitude Dynamics}

In order to make analytical progress on our one-loop
direct theory we assume that our 
long-time solution is of the form
\be
f(Q,T)=f_{0}(Q,T)+f_{p}(Q,T)
\label{eq:167}
\ee
where
\be
f_{0}(Q,T)=e^{-B_{0}Q^{2}}
\ee
and $f_{p}$ is given by Eq.(\ref{eq:210}).
$B$ and $B_{0}$ are large and $A$
grows or decreases with time depending on whether
we are in the stable or unstable phase.  $Q_{0}$ is
a fixed wavenumber characterizing the position of the peak.
Let us define
\be
\phi_{0}(T)=\sqrt{\frac{\pi}{B_{0}(T)}}
\ee
and
\be
\phi (T)=A(T)\sqrt{\frac{\pi}{B(T)}}
~~~.
\ee
Eq.(\ref{eq:167}) is then of the form
\be
f(Q,T)=\phi_{0}(T)\Delta _{B_{0}}(Q,T)
+\phi (T)\Delta_{B} (Q-Q_{0},T)
\ee
where
\be
\Delta _{B}(Q,T)=\sqrt{\frac{B(T)}{\pi}}e^{-B(T)Q^{2}}
\ee
and
\be
\lim_{T\rightarrow\infty}\Delta _{B}(Q,T)=\delta (Q)
~~~.
\ee

We then substitute the assumed solution, Eq.(\ref{eq:167}), into the kinetic 
equation
\be
\frac{\partial f(Q,T)}{\partial T}=-Q^{2}f(Q,T)
+Q^{4}\int_{0}^{T}dS N(Q,T-S)f(Q,S)
~~~.
\ee 
and look for self-consistency.  A key assumption is that the length squared $B$
is arbitrarily large. We work here in three dimensions. 

The peak near $Q=0$ is simpler to treat due to the explicit $Q$ dependence
in the kinetic equation.  For small enough $Q$ one can drop the interaction
term and one has
\be
\frac{\partial f_{0}(Q,T)}{\partial T}=-Q^{2}f_{0} (Q,T)
\ee 
with the solution
\be
f_{0}(Q,T)=e^{-Q^{2}T}
\ee
which gives $B_{0}=T$.

Next we focus on the peak near $Q_{0}$.  We can write
\be
\frac{\partial f_{p}(Q,T)}{\partial T}=-Q^{2}_{0}f_{p} (Q,T)
+Q^{4}_{0}\int_{0}^{T}dS N_{p}(Q_{0},T-S)f_{p}(Q,S)
\ee 
and we need to evaluate 
the memory kernel
\be
N_{p}(Q,T)=g\int\frac{d^{3}K}{(2\pi )^{3}}
f_{p}(K,T)f_{p}({\bf Q}-{\bf K},T)
\label{eq:174}
\ee
for $Q=Q_{0}$.  Assuming the $\delta$-function form
\be
f_{p}(Q,T)=\phi (T)\delta (Q-Q_{0})
~~~,
\label{eq:182}
\ee
one can do the ${\bf K}$ itegration in Eq.(\ref{eq:174}) with
the result
\be
N_{p}(Q_{0},T)
=\frac{gQ_{0}}{2\pi^{2}}\phi^{2}(T)
~~~.
\label{eq:193}
\ee
 We are left with the kinetic equation valid near $Q=Q_{0}$
 \be
 \frac{\partial f_{p}(Q,T)}{\partial T}
 =-Q_{0}^{2}f_{p} (Q,T)
 +Q_{0}^{4}\int_{0}^{T}dS N_{p}(Q_{0},T-S)
 f_{p}(Q,S)
 ~~~.
\label{eq:193}
 \ee
 Canceling a common factor of the $\delta$-function, gives
 the equation for the peak amplitude equation
 \be
 \dot{\phi}(T)=-Q_{0}^{2}\phi (T)
 +GQ_{0}^{4}\int_{0}^{T}dS \phi^{2}(T-S)\phi (S)
 \ee
 where
 \be 
 G=\frac{gQ_{0}}{2\pi^{2}}
~~~.
 \ee
 Changing the scaling of time to $t=Q_{0}^{2}T$ we obtain
 \be
 \dot{\phi}(t)=-\phi (t)
 +G\int_{0}^{t}ds \phi^{2}(t-s)\phi (s)
~~~.
\label{eq:249}
 \ee
If one replaces $\phi (s)$ by $\dot{\phi}(s)$ inside the integral
this equation of motion reduces to Leutheussar's equation\cite{Leut}.

We will assume that Eq.(\ref{eq:249}) can be solved as an initial-value
problem with $\phi ( 0)=1$.

We find that self-consistently that we have been able to replace the
kinetic equation, Eq.(\ref{eq:159}), with the amplitude equation.
Eq.(\ref{eq:249}).  We now show that the solutions to Eq.(\ref{eq:249})
shows the same phase structure as found in the numerical solution of
Eq.(\ref{eq:159}).

\subsection{Power-Law Solution}

In terms of Laplace transforms the equation of motion for the
amplitude $\phi$ satisfies
\be
\left[z+i+N(z)\right]\phi (z)=1
\ee
where
\be
\phi (z)=-i\int_{0}^{\infty}dt~e^{izt}\phi (t)
\ee 
and
\be
N(z)=-iG\int_{0}^{\infty}dt~e^{izt}\phi ^{2} (t)
~~~.
\ee
We want to show in the long-time limit and near the critical
coupling $G^{*}$ there is a power-law solution to Eq.(\ref{eq:249})
of the form
\be
\phi (t)=\frac{A_{0}}{(t+t_{0})^{\alpha}}
~~~.
\ee 
We want to
determine the exponent $\alpha$.

The first step is to work out the Laplace transforms for $\phi$ and
$N$ for the trial solution.  We have
\be
\phi (z)=-i\int_{0}^{\infty}dt~e^{izt}\frac{A_{0}}{(t+t_{0})^{\alpha} }
\nonumber
\ee 
\be
=-iA_{0}\int_{t_{0}}^{\infty}\frac{dx}{x^{\alpha}}e^{iz(x-t_{0})}
~~~.
\ee
Let $y=xz$ in the integral to obtain
\be
\phi (z)=-iA_{0}e^{-it_{0}z}z^{\alpha -1}\sigma (\alpha )
\ee
where the integral reduces to
\be
\sigma (\alpha )=\int_{t_{0}z}^{\infty}\frac{dy}{y^{\alpha}}e^{iy}
=\sigma_{0}-\frac{(t_{0}z)^{1-\alpha}}{(1-\alpha)}+\ldots
\ee
and
\be
\sigma_{0} (\alpha )=\int_{0}^{\infty}\frac{dy}{y^{\alpha}}e^{iy}
~~~.
\ee

Next look at the memory
kernel given by
\be
N (z)=-i\int_{0}^{\infty}dt~e^{izt}G\frac{A_{0}^{2}}{(t+t_{0})^{2\alpha} }
\nonumber
\ee 
\be
=-iA_{0}^{2}G\int_{t_{0}}^{\infty}\frac{dx}{x^{2\alpha}}e^{iz(x-t_{0})}
\nonumber
\ee
\be
=e^{-izt_{0}}[N(0)+\Delta N(z)]
\label{eq:207}
\ee
where
\be
N(0)=-iA_{0}^{2}G\int_{t_{0}}^{\infty}\frac{dx}{x^{2\alpha}}
\nonumber
\ee
\be
=\frac{-iA_{0}^{2}G}{2\alpha -1}\frac{1}{t_{0}^{2\alpha -1}}
\ee 
and we have assumed that $\alpha > 1/2$. In Eq.(\ref{eq:207})
\be
\Delta N(z)=-iA_{0}^{2}G\int_{t_{0}}^{\infty}\frac{dx}{x^{2\alpha}}
[e^{izx}-1]
~~~.
\ee

To lowest order in $z$
the power-law solution corresponds to the cancellation of terms in the
kinetic equation
\be
+i+N(0)=0
\ee
which gives the result
\be
\frac{A_{0}^{2}G}{2\alpha -1}\frac{1}{t_{0}^{2\alpha -1}}=1
\label{eq:375}
\ee
which depends explicitly on the time cutoff.  We turn to the next order term
in the small $z$ expansion of $i+N(z)$ which is given by 
\be
\Delta N(z)=-iA_{0}^{2}z^{2\alpha -1}G\int_{t_{0}z}^{\infty}\frac{dy}{y^{2\alpha}}
[e^{iy}-1]
\nonumber
\ee
\be
=-iA_{0}^{2}z^{2\alpha -1}G J(\alpha )
\ee
where we have the remaining integral
\be
J (\alpha )=\int_{t_{0}z}^{\infty}\frac{dy}{y^{2\alpha}}
[e^{iy}-1]
=J_{0} (\alpha )-i\frac{(t_{0}z)^{2(1-\alpha)}}{2(1-\alpha)}+\ldots
\ee
where
\be
J_{0} (\alpha )=\int_{0}^{\infty}\frac{dy}{y^{2\alpha}}
[e^{iy}-1]
~~~.
\ee

Assuming $0.5 < \alpha < 1$ the integrals $\sigma_{0}$ and $J_{0}$ can be
evaluated.  The kinetic equation then takes the form
\be
[z+i+e^{-izt_{0}}[N(0)-iA_{0}^{2}z^{2\alpha -1}G J(\alpha )]
[-iA_{0}e^{-it_{0}z}z^{\alpha -1}\sigma (\alpha )]=1
~~~.
\ee
Since $z^{2\alpha -1}\ll z$ this reduces to
\be
[-iA_{0}^{2}z^{2\alpha -1}G J_{0}(\alpha )]
[-iA_{0}e^{-it_{0}z}z^{\alpha -1}\sigma_{0} (\alpha )]=1
\ee
which requires
\be
2\alpha -1+\alpha -1 =0
\ee 
or $\alpha =2/3$ which is a self consistent value.  We are left with the
equation 
\be
-A_{0}^{3}GJ_{0}(2/3) \sigma_{0} (2/3)=1
\ee
It is left to appendix C to show that
\be
\sigma_{0} (2/3)J_{0}(2/3)=-2\pi\sqrt{3}
~~~.
\ee
We then  have the constraint on the solution
\be
2\pi\sqrt{3}A_{0}^{3}G=1
\label{eq:386}
~~~.
\ee
Notice that the results for $\alpha$ and $A_{0}$ do not depend on the 
short-time cutoff.

\subsection{Numerical analysis: One-loop Case.}

We can numerically solve the amplitude equation, Eq.(\ref{eq:249}).
We first determine  the times, $t^{*}$, when in unstable runs,
$\phi$  hits its
minimum versus $G$. A power-law fit assuming $t^{*}$ 
goes to infinity as $G$ goes to $G^{*}$
gives an estimate $G^{*}=0.79992\ldots$.

\begin{figure}
\begin{center}\includegraphics[width=6in,scale=1.0]{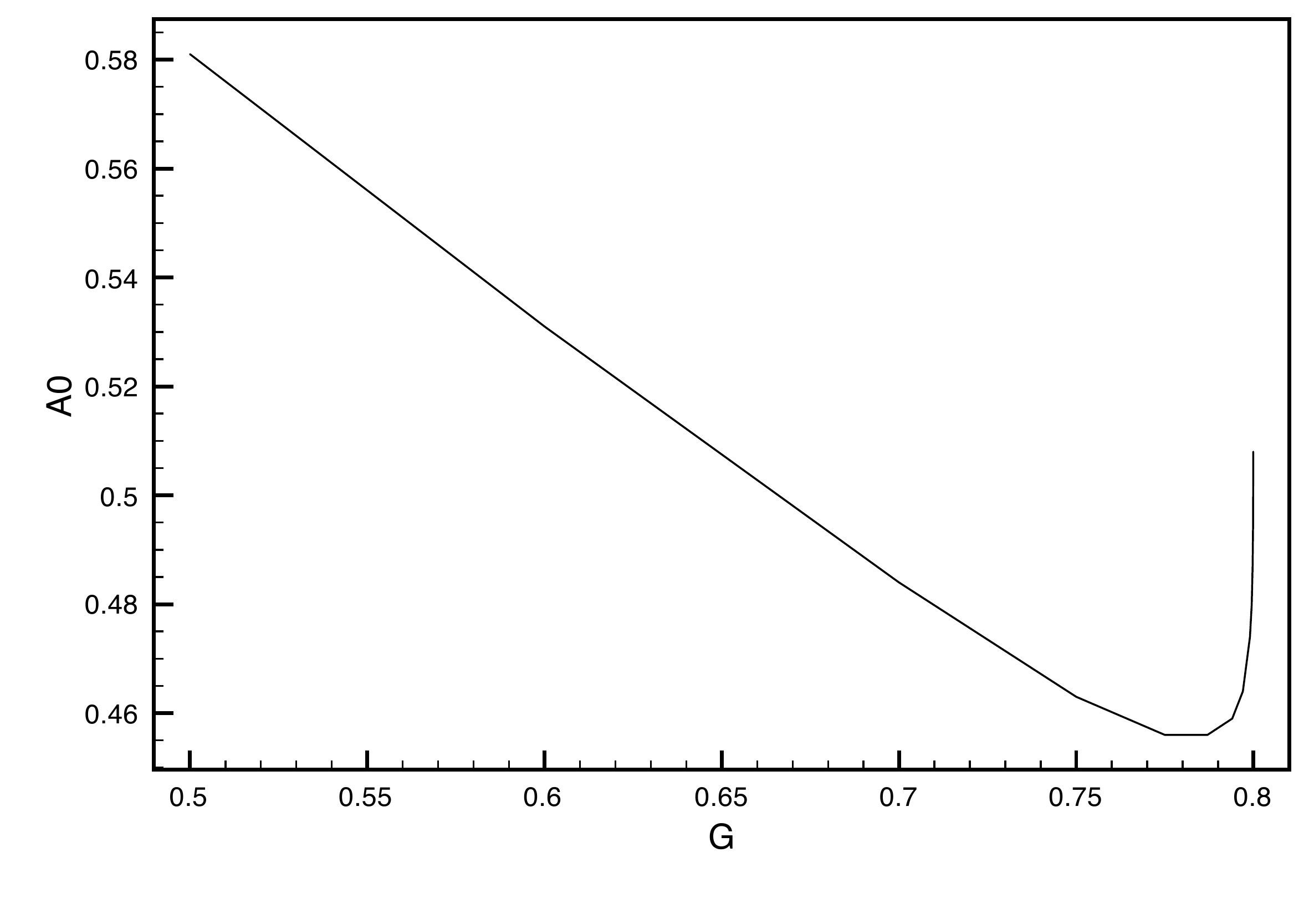}\end{center}
\caption{\small  Plot  of $A_{0}$, defined by Eq.(\ref{eq:221}),
as a function of $G$.}
\label{fig:30}
\end{figure}
\begin{figure}
\begin{center}\includegraphics[width=6in,scale=1.0]{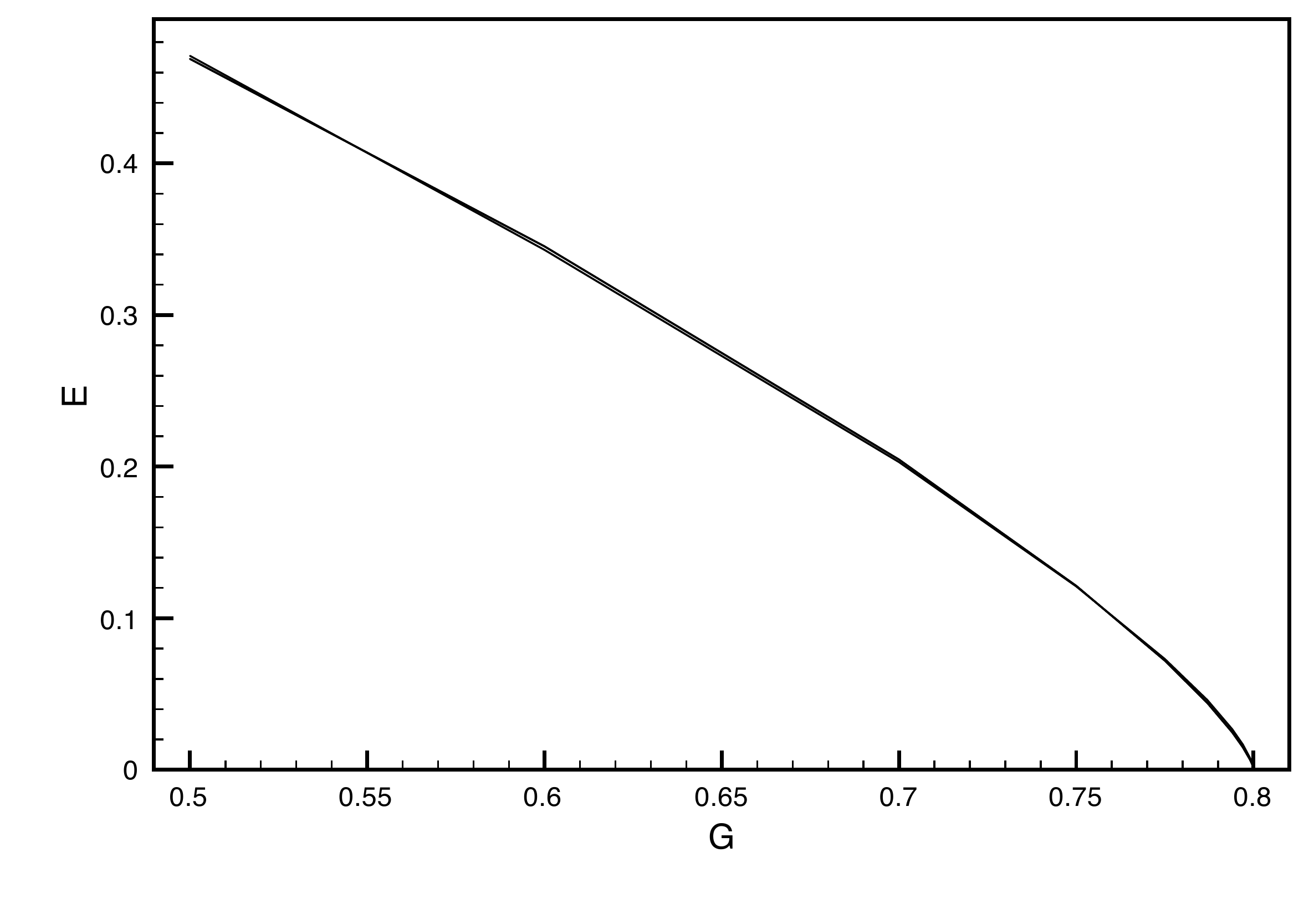}\end{center}
\caption{\small  Plot  of $E$   as a function of $G$.
Fit is to form $E_{0}(G^{*}-G)^{x_{E}}$ with values
$E_{0}=1.16$, $G^{*}=0.79992$ and $x_{E}=0.756$.}
\label{fig:31}
\end{figure}
\begin{figure}
\begin{center}\includegraphics[width=6in,scale=1.0]{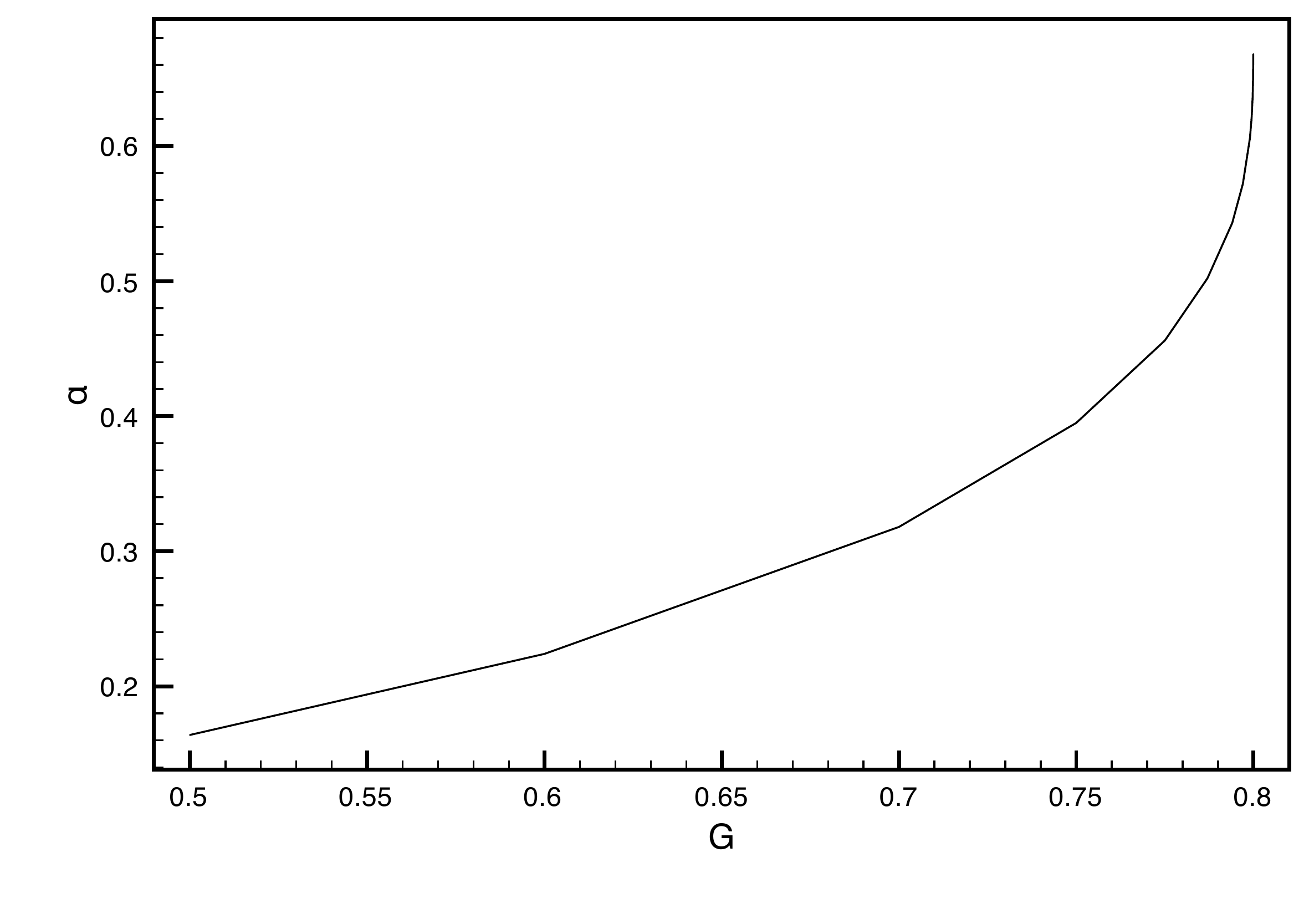}\end{center}
\caption{\small Plot of $\alpha$ as a function of $G$: One-loop case}
\label{fig:33}
\end{figure}

In the stable regime, $G < G^{*}$, the peak amplitude decay can be fit to the form
\be
\phi (t)=A_{0}\frac{e^{-Et}}{(t+t_{0})^{\alpha}}
~~~.
\label{eq:221}
\ee
In Figs.\ref{fig:30}, \ref{fig:31}, and \ref{fig:33} 
we plot the fit parameters $A_{0}$, $E$, and $\alpha$ as 
functions of $G$.

Numerically we find at the critical point, where $E=0$,
$A_{0}=0.51$, $t_{0}=0.185$, and
the exponent $\alpha =0.668$.  The analytic result, Eq.(\ref{eq:386}),
with $G^{*}=0.79992$ gives $A_{0}=0.508$. The analytic results 
agree with the numerical
results.  We can also compute the relaxation time $\tau$
as a function of $G$.  We obtain a very good fit to the data
with the form: $\tau =\tau_{0}/(G^{*}-G)^{x_{\tau_{1}}}$ with
$\tau_{0}=1.387$, $G^{*}=0.799925$ and $x_{\tau_{1}}=0.2197$.
For $G=G^{*}$ we find $\tau (t)=\tau_{1}t^{x_{\tau_{2}}}$
where $\tau_{1}=2.156$ and $x_{\tau_{2}}=0.2718$.

\subsection{Two-Loop Amplitude Contribution}

We now work out the results of the projection onto the structural peak
solution at two-loop order.
We begin with the two-loop expression for the dynamic part of the 
memory function in the structureless approximation in terms of the dimensionless
parameters introduced at one-loop order:
\be
\Gamma^{(4)}_{D}({\bf Q};T)=4\frac{g^{2}}{\tau^{2}}Q^{4}\tilde{C}
\int \frac{d^{d}K_{1}}{(2\pi )^{d}}
\frac{d^{d}K_{3}}{(2\pi )^{d}}K_{1}^{2}K_{3}^{2} \int_{0}^{T}dT_{1}
\int_{0}^{T_{1}}dT_{2}
\nonumber
\ee
\be
\times
f(K_{1},T-T_{1})
f(Q-K_{1},T-T_{2})
f(K_{3},T_{2})
f(Q+K_{3},T_{1})
f(-Q+K_{1}-K_{3},T_{1}-T_{2 })
\ee
We again assume a trial solution given by Eq.(\ref{eq:182})
where the unstable wavenumber $Q_{0}$ is time independent.
We
restrict the analysis here to three dimensions.   We need the
memory function evaluated at $Q=Q_{0}$ and
\be 
K^{(d,4)}(Q_{0},T)=4\frac{g^{2}}{\tau^{2}}Q^{4}_{0}J(Q_{0},T)
\ee
where
\be
J(Q_{0},T)=
 \int_{0}^{T}dT_{1}\int_{0}^{T_{1}}dT_{2}\phi (T-T_{1})
\phi (T-T_{2})\phi (T_{1})\phi (T_{2})\phi (T_{1}-T_{2})
 \int \frac{d^{3}K_{1}}{(2\pi )^{3}}
\frac{d^{3}K_{3}}{(2\pi )^{3}}K_{1}^{2}K_{3}^{2}
\nonumber
\ee
\be
\times \delta (K_{1}-Q_{0})\delta (|Q-K_{1}|-Q_{0})
\delta (K_{3}-Q_{0})
\delta (|Q+K_{3}|-Q_{0})\delta (|-Q+K_{1}-K_{3}|-Q_{0})
~~~.
\ee 
One can then do the integrations over $K_{1}$, $K_{3}$, 
$u_{1}=\hat{K}_{1}\cdot\hat{Q}$, $u_{3}=\hat{K}_{3}\cdot\hat{Q}$ and the
azimuthal angles with the result:
\be
J(Q_{0},T)
=\frac{Q_{0}^{5}}{\sqrt{2}}\frac{8\pi }{(2\pi )^{6}}\int_{0}^{T}dT_{1}\int_{0}^{T_{1}}dT_{2}
\phi (T-T_{1})\phi (T_{2})\phi (T-T_{2})\phi (T_{1})\phi  (T_{1}-T_{2})
~~~.
\ee
The contribution to the memory kernel at two-loop order is given by
\be
K^{(d,4)}(Q_{0},T)=4\frac{g^{2}}{\tau^{2}}
\frac{Q_{0}^{9}}{\sqrt{2}}\frac{8\pi }{(2\pi )^{6}}
\int_{0}^{T}dT_{1}\int_{0}^{T_{1}}dT_{2}
\phi (T-T_{1})\phi (T_{2})\phi (T-T_{2})\phi (T_{1})\phi  (T_{1}-T_{2})
~~~.
\ee

The two-loop peak-amplitude model is given by
\be
\frac{d}{dt}\phi (t) =-Q_{0}^{2}\phi (t) +\int_{0}^{t}dsN(t-s)\phi (s)
\ee
wiith the memory kernel
\be
N(t)=GQ_{0}^{4}\phi^{2} (t)+G_{1}Q_{0}^{8}\int_{0}^{t}dt_{1}\int_{0}^{t_{1}}dt_{2}
\phi (t-t_{1})\phi (t_{1})\phi (t-t_{2})\phi (t_{2})\phi (t_{1}-t_{2})
\ee
and we have the couplings
\be
G=\frac{gQ_{0}}{2\pi^{2}}
\ee
\be
G_{1}=4 g^{2}\frac{Q_{0}}{\sqrt{2}}\frac{8\pi}{(2\pi )^{6}}
=\frac{\sqrt{2}G^{2}}{\pi Q_{0}}
~~~.
\ee
We then rescale times $t=Q_{0}^{2}T$ and have
\be 
\frac{d}{dT}\phi (T) =-\phi (T) +\int_{0}^{T}dsN(T-S)\phi (S)
\ee
and
\be
N(T)=G\phi^{2} (T)+G_{1}\int_{0}^{T}dT_{1}\int_{0}^{T_{1}}dT_{2}
\phi (T-T_{1})\phi (T_{1})\phi (T-T_{2})\phi (T_{2})\phi (T_{1}-T_{2})
~~~.
\label{eq:295}
\ee
This model can be solved numerically, first we look at the power-law
solution at two-loop order.

\subsection{Power-Law Solution at Two-Loop Order}

We insert the trial solution (changing notation from $T$ to $t$)
\be
\phi (t)=\frac{A_{0}}{(t+t_{0})^{\alpha}}
\ee
into the two-loop contribution in Eq.(\ref{eq:295}) with the result:
\be
N^{(4)}(t)=G_{1}W(t)
\ee
where
\be
W(t)=A_{0}^{5}\int_{0}^{t}dt_{1}\int_{0}^{t_{1}}dt_{2}
(t-t_{1}+t_{0})^{-\alpha}(t_{1}+t_{0})^{-\alpha}
(t-t_{2}+t_{0})^{-\alpha}(t_{2}+t_{0})^{-\alpha}
(t_{1}-t_{2}+t_{0})^{-\alpha}
~~~.
\ee
After making the change of variables
$t_{1}=tx$, $t_{2}=ty$, $\epsilon =t_{0}/t$, then
\be
W(t)=A_{0}^{5}t^{2-5\alpha}\tilde{W}(\epsilon )
\ee
where
\be
\tilde{W}(\epsilon )=\int_{0}^{1}dx\int_{0}^{x}dy\frac{1}{(1-x+\epsilon )^{\alpha}}
\frac{1}{(x+\epsilon )^{\alpha}}
\frac{1}{(1-y+\epsilon )^{\alpha}}
\frac{1}{(y+\epsilon )^{\alpha}}
\frac{1}{(x-y+\epsilon )^{\alpha}}
~~~.
\ee
It is important to note that the exponent governing the long time
dependence is given by $2-5\alpha =2-5(2/3)=-4/3=-2\alpha$ and the two terms
contributing to $N(t)$ in Eq.(\ref{eq:295})
have the same power in time.  One can then expect that
$\tilde{W}(\epsilon )$ is logarithmic in $\epsilon$
as $\epsilon$ goes to zero.  A significant amount of work is needed to
show that
\be
\tilde{W}(\epsilon )
=W_{0}~ln~(2\epsilon)^{-1} +W_{1}
\ee
where
\be
W_{0}=2\frac{\Gamma^{2}(1/3)}{\Gamma  (2/3)}
\ee
and the constant $W_{1}$ could be worked out numerically.
The memory kernel is given for long times by
\be
N(t)=\frac{GA_{0}^{2}}{t^{4/3}}
+\frac{G_{1}A_{0}^{5}}{t^{4/3}}\left(W_{0}~ln~(2\epsilon)^{-1} +W_{1}\right)
\nonumber
\ee
\be
=\frac{GA_{0}^{2}}{t^{4/3}}\left(1+\delta ~ln~(2\epsilon)^{-1}+\ldots\right)
\ee
where
\be
\delta =\frac{G_{1}A_{0}^{3}W_{0}}{G}
~~~.
\ee
Exponentiating we have
\be
N(t)=\frac{GA_{0}^{2}}{(t+t_{0})^{4/3}}
\left(\frac{t+t_{0}}{2t_{0}}\right)^{\delta}
(1+\ldots )
\ee
We assume that this result will induce a change in the power-law
governing the peak amplitude
\be
\phi (t)=\frac{A_{0}}{(t+t_{0})^{2/3}}
\left(\frac{t+t_{0}}{2t_{0}}\right)^{\nu}
\ee
and we need to determine $\nu$. If $N(t)$ is characterized by the
exponent $\beta =4/3-\delta$, and $\phi (t)$ by 
$\tilde{\alpha}=2/3-\nu$.  Following the same steps as at one-loop
order we find that the exponents satisfy
\be
\tilde{\alpha}-1+\beta -1=0
\ee
gives the result
$\nu =-\delta$
and
\be
\tilde{\alpha}=2/3+\delta
~~~.
\ee
In  evaluating 
$\delta$ we need the results from the one-loop analysis:
$A_{0}^{3}=1/(2\pi\sqrt{3}G)$, the value of $Q_{0}$ with
$G_{1}=\sqrt{2} G^{2}/(\pi Q_{0})$.  One then finds
\be
\delta =4\frac{W_{0}}{(2\pi )^{2}\sqrt{6}Q_{0}}
=4\frac{\sqrt{2}\Gamma^{3}(1/3)}{(2\pi)^{3}Q_{0}}
=0.488\ldots
~~~.
\ee
Notice that this does no depend on the value of $G^{*}$.
The exponent is increased substantially in going from one to two-loop
order.  More importantly the two-loop theory serves as a controlled 
correction to the one-loop theory.

\subsection{Numerical analysis: Two-Loop Theory}

We can numerically solve the two-loop amplitude equation rather easily.  We
expect the analytic solution of the last section to hold at the critical
point.  We first determine the time $t^{*}(G)$, when in an unstable run,
$\phi (t)$ hits a
minimum and $A_{min}=\phi (t^{*})$. 
We find outstanding fits: $t^{*}=1.308/(G-0.7032)^{0.571}$
and $A_{min}=0.756 (G-0.7030)^{0.4705}$ which gives a good first
estimate for $G^{*}=0.7031\ldots$.  Next we work in the stable phase
and compute
\be
\tau (G)=\int_{0}^{\infty}dt~\phi (t)
\ee
as a function of $G$.  We find a very good fit to
$\tau =1.482/(0.703235 - G)^{0.1796}$ which gives an accurate
determination of $G^{*}=0.703235$.  We can then determine
$\phi (t)$ for $G=G^{*}$.  The resulting data can be fit to 
the form given by 
\be
\phi (t)=A_{0}\frac{e^{-Et}}{(t+t_{0})^{\alpha}}
\label{eq:551}
\ee
and we find the outstanding fit 
with $A_{0}=0.498$, $t_{0}=0.206$ , $\alpha =0.7377$, and
$E=-0.00019$.  The fit is over the time range $0\leq t\leq 2000$.
At $G=G^{*}$ we determine
\be
\tau (t)=\int_{0}^{t} dx \phi (x)=\tau_{0}t^{x_{\tau}}
\ee
where $\tau_{0} =1.040$ and $x_{\tau}=0.3399.$

The two-loop theory is very similar to the one-loop theory.
The analytic work suggests a larger shift in the exponent $\alpha$
than is found numerically.
One may need to use a self-consistent method to obtain more
quantitative analytical results.

\newpage

\section{Kawasaki Rearrangement}

\subsection{General Discussion}

We discuss here an approach, due to Kawasaki, which allows one to
reinterpret perturbation theory such that one obtains  an
ergodic-nonergodic transition at one-loop order.  After establishing
and exploring this result at one-loop order we investigate the stability
of this solution at two-loop order.

The kinetic equation for the Laplace transformed correlation function
$C(z)$ is given by (suppressing the wavenumber dependence in this
section)
\be
[z+K^{(s)}+K^{(d)}(z)]C(z)=\tilde{C}
\label{eq:118a}
\ee
where our convention for Laplace transforms is given by
\be
{\cal L}_{z}(C(t))=-i\int_{0}^{\infty}dt e^{izt}C(t)~~,
\ee
for convolutions we have
\be
{\cal L}_{z}(\int_{0}^{t}dsA(t-s)B(s))
=i{\cal L}_{z}(A(t)){\cal L}_{z}(B(t))
~~~.
\ee
and for time derivatives:
\be
{\cal L}_{z}(\dot{C}(t))=-i[zC(z)-\tilde{C}]
~~~.
\ee
With these results it is easy to see that the inverse Laplace transform
of Eq.(\ref{eq:118a}) is given by
\be
\dot{C}(t)-iK^{(s)}C(t)-\int_{0}^{t}ds~K^{(d)}(t-s)C(s)=0
~~~.
\label{eq:556}
\ee
Eq.(\ref{eq:556}) is not of the conventional mode coupling form.
Kawasaki\cite{KKA} suggested that the kinetic equation, 
Eq.(\ref{eq:118a}),
be rewritten in the form
\be
\left(z+\frac{K^{(s)}}{1+K^{(s)}N(z)}\right)C(z)=\tilde{C}
\label{eq:318}
~~~.
\ee
comparing with Eq.(\ref{eq:118a}) we can solve for $N(z)$  to obtain
\be
N(z)=-\frac{K^{(d)}(z)}{K^{(s)}(K^{(s)}+K^{(d)}(z))}
\nonumber
~~~.
\ee
If we define
\be
N_{0}(z)=-\frac{K^{(d)}(z)}{(K^{(s)})^{2}}
~~~,
\label{eq:127}
\ee
we can write
\be
N(z)=\frac{N_{0}(z)}{1-K^{(s)}N_{0}(z)}
~~~.
\label{eq:559}
\ee
Eq.(\ref{eq:118a}) can then be written in the form
\be
(1+K^{(s)}N(z))(zC(z)-\tilde{C})+K^{(s)}C(z)=0
~~~.
\label{eq:129a}
\ee
Taking the inverse Laplace transform gives
\be
\dot{C}(t)=-L_{0}C(t)-L_{0}\int_{0}^{t}ds~N(t-s)\dot{C}(s)
\label{eq:128}
\ee
where
\be
L_{0}=-iK^{(s)}
\ee
sets the time scale.
Eq.(\ref{eq:128}) is of the conventional mode-coupling form.

We can  develop 
perturbation theory in the dimensionless coupling $g$ as in the development
above.
First we determine $K^{(d)}(z)$ in a power series in $g$ as in
previous sections.  We insert this result in Eq.(\ref{eq:559}):
\be
N(z)=[N_{0}^{(2)}(z)+N_{0}^{(4)}(z)+\ldots ][1+K^{(s)}N_{0}^{(2)}(z)+\ldots ]
\nonumber
\ee
\be
=N^{(2)}(z)+N^{(4)}(z)+\ldots
\ee
with the lowest order approximation given by
\be
N^{(2)}(z)=N_{0}^{(2)}(z)
\ee
and 
at second order in $g$ we have
\be
N^{(4)}(z)=N_{0}^{(4)}(z)+K^{(s)}(N_{0}^{(2)}(z))^{2}
~~~.
\ee

\subsection{One-Loop bare theory}

At one-loop order the mode-coupling kernel is given by
\be
N(z)=
N_{0}(z)=-\frac{K^{(d,2)}(z)}{(K^{(s)})^{2}}
~~~.
\ee
In the time regime, putting in the wavenumber dependence,
\be
N_{0}(q,t)=\frac{\Gamma^{(2)}(q,t)}{q^{4}\bar{D}^{2}\chi^{2}(q)\tilde{C}(q)}
\ee
where
\be
\Gamma^{(2)}(q,z)=\frac{1}{2}D_{1}^{2}q^{4}\beta^{-2}
\int^{\Lambda} \frac{d^{d}k}{(2\pi )^{d}}
\frac{1}{z+i\bar{D}r[k^{2}+({\bf q}-{\bf k})^{2}]}
~~~.
\ee
In bare perturbation theory we have in the 
structureless approximation
\be
K^{(d,2)}(Q,T)=\frac{Q^{4}}{\tau^{2}}N_{0}(Q,T)
\ee
where
\be
N_{0}(Q,T)=g\int\frac{d^{d}K}{(2\pi )^{d}}
e^{-K^{2}T}e^{-({\bf Q}-{\bf K})^{2}T}
~~~.
\ee
In the absence of a cut-off this can be integrated to obtain
\be
N_{0}(Q,T)=ge^{-2Q^{2}T}
\frac{K_{d}}{2}\left(\frac{2}{T}\right)^{\frac{d}{2}-1}
\Gamma (d/2)
~~~.
\ee

\subsection{One-Loop self-consistent theory}

From the work above  we have the self-consistent result at one-loop
order:
 \be
 K^{(d,2)}(Q,T)
=Q^{4}\frac{1}{\tau^{2}}N_{0}(K,T)
\ee
where
\be
N_{0}(Q,T)=g\int\frac{d^{d}K}{(2\pi )^{d}}f(K,T)f({\bf Q}-{\bf K},T)
\ee
The kinetic equation in this case is given by
\be
\dot{ f}(Q,T)=-Q^{2}f (Q,T)
+Q^{2}\int_{0}^{T}dS N_{0}(Q,T-S)\dot{ f}(Q,S)
~~~.
\label{eq:577}
\ee
Consider the difference between the MCT expression, 
Eq.(\ref{eq:577}) and the direct solution given
by Eq.(\ref{eq:159}).  If we use
\be
\dot{f}=-Q^{2}f 
~~~,
\ee
valid at lowest order in $g$,
on the right-hand side of Eq.(\ref{eq:577}) we return
to Eq.(\ref{eq:159}).

\subsection{ENE Transition}

\begin{figure}
\begin{center}\includegraphics[width=6in,scale=1.0]{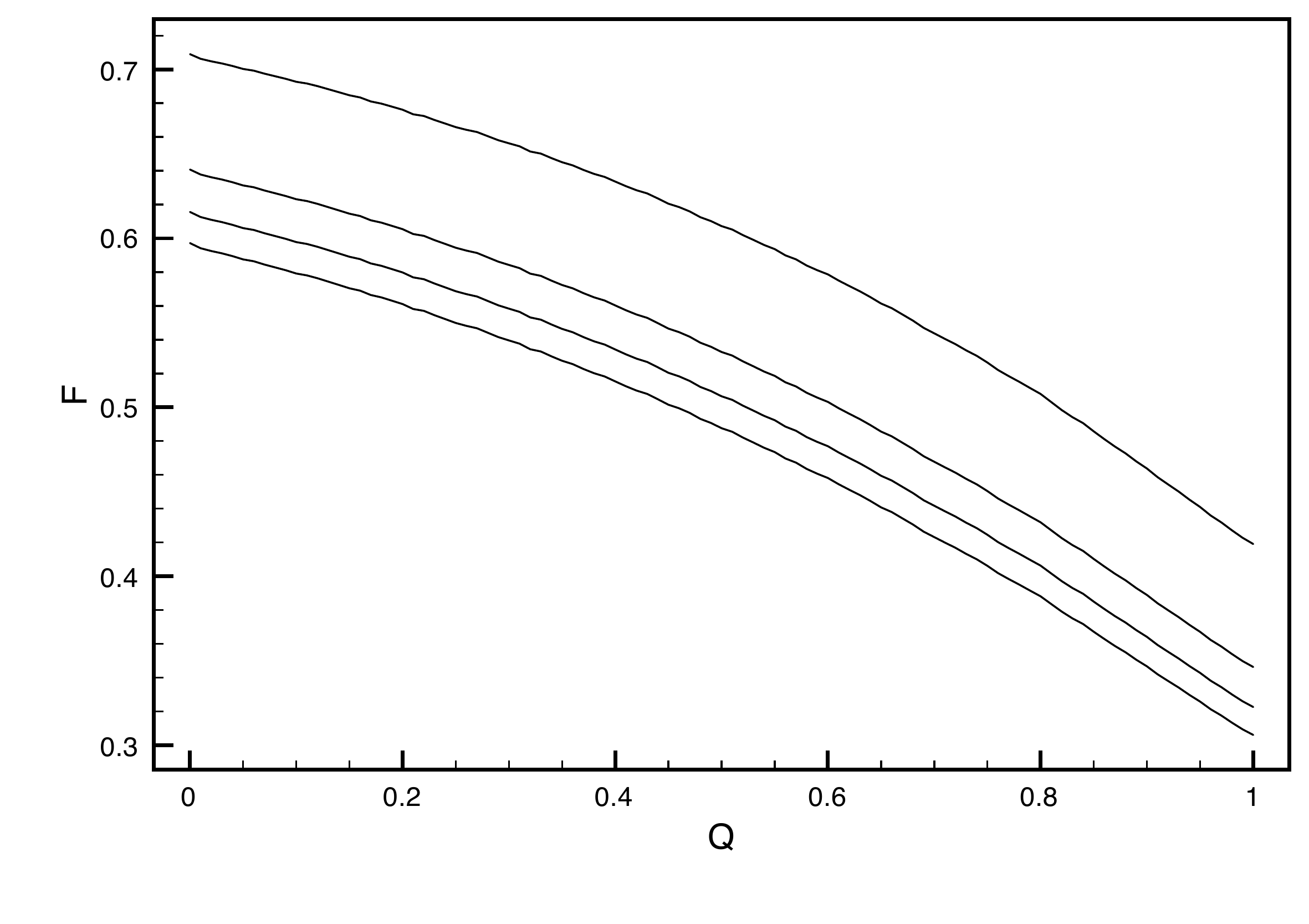}\end{center}
\caption{\small  Plot  of amplitude of nonergodic factor $F(Q)$ for 
$g=9.41, 9.43, 9.5, 10.0$ from the bottom.}
\label{fig:39}
\end{figure}

The solution for the non-ergodic phase can be separated out
as follows.
In Laplace transform space we have
Eq.(\ref{eq:318})at one-loop order and in terms of dimensionless variables:
\be
\left[z+\frac{iQ^{2}\tau^{-1}}{1+iQ^{2}\tau^{-1}N_{0}(Q,z)}\right]
f(Q,z)=1
\label{eq:579}
\ee
where
\be
N_{0}(Q,z)=-i\int_{0}^{\infty}dte^{izt}g
\int\frac{d^{d}K}{(2\pi )^{d}}f(K,t)f(Q-K,t)
~~~.
\ee
In the nonergodic phase to leading order for small $z$,
\be
f(Q,z)=\frac{F(Q)}{z}
\ee
and
\be
N_{0}(Q,z)=\frac{H(Q)}{z}
~~~.
\ee
Inserting these results into Eq.(\ref{eq:579})
and taking the small $z$ limit leads to the result:
\be
\left[1+\frac{1}{H(Q)}\right]F(Q)=1
\ee
and
\be
H(Q)=g
\int\frac{d^{d}K}{(2\pi )^{d}}F(K)F(Q-K)
\ee
This set of equations can be solved iteratively.  Using comparable
numerical methods as used to treat the direct approach, 
we can solve for $F(Q)$
with the results
as shown in Fig.\ref{fig:39}.  The critical coupling is given by 
$g_{mct}^{*}=9.41$.
Notice that the wavenumber dependence is monotomic.

\subsection{Two-loop theory}

At two-loop order we need the result
\be
N_{0}^{(4)}(z)=-\frac{K^{(d,4)}(z)}{(K^{(s)})^{2}}
\nonumber
\ee
\be
=\frac{\Gamma^{(4)}(z)}{q^{4}\bar{D}^{2}\chi^{2}\tilde{C}}
\label{eq:289}
\ee
where $\Gamma^{(4)}=\Gamma_{D}^{(4)}$ is given by Eq.(\ref{eq:205}).

In the nonergodic phase we have the result
\be
C(Q,\omega )=F(Q)\tilde{C}(Q) 2\pi \delta (\omega )+C_{R}(Q, \omega )
\ee
where $C_{R}$ is regular for small $\omega$.  Inserting this result
into Eq.(\ref{eq:289}) we have as a leading contribution for small
$z$:
\be
\Gamma_{D}^{(4)}(z)=\frac{\Gamma_{NE}^{(4)}}{z^{3}}
\ee
where $\Gamma_{NE}^{(4)}$ is independent of $z$.
This
leads to the result
\be
N_{0}^{(4)}=\frac{n_{4}}{z^{3}}
\ee
where $n_{4}$ is independent of $z$.
Checking order by order we have for the nonergodic phase:
\be
N^{(2)}(z)=N_{0}^{(2)}(z)=\frac{n_{2}}{z}
\ee
where
\be
N^{(4)}(z)=K^{(s)}\frac{n_{2}^{2}}{z^{2}}+\frac{n_{4}}{z^{3}}
~~~.
\ee
Clearly as $z\rightarrow 0$ the $N^{(4)}$ term dominates the
second-order term $N^{(4)}$.  Clearly the ENE transition is not a solution
at  two-loop order.

\newpage

\section{Conclusions}

We have introduced a simple dynamic model for a system under 
going diffusive dynamics with a density
dependent diffusion coefficient.  In the case where the diffusion 
coefficient has constant and linear times in the density, we
set up perturbation theory in terms of the coefficient of
the linear term.  For the dynamic structure factor we have worked out
the associated memory function to fourth order.  Analysis of this
perturbation theory led us to the following conclusions in the
simplest case where the static structure factor is a constant up
to a cutoff:

(1).  As one increases the dimensionless coupling one finds significant
slowing down.  The observed diffusion coefficient is not 
modified by higher-order terms in perturbation theory.
It decreases with increasing density if $D_{1}$ is negative. 
This gives a mechanism for making the coupling $g$ large.

(2).  For large enough coupling there is a transition where the
system goes from stable to unstable.

(3).  Near but below the transition, a slow Fourier component
appears that sharpens to a $\delta$-function but with an
algebraically decaying amplitude.

(4).  The sharpening of this structural peak corresponds to
a new length in the problem which grows algebraically with time.

(5).  Near, but above the transition, the system is metastable
with a slow increase with time of the amplitude of the peak.
Eventually the peak grows exponentially with time and the system
is rendered unstable.

(6).  The kinetics of the peak amplitude can be investigated
by assuming the peak can be approximated by a gaussian with a
narrowing width.  This leads to a zero-dimensional model
analogous to the Leutheussar model\cite{Leut} in MCT.  This model
can be studied analytically near the transition for both
one and two loop models.  Similarly this model can be studied
numerically.  The emerging picture of power-law decay near
the transition is consistent with the picture found for the
full field theory.

(7).  We show, for this model, that the ergodic-nonergodic transition,
supported at one-loop order is not a solution at two-loop order.

While we have worked out the perturbation theory for a general
static structure factor, we have explicit results for the
simplifying structureless approximation.  This 
corresponds to a coarse-grained model restricted to wavenumbers
below the first structure factor peak.  The resulting kinetic
model depends on a single dimensionless parameter the coupling
$g$.  At one-loop order, as we increased $g$, we found a critical
coupling $g^{*}=93.0\ldots$ which appears not to be a small
parameter!  However, when we look at two-loop corrections in
bare perturbation theory for small $Q$ and $z$, we find a correction,
compared to $1$, given by ${\cal C}_{d}g$ and in three dimensions
${\cal C}_{3}=0.00946\ldots$.  At the critical coupling this gives
a correction of $0.45$ which is acceptable.  One explanation for
the robustness of perturbation theory is that one could introduce
the effective coupling $g_{eff}=g/(2\pi)^{d}$ which corresponds
to a critical coupling $g_{eff}^{*}=0.45\ldots$.

This model is too simple to compare directly with experiment.
This is because one needs to include the physics at the length
scale of the structure factor maximum.  One then expects an
interplay between the mechanism discussed here which controls
the generation of a metastable structural peak and the peak in
the static structure factor.

The calculation here was carried out in equilibrium.  The same
techniques can be used to treat the associated nonequilibrium
quench problem.  Also, a similar calculation can be carried
out for models with density and momentum fields.  That case should
be interesting since the memory function is of the MCT form
without rearrangement. Finally this model is sufficiently
general, diffusion with field dependent diffusion coefficient,
that there should be additional applications beyond colloids.

Acknowledgements: This work was supported
by the National Science Foundation under Contract
No. DMR-0099324. 

\appendix 

\newpage	

\section{$R_{0}(z)\phi_{k_{1}}\phi_{k_{2}}\ldots \phi_{k_{n}}$}

In developing perturbation theory for time-correlation functions,
we need to work out the effect of the zeroth-order resolvant
operator acting on products of fields.  We need to evaluate
\be
W(12\ldots n)=R_{0}(z)\phi_{k_{1}}\phi_{k_{2}}\ldots \phi_{k_{n}}
~~~.
\ee  
We determine this quantity using the identity
\be
zW(12\ldots n)=\phi_{k_{1}}\phi_{k_{2}}\ldots \phi_{k_{n}}
-R_{0}(z)i\tilde{D}_{\phi}^{(0)}\phi_{k_{1}}\phi_{k_{2}}\ldots \phi_{k_{n}}
~~~.
\label{eq:A2}
\ee
It is not difficult to show that
\be
\tilde{D}_{\phi}^{(0)}\phi_{k_{1}}\phi_{k_{2}}\ldots \phi_{k_{n}}
=\sum_{i=1}^{n}L_{0}(i)\phi_{k_{1}}\phi_{k_{2}}\ldots \phi_{k_{n}}
-\hat{S}_{P}\left(\gamma (12)\phi_{k_{2}}\ldots \phi_{k_{n}}\right)
\label{eq:A3}
\ee
where $L_{0}(1)=L_{0}({\bf k}_{1})$ is defined by Eq.(\ref{eq:84}) and
\be
\gamma (12)=
2\beta^{-1}\Gamma_{0}({\bf k}_{1},{\bf k}_{2})=-2\beta^{-1}
D_{0}{\bf k}_{1}\cdot{\bf k}_{2}(2\pi )^{d}
\delta ({\bf k}_{1} +{\bf k}_{2})
\nonumber
\ee
\be
=(L_{0}(1)+L_{0}(2))\tilde{C}(12)
~~~,
\ee
and $\hat{S}_{P}$ is an operator which symmetrizes the product
it acts on such that $\gamma (ij)$ appears with all possible
pairs.  Using Eq.(\ref{eq:A3}) in Eq.(\ref{eq:A2}) gives
\be
zW(12\ldots n)=\phi_{k_{1}}\phi_{k_{2}}\ldots \phi_{k_{n}}
-i\sum_{i=1}^{n}L_{0}(i)W(12\ldots n)
+\hat{S}_{P}\left(i\gamma (12)W(34\ldots n)\right)
~~~.
\ee
This can be put in the form
\be
W(12\ldots n)=T_{0}(12\ldots n)
\left[\phi_{k_{1}}\phi_{k_{2}}\ldots \phi_{k_{n}}
+\hat{S}_{P}\left(i\gamma (12)W(34\ldots n)\right)\right]
\ee
where
\be
T_{0}(12\ldots n)=\frac{1}{[z+i\sum_{i=1}^{n}L_{0}(i)]}
~~~.
\ee
This allows the $W's$ to be determined recursively.  We need
$W(1)$ through $W(1234)$:
\be
W(1)=T_{0}(1)\phi (k_{1})
\ee
\be
W(12)
=T_{0}(12)[\phi_{k_{1}}\phi_{k_{2}}
-\tilde{C}({\bf k}_{1},{\bf k}_{2})]
+\frac{\tilde{C}({\bf k}_{1},{\bf k}_{2})}{z}
~~~.
\label{eq:172}
\ee
\be
W(123)=
T_{0}(123)[
\phi_{k_{1}}\phi_{k_{2}}\phi_{k_{3}}
-\tilde{C}(23)\phi_{k_{1}}
-\tilde{C}(13)\phi_{k_{2}}
-\tilde{C}(12)\phi_{k_{3}}]
\nonumber
\ee
\be
+[T_{0}(1)
\phi_{k_{1}}\tilde{C}(23)
+T_{0}(2)
\phi_{k_{2}}\tilde{C}(13)
+T_{0}(3;z)
\phi_{k_{3}}\tilde{C}(12)]
\ee
\be
W(1234)=
T_{0}(1234)
[\phi_{k_{1}}\phi_{k_{2}}\phi_{k_{3}}\phi_{k_{4}}
-\langle \phi_{k_{1}}\phi_{k_{2}}\phi_{k_{3}}\phi_{k_{4}}\rangle
]
\nonumber
\ee
\be
+B_{s}(1234)]
+\frac{\langle \phi_{k_{1}}\phi_{k_{2}}\phi_{k_{3}}\phi_{k_{4}}\rangle}{z}
\ee
where
\be
B_{s}(1234)=B(12,34)+B(13,24)+B(14,23)+
=B(23,14)+B(24,13)+B(34,12)
\ee
and
\be
B(12,34)=(T_{0}(34)-T_{0}(1234))\tilde{C}(12)
\left[\phi_{k_{3}}\phi_{k_{4}}-\tilde{C}(34)\right]
~~~.
\ee

\section{$\tilde{D}_{\phi}^{(I)}\phi({\bf k}_{1})\phi({\bf k}_{2})$}

We need to evaluate
\be
i\tilde{D}_{\phi}^{I}\phi({\bf k}_{1})\phi({\bf k}_{2})
\nonumber
\ee
\be
=i\int d^{d}x \int d^{d}y
\left[\frac{\delta {\cal H}_{\phi}}{\delta \phi ({\bf x})}
-k_{B}T\frac{\delta}{\delta \phi ({\bf x})}\right]
\Delta\Gamma_{\phi}({\bf x},{\bf y})
\frac{\delta }{\delta \phi ({\bf y})}
\phi({\bf k}_{1})\phi({\bf k}_{2})
\nonumber
\ee
\be
=v^{(I)}({\bf k}_{1})\phi({\bf k}_{2})
+v^{(I)}({\bf k}_{2})\phi({\bf k}_{1})+S({\bf k}_{1},{\bf k}_{2})
\ee
where
\be
S({\bf k}_{1},{\bf k}_{2})=-i\int d^{d}x \int d^{d}y
k_{B}T\Delta\Gamma_{\phi}({\bf x},{\bf y})
\frac{\delta}{\delta \phi ({\bf x})}\frac{\delta }{\delta \phi ({\bf y})}
\phi({\bf k}_{1})\phi({\bf k}_{2})
\nonumber
\ee
\be
=i\int d^{d}x \int d^{d}y k_{B}T\Delta\Gamma_{\phi}({\bf x},{\bf y})
[e^{i{\bf k}_{1}\cdot {\bf x}}e^{i{\bf k}_{2}\cdot {\bf y}}+
e^{i{\bf k}_{2}\cdot {\bf x}}e^{i{\bf k}_{1}\cdot {\bf y}}]
\nonumber
\ee
\be
=i\int d^{d}x \int d^{d}y k_{B}TD_{1}\delta\phi ({\bf x})
\delta ({\bf x}-{\bf y})2{\bf k}_{2}\cdot{\bf k}_{1}
e^{i({\bf k}_{1}+{\bf k}_{2})\cdot {\bf x}}
\nonumber
\ee
\be
=2i\beta^{-1}
D_{1}{\bf k}_{1}\cdot{\bf k}_{2}\phi({\bf k}_{1}+{\bf k}_{2})
\nonumber
\ee
\be
=\int \frac{d^{d}k_{3}}{(2\pi )^{d}}
\tilde{S}({\bf k}_{1},{\bf k}_{2},{\bf k}_{3})\phi ({\bf k}_{3})
\ee
and
\be
\tilde{S}({\bf k}_{1},{\bf k}_{2},{\bf k}_{3})=
2i\beta^{-1}D_{1}{\bf k}_{1}\cdot{\bf k}_{2}
(2\pi )^{d}\delta ({\bf k}_{3}-{\bf k}_{1}-{\bf k}_{2})
~~~.
\ee

\section{Integrals $J_{0}(2/3)$ and $\sigma_{0} (2/3)$}

Consider the integrals
\be
\sigma_{0} (\alpha )=\int_{0}^{\infty}\frac{dy}{y^{\alpha}}e^{iy}
\ee
and
\be
J_{0} (\alpha )=\int_{0}^{\infty}\frac{dy}{y^{2\alpha}}(e^{iy}-1)
~~~.
\ee
The second integral can be related to the first via integration by parts:
\be
J_{0} (\alpha )
=\frac{i}{(2\alpha -1)}\sigma_{0} (2\alpha -1)
~~~.
\ee
We have from  Dwight integrals 858.562 and 858.563\cite{dwight}
\be
\sigma_{0} (\alpha )=\left[\frac{\pi}{2\Gamma (\alpha )}\right]
\left[\frac{1}{cos(\alpha\frac{\pi}{2})}
+i\frac{1}{sin(\alpha\frac{\pi}{2})}\right]
\nonumber
\ee
\be
=\left[\frac{\pi i}{\Gamma (\alpha )sin(\alpha \pi )}\right]
e^{-i\alpha\frac{\pi}{2}}
~~~.
\ee
Then
\be
J_{0}(\alpha )\sigma_{0} (\alpha )
=-\frac{i\pi^{2}}{2\alpha -1}\frac{1}{\Gamma (2\alpha -1 )\Gamma (\alpha )}
\frac{e^{-i\frac{\pi}{2}(3\alpha -1}}{sin((2\alpha -1) \pi )sin(\alpha  \pi )}
~~~.
\ee
For the relevant case $\alpha =2/3$, we have
\be
J_{0}(2/3 )\sigma_{0} (2/3 )= -\frac{3\pi^{2}}{\Gamma (1/3)\Gamma (2/3)}
\frac{1}{sin(\pi/3) sin(2\pi/3)}
~~~.
\ee
Using  $sin (\pi/3)=\sqrt{3}/2$, $sin (2\pi/3)=\sqrt{3}/2$, and
\be
\Gamma (1/3)\Gamma (2/3)=\frac{\pi}{sin (\pi/3)}=\frac{2\pi}{\sqrt{3}}
~~~,
\ee
we have finally
\be
J_{0}(2/3 )\sigma_{0} (2/3 )= -2\pi\sqrt{3}
~~~.
\ee


\begin{thebibliography}{99}
\bibitem{loop}  In the standard MSR field-theoretic description 
perturbation theory is naturally  carried out in terms of
Feynman diagrams with a loop structure.  One loop order is second order 
in $D_{1}$ while two-loop order is of fourth order.
\bibitem{Scom}  This requires a certain amount of coarse graining since $S$ is infinite 
in the microscopic case.
\bibitem{1PI} Due to the one-particle irreduceable property of the
memory function, we can replace $v\rightarrow ~v^{I}$ in
Eqs.(\ref{eq:19}) and (\ref{eq:21}).
\end{thebibliography}
\end{document}